\documentclass[preprint,12pt]{elsarticle}

\usepackage{amssymb}
\usepackage{amsmath}

\usepackage[table,xcdraw]{xcolor}
\usepackage{multirow}
\usepackage{titlesec}
\usepackage{algorithm}
\usepackage{algpseudocode}
\usepackage{cancel}
\usepackage[normalem]{ulem}
\usepackage[acronym]{glossaries}
\usepackage{calc}
\newsavebox\CBox
\newcommand\hcancel[2][0.5pt]{%
  \ifmmode\sbox\CBox{$#2$}\else\sbox\CBox{#2}\fi%
  \makebox[0pt][l]{\usebox\CBox}%
  \rule[0.5\ht\CBox-#1/2]{\wd\CBox}{#1}}

\titleformat{\paragraph}
{\normalfont\normalsize\bfseries}{\theparagraph}{1em}{}
\titlespacing*{\paragraph}
{0pt}{3.25ex plus 1ex minus .2ex}{1.5ex plus .2ex}
\journal{Applied Soft Computing}

\begin{document}

\begin{frontmatter}



\title{Generalizing Supervised Deep Learning MRI Reconstruction to Multiple and Unseen Contrasts using Meta-Learning Hypernetworks}

\author[iitm,htic]{Sriprabha Ramanarayanan}
\author[iitm,htic]{Arun Palla}
\author[htic]{Keerthi Ram}
\author[iitm,htic]{Mohanasankar Sivaprakasam
}
\affiliation[iitm]{organization={Department of Electrical Engineering, Indian Institute of Technology Madras (IITM)},
            country={India}}

\affiliation[htic]{organization={Healthcare Technology Innovation Centre},
            addressline={IITM}, 
            country={India}}




\begin{abstract}
Meta-learning has recently been an emerging data-efficient learning technique for various medical imaging operations and has helped advance contemporary deep learning models. Furthermore, meta-learning enhances the knowledge generalization of the imaging tasks by learning both shared and discriminative weights for various configurations of imaging tasks during training. However, existing meta-learning models attempt to learn a single set of weight initializations of a  neural network that might be fundamentally restrictive under the heterogeneous (multimodal) data scenario. This work aims to develop a multimodal meta-learning model for image reconstruction, which augments meta-learning with evolutionary capabilities to encompass diverse acquisition settings of heterogeneous data. Our proposed model called KM-MAML (\textbf{K}ernel \textbf{M}odulation-based \textbf{M}ultimodal \textbf{M}eta-\textbf{L}earning), has hypernetworks (auxiliary learners) that evolve to generate mode-specific (or context-specific) weights. These weights provide the mode-specific inductive bias for multiple modes by re-calibrating each kernel of the base network for image reconstruction via a low-rank kernel modulation operation. Furthermore, we incorporate gradient-based meta-learning (GBML) in the contextual space to update the weights of the hypernetworks based on different modes. The hypernetworks and the base reconstruction network in the GBML setting provide discriminative mode-specific features and low-level image features, respectively. We extensively evaluate our model for multi-contrast magnetic resonance image reconstruction considering the essential research directions in fastMRI for multimodal and rich transfer learning capabilities across various MRI contrasts. Our comparative studies show that the proposed model (i) exhibits superior reconstruction performance over joint training, other meta-learning methods, and various context-specific MRI reconstruction architectures, and (ii) better adaptation to 80\% and 92\% of unseen multi-contrast data contexts with improvement margins of 0.1 to 0.5 dB in PSNR and around 0.01 in SSIM, respectively.
Besides, a representation analysis with U-Net as the base network shows that kernel modulation infuses 80\% of mode-specific representation changes in the high-resolution layers. Our source code is available at https://github.com/sriprabhar/KM-MAML/. 
\end{abstract}


\begin{graphicalabstract}
\centering
\includegraphics[width=0.9\linewidth]{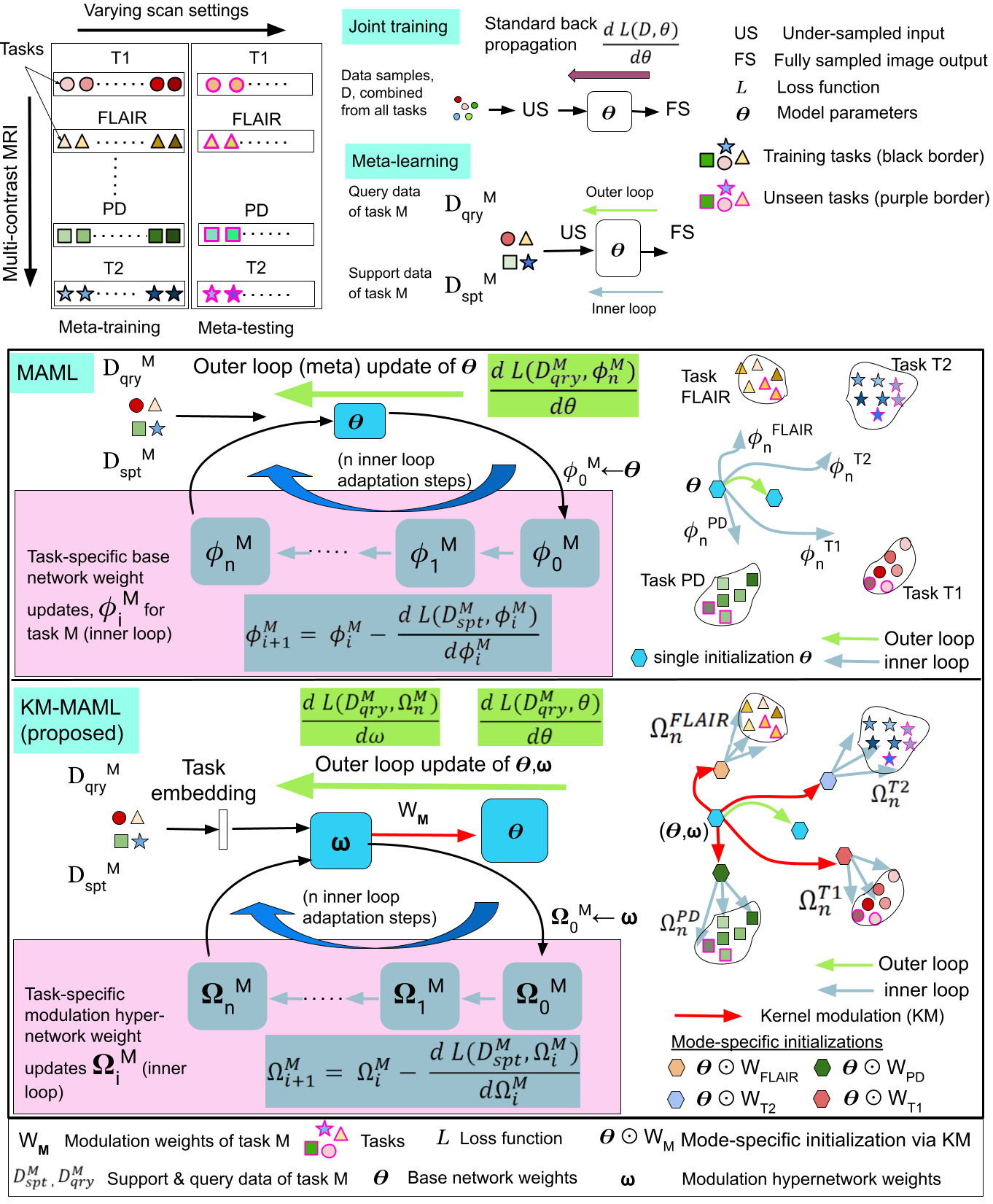}
\end{graphicalabstract}

\begin{highlights}
\item A multi-modal meta-learning model for image reconstruction that provides mode-specific inductive bias closer to the target data distribution with deviated acquisition settings.
\item Provides evolutionary capabilities of hypernetworks to dynamically predict weights that modulate the base reconstruction network weights for multimodal image reconstruction.
\item Gradient-based meta-learning to optimize the kernel modulation hypernetworks in the contextual space.
\item Extensive experimentation for multi-contrast MRI reconstruction to showcase the superior adaptation capabilities on-the-fly and via fine-tuning in a few gradient steps to unseen multi-contrast datasets.
\item Superior reconstruction performance over joint training, other meta-learning methods, and various context-specific  MRI reconstruction networks.
\item Representational analysis showing that kernel modulation induces maximum mode-specific features in the high-resolution layers of the encoder-decoder base network.
\end{highlights}

\begin{keyword}
hypernetwork \sep evolution \sep meta-learning \sep kernel modulation \sep mode-specific inductive bias \sep multi-contrast MRI reconstruction
\end{keyword}

\end{frontmatter}


\section{Introduction}
\label{sec:intro}
Deep learning methods have seen remarkable improvements in various imaging tasks like image classification, recognition, and reconstruction. The success of these methods has vastly been in scenarios where the model is trained and tested on a homogeneous dataset representing a specific concept like similar objects, characteristic features, image structures, and contrast levels. The homogeneous conditions across training and test data limit the robustness of the model when the image data deviates due to shifts in scan settings and contrasts \cite{ontheflytta}. For instance, in diverse medical imaging systems like magnetic resonance imaging (MRI), devising robust models that generalize to multi-scanner data is crucial for transferring these models into clinical practice \cite{meta_mihi}. Recent advances aim to improve the robustness of the model by exploring in two promising directions \cite{multimodameta}, (i) At the data level, obtaining heterogeneous modalities (multimodal image data) with multiple imaging settings to learn from diverse data (ii) At the model level, incorporating adaptive learning mechanisms to improve the generality of the model to varying imaging conditions. 
To handle challenging scenarios with multi-modal data distributions encompassing diverse acquisition settings, an adaptive learning model that can provide the shared knowledge and discriminative features of varied modalities is required \cite{fastnflexible, controllable, gnldecouple}.

\begin{figure}[t!]
    \centering
    \includegraphics[width=0.9\linewidth]{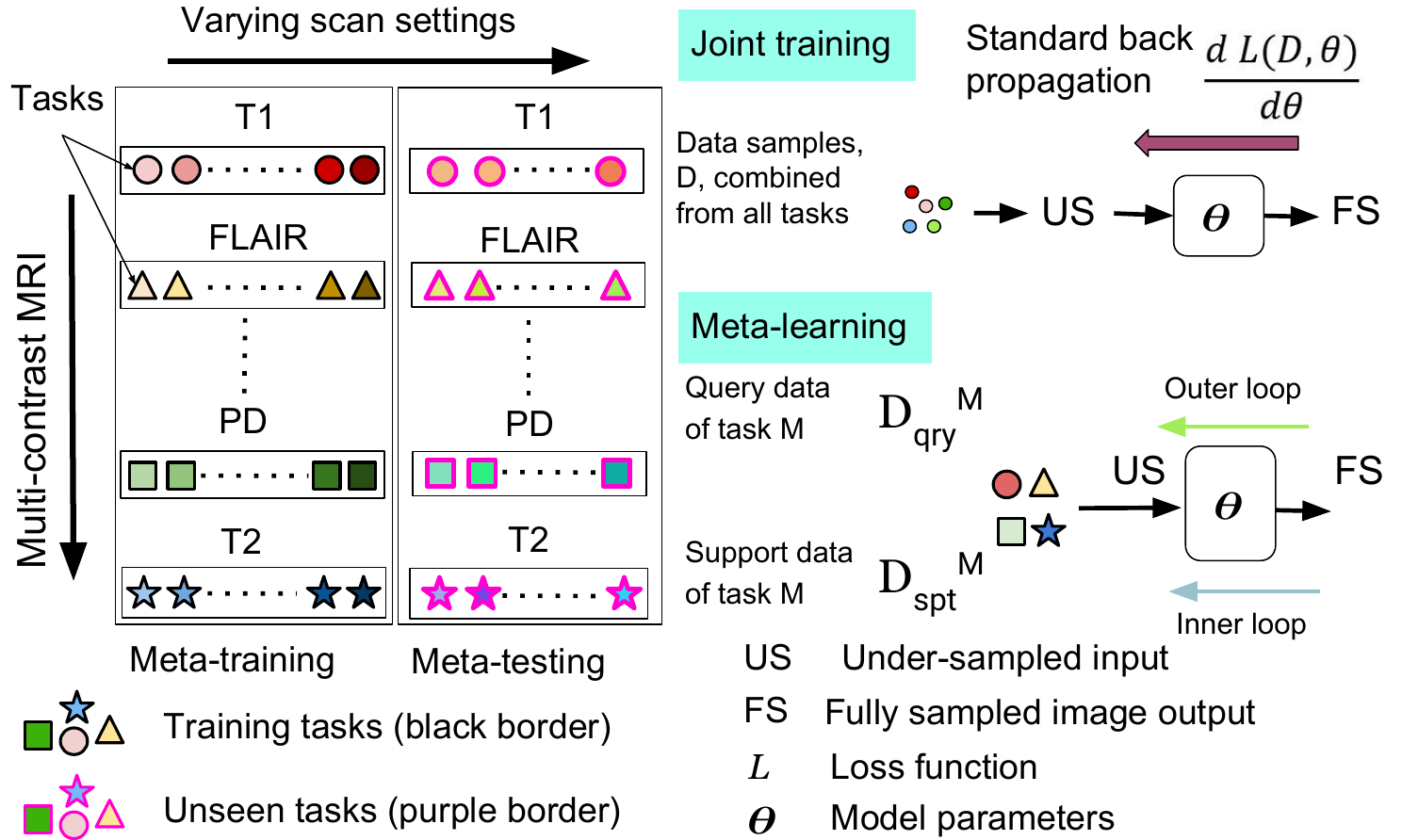}
    \caption[Short caption]{
    Graphical abstract of MRI data sources and learning methods (best viewed in color). Left: The multimodal MRI data sources (datasets), each represented by various shapes and colors. Each row corresponds to a specific mode of the multimodal MRI data and covers a specific contrast, with different acquisition settings like different types and amounts of under-sampling degradation. Right: Joint training has a single level of optimization of the base network by combining the training samples from the datasets. Meta-learning has inner (gray arrow) and outer (green arrow) levels of optimization. We categorize each dataset corresponding to each contrast and scan setting as a training task for the neural network. Each task consists of support and query partitions of the dataset. The inner and outer levels of optimization use the support and query partitions of each task data. 
    } 
    \label{fig:multimodaldatasource}
\end{figure}

\begin{figure}[t!]
    \centering
    \includegraphics[width=0.97\linewidth]{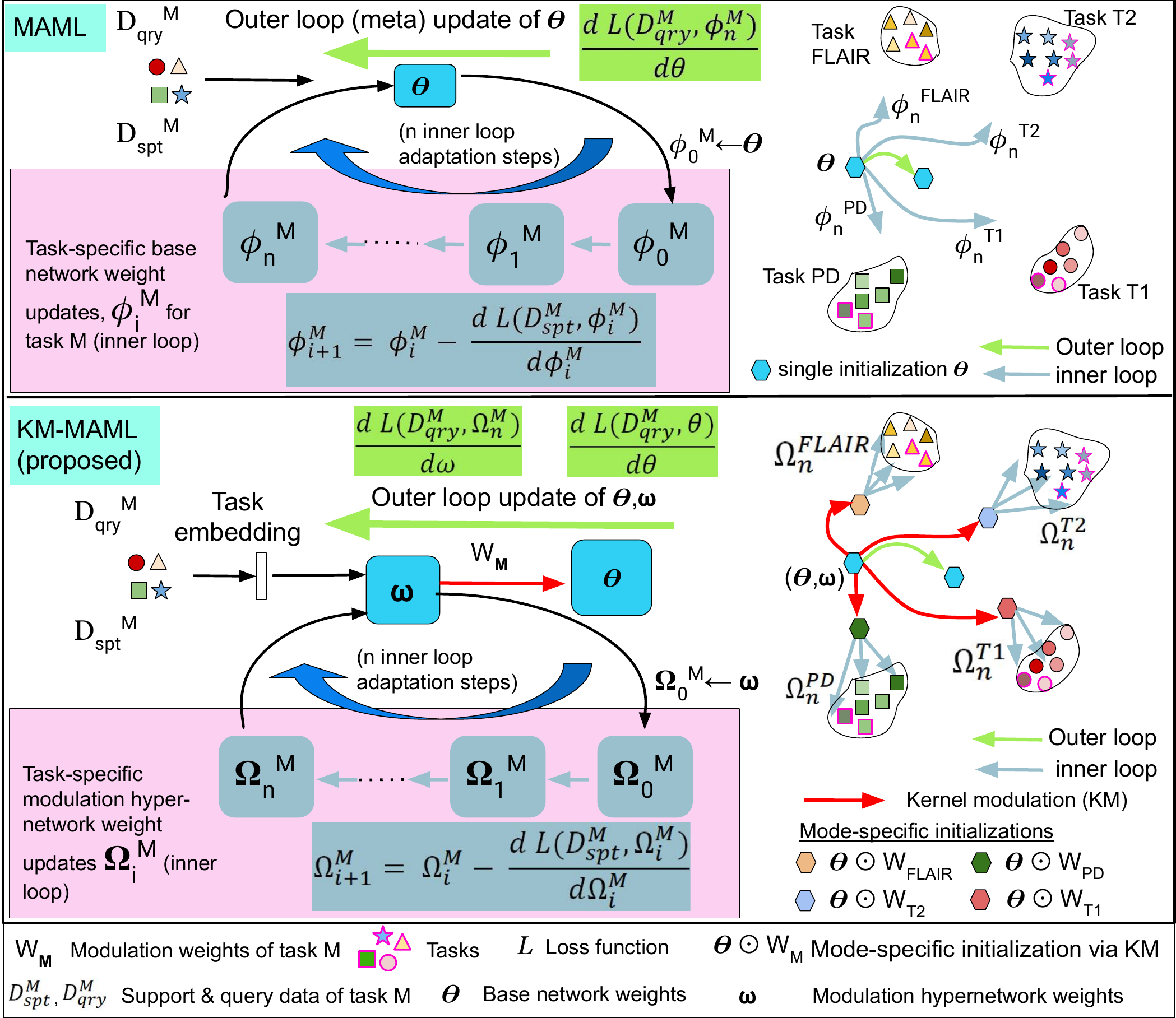}
    \caption[Short caption]{
    Concept diagram of 
    the bi-level optimization of MAML and KM-MAML (best viewed in color). Left: MAML involves bi-level optimization (gray arrows for the inner loop and green arrows for the outer loop) of the base learner (blue box). Different from MAML, KM-MAML involves two networks - a base learner for the imaging task and an auxiliary network called the kernel modulation (KM) network to infer mode-specific weights. The two levels of optimization are 1) outer loop or meta-weight updates of the base network $\theta$ and the KM network $\omega$ shown as green arrows backward. 2) The inner loop or task-specific adaptation via a few gradient updates of the KM network (gray arrows within the pink box). The meta-parameters of the KM network infuse mode-specific knowledge to the base learner via kernel modulation to create an improved base learner. (Right) Weight update process in the task space. KM-MAML provides mode-specific initializations (red arrows pointing to multi-colored hexagons) that coarsely capture the target MRI data distribution, unlike MAML, which has a single meta-initialization. Fine-tuning the base network or the hypernetwork by a few gradient steps (gray arrows) refines the model closer to the target data distribution (Algorithm \ref{kMAMLalgo} shows the training details).
    } 
    \label{fig:graphabsmaml}
\end{figure}

Multimodal image data refers to the outputs of image acquisition devices (like sensors or acquisition technologies) with multiple intensity representations of visual concepts \cite{multimodameta} (Figure \ref{fig:multimodaldatasource}(left)).  
Integrating multiple perspectives of concepts using multimodal data enhances the knowledge generalization of the concept by learning a shared representation across modalities, eliminating models trained independently for each modality \cite{univusmri, mac}. The learning approach commonly adopted to integrate multimodal data is joint training using stochastic gradient descent (SGD) (Figure \ref{fig:multimodaldatasource} (right)), Joint training). However, due to statistical shifts across various modes \cite{univusmri}, the shared knowledge gained by joint training from image samples alone might be inadequate and could underfit modalities with deviations in the acquisitions settings from the training data \cite{univusmri, fastnflexible}. Secondly, the learning process does not consider discriminative features, meaning that it lacks an adaptive mechanism of learning unique sets of weights for each modality and encapsulating relationships across modalities in a common weight space \cite{gnldecouple}.

Recently, meta-learning methods \cite{metaperspective} have emerged as a data-efficient adaptive learning technique for regression and classification methods, both at the task (a unique mode) and the sample level. Model-agnostic meta-learning (MAML) \cite{maml} learns a common prior knowledge across tasks and the task-specific parameters of a network in a bi-level interleaved optimization process (Figure \ref{fig:multimodaldatasource} (right bottom)), enabling faster adaptation to new tasks using the learned prior (Figure \ref{fig:graphabsmaml}, MAML). MAML outperforms conventional learning \cite{metalinconceptspace}; however, when the data distribution of tasks is multimodal, the knowledge gained by MAML is fundamentally restricted by a single initialization that is learned by a single  (base) network \cite{caml}  performing an imaging operation. For instance, it would be infeasible to seek a common initialization for an imaging task with continuously varying acquisition settings. 

Within the premises of heterogeneous data, the data samples occupy different regions in the same high dimensional space as clusters that correspond to the modes of the data. However, the clusters might not lie closer to each other as the degree of similarity between them might be vastly different. Due to variations in the acquisition settings, different modes are endowed with complementary properties that might contribute differently to the learning process. 
This motivates the need for adaptive context-aware (or mode-specific) initializations to provide reliable and selective meta-knowledge closer to the target data distribution with deviated acquisition settings.
We consider the problem of image restoration using a single model that can scale to multiple acquisition settings of multi-modal data by generating mode-specific initializations for rapidly adapting to unseen data.

Task-aware modulation of the model weights using Multi-modal meta-learning (MMAML) \cite{mmaml} is a promising direction that infers mode-specific latent representations to capture the features corresponding to the modes of heterogeneous data distribution.
Our method uses auxiliary networks to execute mode-specific modulation of the weights of a base convolutional neural network (CNN) using the MMAML approach, in contrast to adaptive strategies that adjust the activations of the base CNN. In order to improve upon MMAML, our focus lies on how optimally the two networks learn together with two objectives. The objective of the auxiliary networks is to provide the high-level inductive bias \cite{maml} of multiple modes of heterogeneous data, and the objective of the base network is to solve the imaging tasks. Unlike previous MMAML methods \cite{mmaml, caml}, which infuse adaptive mechanism only on the base network via meta-learning, our approach infuses adaptive learning on both the networks at the model and optimization levels to achieve the two objectives.

(i) \textit{Model level: }Inspired by the evolutionary computing \cite{ecipsurvey,eccompsurvey} capabilities of weight-learning networks \cite{metal_survey,mac,gnldecouple}, we propose adaptive hypernetworks \cite{hypernetworks}, called kernel modulation (KM) hypernetworks, consisting of a low-rank KM layer to provide mode-specific inductive bias to the base CNN. The human brain exhibits attentional modulation, wherein modulatory signals representing behavioral goals evolving in brain regions are sent to sensory neurons in the cortex for perception and contextual modulation \cite{neuralbasis}. Similarly, the proposed KM hypernetworks evolve to generate different weights to modulate kernels in the base network for dynamically re-calibrating the behavior of the base network.

(ii) \textit{Optimization level:} We categorize the hypernetworks' weights as context-based, and the base network's weights as image feature-based. The KM hypernetworks are optimized using bi-level gradient-based meta-learning inspired by the learning-to-learn \cite{metal_survey} approach in the contextual space. The base network is optimized via both gradient updates, and mode-specific kernel modulation parameters from hypernetworks \cite{hypermaml} (Figure \ref{fig:graphabsmaml} KM-MAML). At test time, the base network is adapted on-the-fly or fine-tuned using KM for multi-modal imaging tasks.

We demonstrate the efficacy of our approach in multi-contrast magnetic resonance image (MC-MRI) reconstruction \cite{multicontrastcmpb1}. MC-MRI captures diverse and complementary perspectives of a single subject using multiple MRI contrasts, each representing a mode. 
As charted out in the fastMRI reconstruction challenge results \cite{fastmri}, the pre-existing MC-MRI deep learning models \cite{disn, dudor, dsformer} lack scalability to different contrasts and exhibit inadequate transfer capabilities across MRI modalities. We take a kernel modulation-based meta-learning approach to identify the modes of each MRI contrast and scale to multiple contrast-specific models.
We summarize our contributions as,

\begin{enumerate}
\item We propose a meta-learning model called KM-MAML, that learns via context-aware kernel modulation for multimodal image reconstruction. The proposed model has a base reconstruction network and layer-wise hypernetworks, which dynamically modulate each kernel of a base reconstruction network based on the mode of the multimodal data.



\item The proposed model is optimized with two training objectives, extracting (i) the mode-specific inductive bias from hypernetworks by adopting learning-to-learn in the contextual space of heterogeneous data and (ii) low-level image features from the base network. 


\item Extensive experimentation on MC-MRI reconstruction shows that the proposed model provides superior reconstruction performance over other learning methods, generality to 80\% and 92\% of unseen multi-contrast contextual settings in PSNR and SSIM, with improvement margins of around 0.1 dB in PSNR and 0.01 in SSIM, respectively. Our approach also matches the performance of various context-specific MRI reconstruction architectures.

\item Our analysis of the representational similarity between pre and post kernel modulation features shows that with U-Net as the base network, kernel modulation induces maximum (of 80\%) discriminative mode-specific representations in the top (highest resolution) layers of the encoder and decoder.

\end{enumerate}

This paper is organized as follows. Section \ref{sec:rel} describes the related works. Section \ref{sec:method} provides the material and methods with algorithm and architecture details. Section \ref{sec:expr} and \ref{sec:modelcompare} provides the dataset, implementation details, and results. Section \ref{sec:concl} briefly summarizes our findings and conclusion.

 \section{Related Work}
\label{sec:rel}
The concept of meta-learning is originally developed by the learning-to-learn \cite{metal_survey, metaperspective, meta-nc} approach which is further extended using evolutionary computing (EC) \cite{multiobjec, ES-MAML, shipmeta, eccompsurvey} methods to learn the rules. 
The emerging directions in multimodal learning have shown that the training data acquired from one modality can benefit from the shared knowledge across different modalities \cite{visualqa,imgcaptioning}. The focus of multimodal meta-learning is to effectively learn a mode-specific prior using cross-modal discriminative features to adapt to several unseen multimodal data \cite{multimodameta}.

\subsection{Evolutionary computing and Meta-learning}

Recent works that relate evolutionary computing and deep learning, include  differentiable compositional pattern producing network (DPPN) \cite{DPPN}, hypernetworks \cite{hypernetworks,conthyp}, and population-based meta-learning (PBML) \cite{PBML}. 
DPPN and hypernetworks use a multi-layer perceptron (MLP) network to directly evolve the structure and weights of another neural network.
PBML relates EC with optimization-based meta-learning (MAML) by sharing the paradigm of learning to learn. The authors of PBML show that MAML manifests adaptive evolvability \cite{evolvability} and provides inductive biases that balance exploration and exploitation along various dimensions of mutation functions. In this context, we motivate that ours is an evolutionary deep learning model with weight learning capability and adaptive evolvability using hypernetworks and MAML, respectively, for image reconstruction.

\subsection{Hypernetworks}

Hypernetworks are mainly inspired by HyperNEAT \cite{hyperneat}, a neuroevolution framework wherein a network could be more intuitively evolved, node by node, and connection by connection to produce connectivity patterns. Initially developed for model compression \cite{hypernetworks}, hypernetworks find various imaging applications like video frame prediction \cite{DFN}, point-cloud up-sampling \cite{meta-pu}, and multiple parameterized image restoration operators \cite{gnldecouple, mac}. These are model-based meta-learning methods \cite{metal_survey, metal_universality} wherein the hyper-networks predict all the parameters of the imaging task-oriented base network.  In our work, the KM hypernetworks are not the sole source of base network weights. As a result, our model benefits from context-invariant knowledge learned by the base network. 
In interactive image restoration applications \cite{cfs-net, controllable,hyper-res}, hypernetworks are used to modulate the activations of the base convolution layer. In our model, the hypernetworks provide mode-specific inductive bias via kernel modulation for multi-modal imaging operations.

\subsection{Gradient-based Meta-learning}

Several variants of gradient-based meta-learning (GBML) or MAML \cite{maml_finn, adaptivemaml, l2fmaml, taml, harith} have shown promising results in few-shot learning, owing to its potential for rapid adaptation and feature reuse capabilities to future tasks \cite{featurereuse}. 
 MAML methods that focus on improving inner loop optimization using a part of the network architecture for classification and regression tasks exist in the literature. These methods, namely MeTAL \cite{metataskadaptiveloss}, HyperMAML \cite{hypermaml}, and ALFA \cite{alfa} use hypernetworks to learn task adaptive loss function,  task-specific gradient updates, and inner loop regularization hyperparameters respectively. 
The meta-modulated CNN for snapshot compressive sensing (Meta-SCI) \cite{metasci} uses a set of parameters that are used for KM based on different mask settings during test-time adaptation. The conditional neural adaptive process (CNAP) \cite{fastnflexible} uses a linear classifier as an adaptation network for classification tasks. In context adaptation via meta-learning  (CAVIA) \cite{cavia},  a set of contextual parameters are meta-learned for adaptation to multiple tasks. 

Multi-modal meta-learning methods, task-aware modulation (MMAML) \cite{mmaml}, and contrastive knowledge distillation meta-learning (CAML) \cite{caml} modulate the base network layer with a single mode-specific parameter. Unlike these works, which focus on bi-level optimization of only the base network, our work efficiently utilizes both the networks based on task-specific weight updates via kernel modulation of the base network and bi-level gradient descent updates of our proposed KM hypernetwork. The KM layer outputs modulation weights based \cite{metasci, advarsarial} on each sample to modulate each kernel of the base network. 




\subsection{Hypernetworks for MRI reconstruction}

Methods that use hypernetworks for MRI reconstruction include the Hyper-recon network \cite{wang2021regularizationagnostic,mulpreconsinglehyp} and the MAC-ReconNet \cite{mac}. The MAC-ReconNet provides scalable reconstructions for multiple acquisition settings. The Hyper-recon network provides regularization-agnostic hyperparameters for image reconstruction. 
Unlike these methods, we use both the base network and the modulation networks with GBML to provide a good initialization for adapting to unseen tasks. 

The hypernetwork-based methods aforementioned \cite{mac, controllable}
lack interaction with the base network weights to achieve mode-specific weight initializations. As introduced before, different MRI contrasts exhibit different intensity properties and contribute to the learned features differently. This inspires us to employ MMAML for enhancing the generalization ability of the base neural network using meta-learned hypernetworks.
Existing MMAML-based methods exhibit a resemblance to vanilla MAML as the modulation network is optimized in the outer level of optimization, emphasizing context-invariant learning. Secondly, MMAML infers a single mode-specific value that globally modulates all the kernels of the base network layer. Thirdly, these methods focus on classification tasks and inherently assume the presence of ground truth labels at test time for fine-tuning. 
We address these limitations by employing 1) low-rank kernel modulation offering multiplicative interactions \cite{MultiplicativeIA} by dynamically modulating each kernel of the base network with different heterogeneous context-specific weights 2) two-level optimization of the hypernetworks' weights to lay emphasis on contextual learning. Furthermore, to support fine-tuning and on-the-fly adaptation to unseen contexts, our method employs only the input images to generate the task embeddings for conditioning the hypernetworks.


\section{Material and Methods}
\label{sec:method}

\subsection{Problem Formulation}
The data acquisition forward model of the MR image reconstruction problem \cite{dagan} can be formulated as:
\begin{align}
 \label{Forward_model}
 \mathcal{A}x+\epsilon = y
\end{align}

where, $x$ $\in$ $\mathbb{C}^\emph{N}$ denotes the desired image, $y$ $\in$ $\mathbb{C}^\emph{M'}$ is the under-sampled (US) measurement from the MRI scanner, $\epsilon \in \mathbb{C}^\emph{M'}$ is the noise and $\mathcal{A}: \mathbb{C}^\emph{N} \rightarrow \mathbb{C}^\emph{M'}$ represents the forward operator of the MRI acquisition process that causes aliasing artifacts due to k-space under-sampling. The under-sampled image reconstruction is ill-posed because the problem is under-determined ($M' << N$) and the operator $\mathcal{A}$ is ill-conditioned. The under-sampled or zero-filled (ZF) image is given by, $x_{US} = F_{US}^{H}y$ where $F_{US}$ is the under-sampled Fourier encoding matrix. The reconstruction of the under-sampled image is achieved by introducing an apriori knowledge of $x$ into the unconstrained optimization \cite{dagan} given as:

\begin{align}
 \label{Optimization formulation}
 \underset{x} {\operatorname{min}} \quad ||\mathcal{A}x - y||_{2}^2 + \mathcal{R}(x)
\end{align}

where, $||\mathcal{A}x - y||_{2}^2$ is the data fidelity term \cite{dc_cnn} and $\mathcal{R}$ is a regularization term.

Deep learning-based MRI reconstruction involves training a deep learning (DL) model using a single-level optimization on the average loss of all observed data samples. This supervised joint training can be formulated as:
\begin{align}
 \label{Joint-training formulation}
 \theta^{*} = \underset{\theta} {\operatorname{argmin}}  \displaystyle \mathop{\mathbb{E}}_{(x_{US},x_{FS})\in \bigcup\limits_{i}^{} \mathcal{D}_{i}}[||x_{FS} - f(x_{US};\theta) ||_{2}^2]
\end{align}

where, $\mathcal{D}_{i}$ represents the dataset of the $i^{th}$ configuration of MRI contrast, under-sampling mask type, and acceleration factor, consisting of ground truth or fully sampled (FS) image $x_{FS}$ and its corresponding under-sampled image $x_{US}$. Here, $f$ is the DL model parameterized by $\theta$ with k-space data fidelity. Unlike iterative methods in Eq. \ref{Optimization formulation}, 
the formulation in Eq. \ref{Joint-training formulation} infers an optimal parameter set $\theta^{*}$ \cite{metal_survey}.

In MAML-based MRI reconstruction, we consider each combination of MRI contrast, under-sampling mask type, and acceleration factor for under-sampling as a task. We partition the data of each task $M$ into support ($x^{M}_{US,spt},x_{FS,spt}$) and query ($x^{M}_{US,qry},x_{FS,qry}$) samples. The parameters of the network, $\theta$, in Fig. \ref{fig:graphabsmaml} (MAML) are called \emph{meta-initializations}. For every task \emph{M}, MAML uses the support samples to perform a few gradient-descent steps (\emph{adaptation}) from meta-initializations to obtain task-specific parameters $\phi_{M}$ (Eq. 5). The loss due to task-specific parameters on query data is aggregated over all train tasks ($\mathcal{T}$), to provide supervision for meta-initializations (Eq. \ref{MAML training formulation}). The MAML optimization \cite{metal_survey} is given as:
\begin{align}
\label{MAML training formulation}
 \theta^{*} = \underset{\theta} {\operatorname{argmin}} \sum_{M\in \mathcal{T}} ||x_{FS,qry} - f(x^{M}_{US,qry};
 \phi_{M}) ||_{2}^2 \\ s.t \quad \phi_{M} = \underset{\theta} {\operatorname{argmin}}\, ||x_{FS,spt} - f(x^{M}_{US,spt};\theta) ||_{2}^2 \quad \forall M\in \mathcal{T}
\end{align}

 
 In the proposed KM-MAML, the parameters of the KM hypernetworks are optimized in the bi-level optimization and the parameters of the base network are optimized in the outer loop as shown in Fig. \ref{fig:graphabsmaml} (KM-MAML). The KM-MAML formulation is  given by:
 
 \begin{align}
\label{KM-MAML training formulation}
 \omega^{*}, \theta^{*} = \underset{\omega, \ \theta} {\operatorname{argmin}} \sum_{M\in \mathcal{T}} ||x_{FS,qry} - f(x^{M}_{US,qry};
 \Omega_{M},\theta) ||_{2}^2 \\ s.t \quad \Omega_{M} = \underset{\omega} {\operatorname{argmin}}\, ||x_{FS,spt} - f(x^{M}_{US,spt};\omega,\theta) ||_{2}^2 \quad \forall M\in \mathcal{T}
\end{align}

\begin{algorithm}[t!]
\caption{KM-MAML Training Algorithm}
\label{kMAMLalgo}
\begin{algorithmic}[1]
\Require Learning rates $\eta_{1}$, $\eta_{2}$ and \( p(\mathcal{T}) \): Multimodal train task distribution
\Require \( CE(.) \): Context encoder
\State Randomly initialize $\omega$: modulation network weights and $\theta$: base network weights

\For{$e = 1\ to\ max\_epochs$}
\State Sample a task mini-batch: $\mathcal{T}_{batch}$ $\sim$ \( p(\mathcal{T}) \)
\State $\mathcal{L}_{meta} = 0$
\For{each training task $M$ in \( \mathcal{T}_{batch} \)}
\State Sample a mini-batch of support data: $D^{M}_{spt}=\{ x_{US,n}^{M},x_{FS,n} \}_{n=1}^{N_{spt}}$
\State Initialize $\Omega_{0}^{M} \leftarrow \omega$
\For{$u = 0 \ to \ U-1$}
\State Context embedding: $\gamma$ = $CE(x_{US}^{M})$
\Comment{Support data 

\hskip20.7emmini-batch input
}
\State $\alpha$, $\beta$ = KM Hypernetwork($\gamma; \Omega_{u}^{M}$)
\State $W_{M}$ = $\beta \otimes_{outer} \alpha$
\Comment{Modulation weights of task M
}
\State $\theta_{mod}$ = $\theta \odot W_{M}$
\Comment{Mode-specific initialization (KM)
}
\State $\mathcal{L}_{1} = L([\Omega_{u}^{M}, \theta_{mod}], D^{M}_{spt})$
\State $\Omega_{u+1}^{M} \leftarrow \Omega_{u}^{M}-\eta_{2} \nabla_{\Omega_{u}^{M}}\mathcal{L}_{1}$ \Comment{mode-specific weight updates 
}
\EndFor
\State Sample a mini-batch of query data: $D^{M}_{qry}=\{ x_{US,n}^{M},x_{FS,n} \}_{n=1}^{N_{qry}}$
\State $\alpha$, $\beta$ = KM Hypernetwork($\gamma, \Omega_{U-1}^{M}$)
\State $W_{M}$ = $\beta \otimes_{outer} \alpha$
\Comment{$\otimes_{outer}$ denotes outer product 
}
\State $\theta_{mod}$ = $\theta \odot W_{M}$
\State $\mathcal{L}_{meta} \leftarrow \mathcal{L}_{meta} + L([\Omega_{U-1}^{M}, \theta_{mod}], D^{M}_{qry})$
\EndFor
\State $\theta \leftarrow \theta - Adam[\eta_{1},\mathcal{L}_{meta}]$
\State $\omega \leftarrow \omega - Adam[\eta_{1},\mathcal{L}_{meta}]$
\Comment{Meta-updates}
\EndFor
\end{algorithmic}
\end{algorithm}
 
The training process of KM-MAML is illustrated in Algorithm \ref{kMAMLalgo}. On every task's support samples, KM-MAML performs adaptation of the KM hypernetworks to result in a task-specific KM hypernetwork model that is characterized by weights $\Omega_{M}$ (Eq. 7), and this constitutes the inner level optimization in the bi-level MAML process (steps 5 to 14 in Algorithm \ref{kMAMLalgo}). Step 13 in Algorithm \ref{kMAMLalgo} describes the $L_1$ image reconstruction loss due to the task-specific parameters $\Omega^{M}_{u}$ at the $u^{th}$ inner step and the modulated base network weights $\theta_{mod}$ using support data $D^{M}_{spt}$ of task M. The loss due to task-specific parameters on query data, $L([\Omega_{U-1}^{M}, \theta_{mod}], D^{M}_{qry})$ is aggregated over all train tasks ($\mathcal{T}$) in step 20, to provide supervision for meta-initializations $\omega$ and $\theta$ (Eq. \ref{KM-MAML training formulation} and steps 22 and 23). The low-rank modulation weights $\alpha$ and $\beta$ and the modulated base network weights $\theta_{mod}$ at the $u^{th}$ inner update are shown in steps 10 to 12 and for the outer loop, in steps 18 to 20 for each task ($\odot$ denotes element-wise multiplication). The outer loop or meta-update of the base network and the hypernetworks weights are shown in steps 22 and 23, respectively. These weights, $\theta$ and $\omega$ form the meta-initializations of the proposed network. 

\subsection{Architecture}
\begin{figure}[t]
    \centering
    \includegraphics[width=\linewidth]{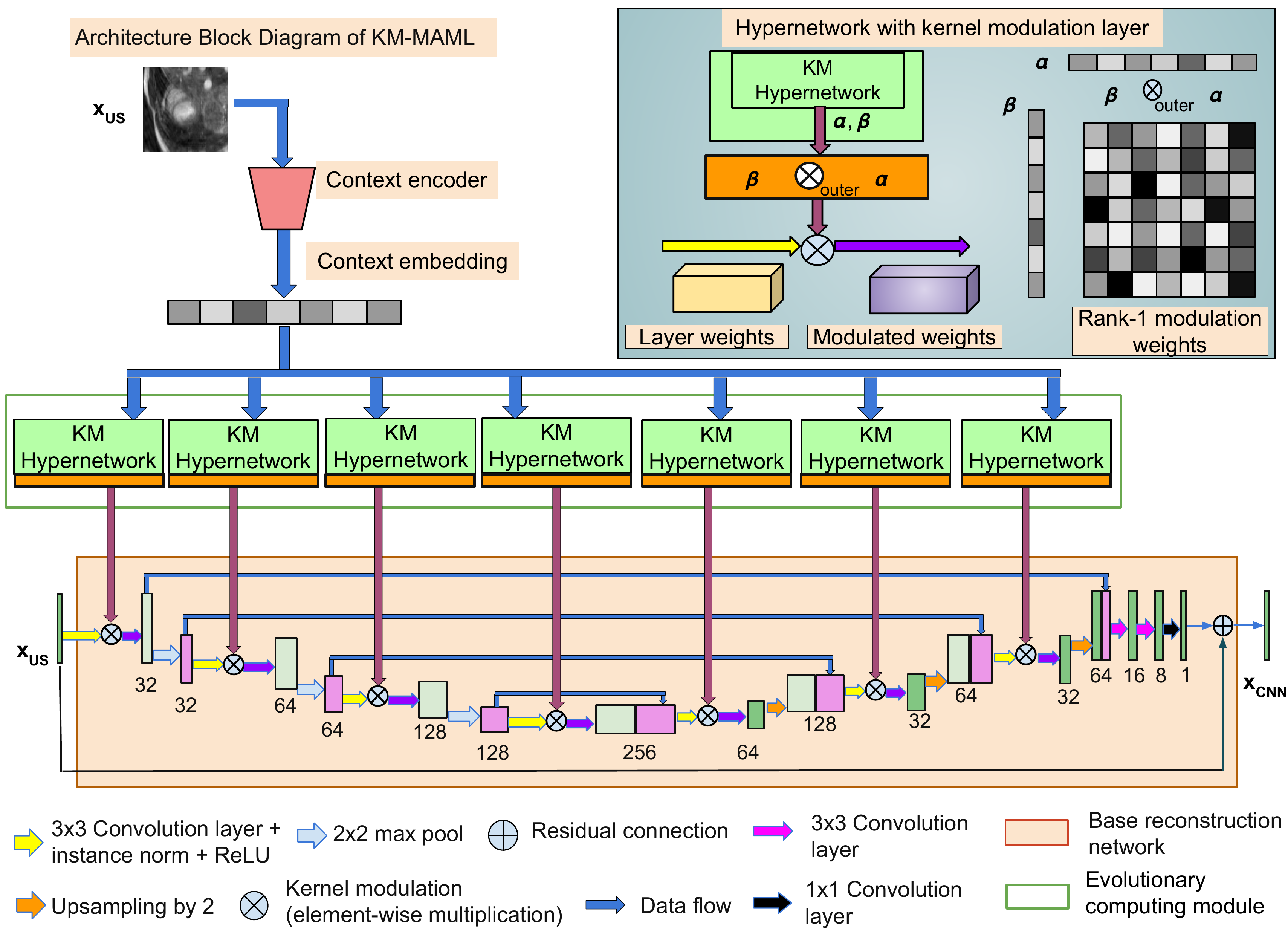}
    \caption{Network architecture of  KM-MAML: The context-encoder CNN provides embedding vectors to represent each mode. The KM hypernetworks constitute the evolutionary computing module which predicts layer-wise weights that modulate the base network weights via low-rank kernel modulation layer (orange block), shown with rank $r = 1$ in the top right. The base CNN performs the multimodal image reconstruction task. $X_{CNN}$ denotes the predicted image of the base network. 
    }
    \label{fig:arch}
\end{figure}

The architecture of KM-MAML consists of a context encoder, a base reconstruction network, and layer-wise KM hypernetworks as shown in Figure \ref{fig:arch}. 
The context encoder is a simple auto-encoder that learns to map the under-sampled MRI input image to output in an unsupervised manner. The latent layer of the context encoder has $c$ channels which are averaged to obtain a latent $c$-dimensional embedding vector which is passed to the KM hypernetworks.
The input and output of the base reconstruction network are the under-sampled, and the predicted fully sampled MRI images, respectively. The base reconstruction network is an encoder-decoder network, U-Net \cite{unet, dc_unet}, a popular benchmark for MRI reconstruction \cite{fastmri}. 
The base network has seven layers with three sub-sampling levels of the encoder and decoder each and a bottleneck layer. 

There are seven KM hypernetworks corresponding to each layer of the base network. Each KM hypernetwork is a linear multi-layer perceptron network with three layers - an input layer with $c$ units, a bottleneck layer, and an output layer with a size equal to the number of filters in the corresponding layer of the base network. The KM hypernetwork of layer $l$ gives two rectangular matrices as output, $\beta^{l} \in R^{{n^{l}_{out}}\times r}$ and $\alpha^{l}\in  R^{r \times{n^{l}_{in}}}$ where $n^{l}_{out}$ and $n^{l}_{in}$  are the number of output and input channels of the base network layer $l$ and $r$ is the rank. The outer product operation between $\beta^{l}$ and $\alpha^{l}$ is computed in the KM layer (shown in steps 11 and 18 in Algorithm \ref{kMAMLalgo}) to obtain modulation weights,  $W^l_{M}$ given as,

\begin{equation}\label{modulationeqn}
 W^l_{M} =  \beta^{l} \otimes_{outer} \alpha^{l}
\end{equation}

The modulation weights are specific to each kernel of the base network and are element-wise multiplied. 
 For instance, if the base network filter size is $64\times32\times3\times3$, then the corresponding KM hypernetwork has $64 + 32$ output neurons, 64 for $\beta$ and 32 for $\alpha$. The outer product at the output gives modulation weights $W^l_{M}$ of size $64\times32$ for kernel modulation. 



\section{Experiments and Results}
\label{sec:expr}

\subsection{Datasets and Implementation Details}
\subsubsection{Datasets}
We use three multi-contrast MRI datasets consisting of axial brain images, namely  \textbf{MRBrainS} \cite{mrbrains_dataset}, \textbf{IXI}\footnote{https://brain-development.org/ixi-dataset/} and \textbf{SRI24 Atlas} \cite{sri24atlas,disn}. 

1) \textbf{MRBrainS} dataset: We consider 336 slices of T1 and FLAIR contrasts from 7 volumes of T1 and FLAIR (fluid-attenuated inversion recovery) with T1: TR = 7.9 ms, TE = 4.5 ms and FLAIR: TR = 11s and TE = 125 ms acquired on a Philips scanner. The repetition and echo times, TR, and TE denote the contrast-specific MRI settings. All the images have a size of $240\times240$.

2) \textbf{IXI} dataset: We consider 1400 slices of T2 and Proton density (PD) weighted contrasts acquired from 14 volumes  with T2: TR = 5725 ms, TE = 100 ms, and PD: TR = 5725 ms, TE = 8 ms acquired on a Philips scanner. The images are preprocessed by cropping to a size of $240\times240$.

For training the models we consider T1, FLAIR and T2 and PD contrasts from \textbf{MRBrainS} and \textbf{IXI} datasets.

3) \textbf{SRI24 Atlas} dataset: 135 slices of T1, T2, and PD-weighted MRI images with T1: TR = 6.5 ms, TE = 1.54 ms and T2, PD: TR = 10s, TE = 14/98 ms acquired on a GE scanner. All the images have a size of $240\times240$.

For testing the models on unseen scenarios, we use T1, T2, and PD from \textbf{SRI24 Atlas} dataset.

For comparison with other MRI reconstruction architectures, we use the following datasets: 

4)
\textbf{Automated Cardiac Diagnosis Challenge (ACDC) \cite{acdc_dataset} cardiac MRI} dataset which consists of 150 and 50 patient records with 1841 and 1076 slices for training and validation respectively. The 2D slices are extracted and cropped to 150$\times$150. 
The masks for radial under-sampling are taken from RefineGAN\footnote{https://github.com/tmquan/RefineGAN} repository 

5) The \textbf{knee dataset} contains 200 training and validation single channel slices of 10 subjects, each obtained from a 15-element coil knee data\cite{variational} cropped to 320$\times$320.

6) For training on large-scale data, we use the \textbf{fastMRI}  \cite{fastmri} dataset with 7040 training and 3317 validation slices of 200 and 99 volumes, respectively, of coronal proton-density  without (PD) and with fat suppression (PDFS). We split the validation slices into 437 slices from 18 volumes for adaptation and 2880 slices for evaluation. The slices have multiple resolution levels - 640$\times$368, 640$\times$372, 640$\times$400, 640$\times$454, and mask patterns different from the training data.

\subsubsection{Implementation Details}
\label{impldetails}
\textbf{Training details:} For training and evaluating the models, we use four MRI contrasts - T1, FLAIR, T2, and PD. In addition, we augment the datasets based on two types of under-sampling mask patterns - Cartesian and Gaussian masks with three different acceleration factors for under-sampling namely, 4x, 5x, and 8x. We pose each configuration of contrast, mask type, and acceleration factor as a task for meta-learning. The configurations above are combined to form 24 tasks. 
The US input images are obtained by retrospectively under-sampling the fully sampled k-space \cite{dc_wcnn}. 
Each task $M$ has samples split into support $D^{M}_{spt}$ and query $D^{M}_{qry}$ images for meta-training. 

For evaluating the adaptability to unseen tasks (from SRI24 Atlas), we choose three contrasts - T1, T2, and PD, two mask types, and five accelerations - 5x, 6x, 7x, 8x, and 9x to obtain 30 tasks. 
For adaptation via fine-tuning, we split the test images into test support for fine-tuning and test query images for evaluation.
Table \ref{tab:trainingdatadetails} shows the details of the number of samples used for training the models.
All MAML models are trained with one inner loop gradient step. The batch size for support and query data $(N_{spt}, N_{qry})$ is 10. The task mini-batch size is chosen as 3. The loss function is $L_{1}$ norm between the model's prediction and ground truth.
We have chosen the learning rates $\eta_{1}$, $\eta_{2}$  of 0.001 with 600 epochs for training. For SGD, the learning rate is chosen as 0.001.
All models are implemented using PyTorch and trained on $NvidiaRTX-3090$ GPU with $24$ GB memory. We have chosen the context embedding vector size as $256$ and the rank $r=1$ for kernel modulation. 
Our source code is available at https://github.com/sriprabhar/KM-MAML/. 

\begin{table}[]
\scriptsize
\centering
\caption{Details of cross-scanner datasets, and the number of samples used for training and testing the models. Note that for SGD the support and query samples are mixed.}
\label{tab:trainingdatadetails}
\begin{tabular}{|c|c|c|ccc|}
\hline
\multirow{2}{*}{\textbf{\begin{tabular}[c]{@{}c@{}}Dataset \\ type\end{tabular}}} & \multirow{2}{*}{\textbf{\begin{tabular}[c]{@{}c@{}}Dataset / scanner \\ name\end{tabular}}} & \multirow{2}{*}{\textbf{Contrast}} & \multicolumn{3}{c|}{\textbf{Number of data samples}}                                                          \\ \cline{4-6} 
                                                                                  &                                                                                   &                                    & \multicolumn{1}{c|}{\textbf{Support}} & \multicolumn{1}{c|}{\textbf{Query}} & \textbf{Testing} \\ \hline
\multirow{4}{*}{\begin{tabular}[c]{@{}c@{}}Training \\ datasets\end{tabular}}     & \multirow{2}{*}{MRBrainS / Philips}                                                         & T1                                 & \multicolumn{1}{c|}{3456}                   & \multicolumn{1}{c|}{2304}                 & 2304                \\ \cline{3-6} 
                                                                                  &                                                                                   & FLAIR                              & \multicolumn{1}{c|}{3456}                   & \multicolumn{1}{c|}{2304}                 & 2304                \\ \cline{2-6} 
                                                                                  & \multirow{2}{*}{IXI / Philips
                                                                                  }                                                              & PD                                 & \multicolumn{1}{c|}{4800}                   & \multicolumn{1}{c|}{4800}                 & 24000               \\ \cline{3-6} 
                                                                                  &                                                                                   & T2                                 & \multicolumn{1}{c|}{4800}                   & \multicolumn{1}{c|}{4800}                 & 24000               \\ \hline
\multirow{3}{*}{\begin{tabular}[c]{@{}c@{}}Unseen \\ datasets\end{tabular}}       & \multirow{3}{*}{SRI24 Atlas / GE}                                                      & T1                                 & \multicolumn{1}{c|}{-}                      & \multicolumn{1}{c|}{-}                    & 1350                \\ \cline{3-6} 
                                                                                  &                                                                                   & T2                                 & \multicolumn{1}{c|}{-}                      & \multicolumn{1}{c|}{-}                    & 1350                \\ \cline{3-6} 
                                                                                  &                                                                                   & PD                                 & \multicolumn{1}{c|}{-}                      & \multicolumn{1}{c|}{-}                    & 1350                \\ \hline
\end{tabular}
\end{table}

\textbf{Evaluation metrics:} Our evaluation metrics are Peak Signal-to-Noise Ratio (PSNR), Structural Similarity Index  (SSIM),  to assess the performance of the models, and Centered Kernel Alignment (CKA) \cite{cka} to quantify the \% change in the mode-specific features pre and post kernel modulation.

\textbf{K-space Data fidelity:} The output of the base network is fed to a k-space data fidelity unit to ensure consistency with the acquired k-space \cite{dc_cnn} (\ref{appendix:datafidelity}). 

\textbf{Architecture-based Comparison:} We compare our method with image domain MRI reconstruction CNNs, DAGAN \cite{dagan}, DC-CNN \cite{dc_cnn}, DC-DEN \cite{dc-ensemble}, DC-RDN \cite{recursive_dilated}, DC-UNet \cite{dc_unet}, MICCAN \cite{miccan}, and MAC-ReconNet \cite{mac}. We use the source code of DC-CNN\footnote{https://github.com/js3611/Deep-MRI-Reconstruction}, DAGAN\footnote{ https://github.com/tensorlayer/DAGAN} and MAC-ReconNet\footnote{https://github.com/sriprabhar/MAC-ReconNet} for implementation. 
We have also compared our method with recent state-of-the-art methods that have been proposed for MRI reconstruction. These are OUCR (Over-complete and under-complete CNNs) \cite{oucr}, vision transformer-based methods, ViT \cite{vit}, and SWIN transformer \cite{swin} models, and a knowledge distillation-based MRI reconstruction method, SFT-KD-Recon\footnote{https://github.com/gayathrimatcha/sft-kd-recon} \cite{sftkdrecon}.
For DC-UNet, DC-RDN, MICCAN, DC-DEN, ViT and SWIN transformers, we replace the CNN in DC-CNN with UNet from the fastMRI repository\footnote{https://github.com/facebookresearch/fastMRI}, dilated convolutions with recursive connection, channel attention, dense connection blocks, vision transformers and SWIN transformers respectively \cite{dcrsn}.



\subsection{Results and Discussion}
We extensively evaluate our approach by comparing it with three representative models, joint training using SGD, MAML \cite{maml}, and MMAML \cite{mmaml} for MC-MRI reconstruction. Our experiments showcase (i) the scalability of the models in MC-MRI reconstruction,  (ii) generalization to unseen contrasts via adaptation on the fly and fine-tuning, (iii) the contribution of KM and the benefits of GBML using comparative studies and representational similarity measures, (iv) an ablative study with and without meta-learning, 
(v) performance comparison with several MRI reconstruction architectures for specific and multiple acquisition contexts, and
(vi) fine-tuning to new anatomies and image resolution levels.
  
\subsubsection{Multi-modal MRI reconstruction performance}
\begin{table}[]
\scriptsize
\centering
\caption{Quantitative comparison of ZF, SGD, MAML, MMAML, and KM-MAML for multiple training tasks combining various contrasts, acceleration factors, and mask types. Green and blue colors indicate the best and the second best metrics respectively. The tasks are denoted in short as contrast type - mask type - acceleration factor (T1C8 indicates T1 contrast with Cartesian mask pattern and 8x acceleration). FL denotes FLAIR MRI}
\label{tab:expr1}
\begin{tabular}{|c|c|c|c|c|c|}
\hline
                                & \textbf{ZF}          & \textbf{SGD}                         & \textbf{MAML}                        & \textbf{MMAML}                       & \textbf{KM-MAML}                     \\ \cline{2-6} 
\multirow{-2}{*}{\textbf{Task}} & PSNR / SSIM    & PSNR / SSIM                          & PSNR / SSIM                          & PSNR / SSIM                          & PSNR / SSIM                          \\ \hline
T1C5                            & 31.38 / 0.665  & 35.03 / 0.866                        & {\color[HTML]{3531FF} 35.22 / \color[HTML]{000000} 0.875} & {\color[HTML]{000000} 35.14 / \color[HTML]{3531FF} 0.876} & {\color[HTML]{009901} 35.58 / 0.895} \\ \hline
T1G4                            & 32.24 / 0.545  & 40.07 / 0.913                        & 40.59 / 0.922                        & {\color[HTML]{3531FF} 40.71 / 0.927} & {\color[HTML]{009901} 41.02 / 0.935} \\ \hline
T1G8                            & 29.17 / 0.451  & 35.03 / 0.805                        & 35.33 / 0.820                        & {\color[HTML]{3531FF} 35.51 / 0.841} & {\color[HTML]{009901} 35.96 / 0.859} \\ \hline
FLC4                            & 28.40 / 0.642  & 34.12 / 0.879                        & {\color[HTML]{009901} 34.40 / \color[HTML]{3531FF} 0.885} & 34.23 / 0.882                        & {\color[HTML]{009901} 34.40 / 0.889} \\ \hline
FLC8                            & 26.49 / 0.589  & 30.83 / 0.813                        & {\color[HTML]{3531FF} 31.02 / 0.828} & 30.89 / 0.824                        & {\color[HTML]{009901} 31.20 / 0.841} \\ \hline
FLG8                            & 26.36 / 0.426  & 33.40 / 0.802                        & {\color[HTML]{3531FF} 33.73 / \color[HTML]{000000} 0.811} & {33.56 / \color[HTML]{3531FF} 0.820} & {\color[HTML]{009901} 33.93 / 0.832} \\ \hline
PDC4                            & 25.68 / 0.625  & {\color[HTML]{000000} 32.76 / \color[HTML]{000000} 0.877} & {\color[HTML]{3531FF} 33.06 / 0.882} & 32.96 / 0.879                        & {\color[HTML]{009901} 33.24 / 0.888} \\ \hline
PDC8                            & 23.40 / 0.551  & { \color[HTML]{000000} 29.07 / \color[HTML]{000000} 0.803 } & {\color[HTML]{3531FF} 29.48 / 0.814} & 29.32 / 0.804                        & {\color[HTML]{009901} 29.64 / 0.819} \\ \hline
PDG4                            & 26.88 / 0.541  & 37.78 / 0.908                        & {\color[HTML]{3531FF}37.98 / 0.911} & 37.81 / 0.908                        & {\color[HTML]{009901} 38.04 / 0.914} \\ \hline
T2C4                            & 25.94 / 0.634 & {\color[HTML]{000000} 32.01 / \color[HTML]{000000} 0.873} & {\color[HTML]{3531FF}32.43 / 0.880} & 32.26 / 0.875                        & {\color[HTML]{009901} 32.55 / 0.884} \\ \hline
T2G4                            & 27.16 / 0.543  & {\color[HTML]{000000} 37.27 / \color[HTML]{000000} 0.910} & {\color[HTML]{3531FF} 37.48 / 0.912}                        & 37.21 / 0.906                        & {\color[HTML]{009901} 37.59 / 0.916} \\ \hline
T2G5                            & 25.71 / 0.500  & {\color[HTML]{000000} 34.83 / 0.860} & \color[HTML]{3531FF} 35.06 / \color[HTML]{3531FF} 0.863                        & 34.82 / 0.855                        & {\color[HTML]{009901} 35.24 / 0.870} \\ \hline
\end{tabular}
\end{table}
We evaluate the scalability of the meta-initializations of the model to training tasks. 
Table \ref{tab:expr1} shows the quantitative results for 12 out of 24 tasks.
From the table, our observations are.
(i) MAML-based models enrich the learning process compared to joint training indicated by improved SSIM metrics. 
(ii) KM-MAML consistently performs better than other methods for 96\% of tasks in SSIM and 80\% in PSNR. 
(iii) MMAML performs on par with MAML due to its closer resemblance to MAML, constraining the meta-training process to only the base network. (iv) KM-MAML consistently performs better than MMAML, highlighting the importance of meta-learning the auxiliary network for efficient adaptation. 

\begin{figure}[t!]
    \centering
    \includegraphics[width=\linewidth]{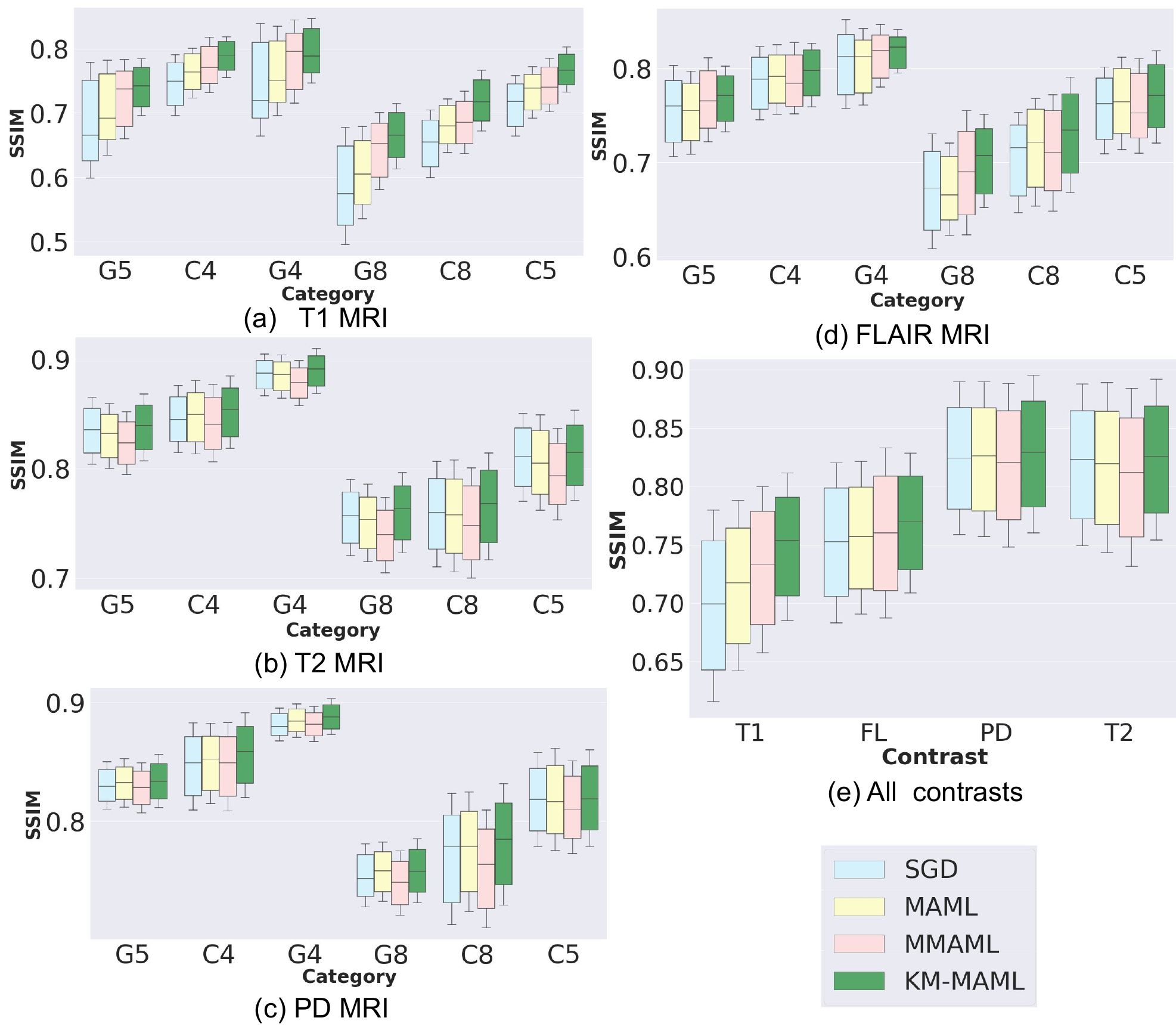}
    \caption{SSIM plots showing the performance of models for training tasks with varying contrasts, mask types, and acceleration factors (a) T1 (b)  T2 (c) PD, and (d) FLAIR respectively. C denotes the Cartesian mask type, and G denotes the Gaussian mask type. For instance, G5 indicates a Gaussian mask pattern with 5x acceleration.  Figure (e) shows the consolidated performance of each contrast. FL indicates FLAIR MRI contrast
    }
    \label{fig:expr1_boxplot}
\end{figure}

\begin{figure}[t!]
    \centering
    \includegraphics[width=\linewidth]{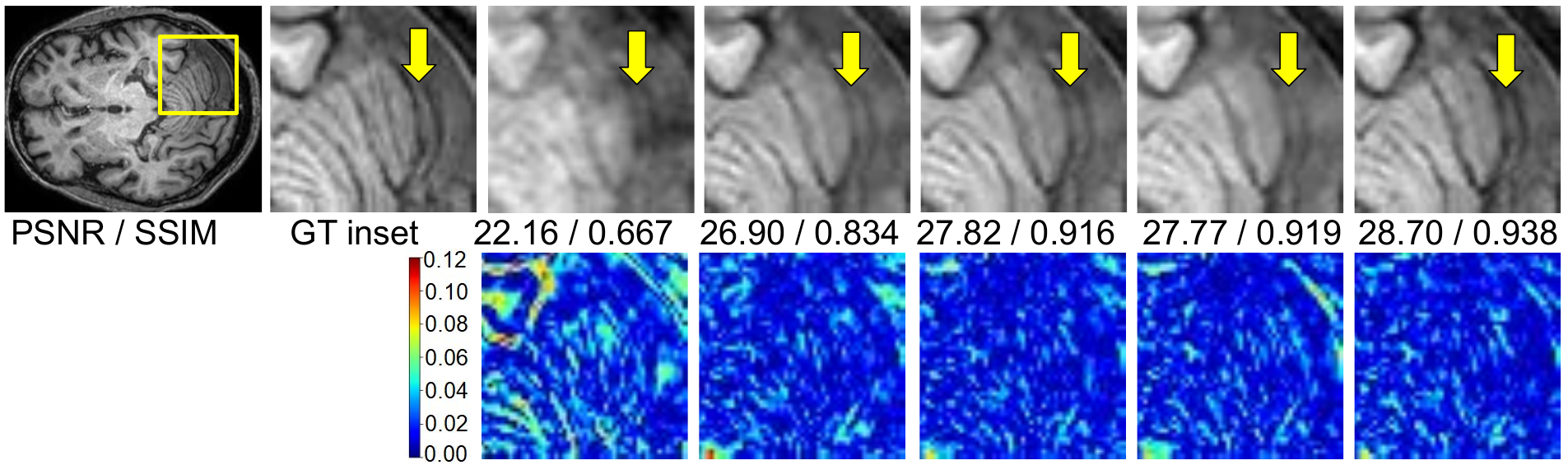}
    \caption{Qualitative comparison of the reconstruction performance for  T1 MRI. From the left: the ground truth (GT) image with the region of interest (ROI) highlighted as yellow box, GT inset, ZF image, joint training, MAML, MMAML, and KM-MAML. The yellow arrows in the image highlight the improved structure recovery of KMMAML. The residual images with respect to the target indicate that KM-MAML recovers details better than other learning methods.  
    }
    \label{fig:expr1_visual_t1}
\end{figure}

\begin{figure}[t!]
    \centering
    \includegraphics[width=\linewidth]{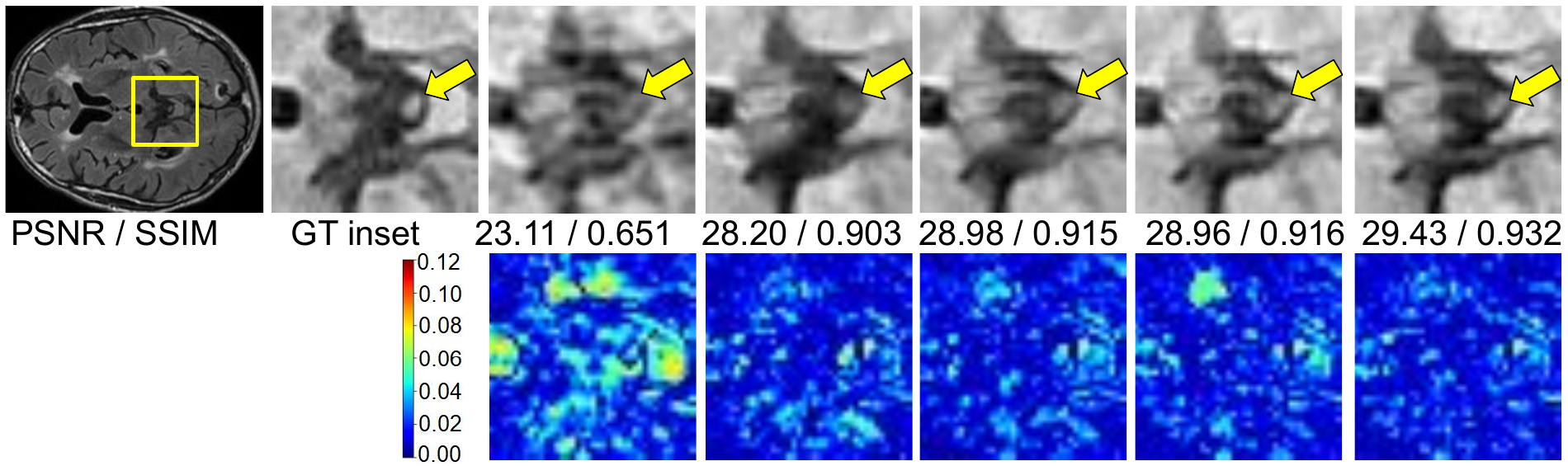}
    \caption{Qualitative comparison of the reconstruction performance for FLAIR MRI. From the left: the ground truth (GT) image with ROI, GT inset, ZF image, joint training, MAML, MMAML and KM-MAML. As pointed out by yellow arrows, KM-MAML is able to recover structures closer to the target better than other learning methods. This is also evident from the residual images. MMAML and MAML provide better reconstruction quality than SGD but with missing structures around the regions of interest.
    }
    \label{fig:expr1_visual_flair}
\end{figure}

The box plots for each configuration of contrast, mask types, and acceleration factors (Figure \ref{fig:expr1_boxplot}(a - d)) and each MRI contrast separately (Figure \ref{fig:expr1_boxplot} (e)) together show that KM-MAML can provide improved accuracy metrics for multiple modes. 
The reliability of KM-MAML is demonstrated with relatively lesser deviations in box plots as compared to other methods. 
The comprehensive view within each contrast shows the fidelity of our model across mask patterns and different amounts of under-sampling. 

Figures \ref{fig:expr1_visual_t1} and \ref{fig:expr1_visual_flair} provides the qualitative reconstruction results for  
 T1 and FLAIR MRI reconstruction. The T1 target image shows the cerebellum region in the hindbrain while the FLAIR image shows regions around the corpus callosum. The visual results show that (i) KM-MAML is able to recover fine structures much better when compared with other methods for both contrasts and exhibits the least residual error. (ii) MMAML prediction suffers from a blur in the T1 highlighted region as compared to other methods. (iii) The residual images of MAML-based methods show lesser errors than joint training. 
These observations emphasize the importance of discriminating the representations at the contextual and image levels.  The KM hypernetworks capture the semantic relationship between various modes, while the base network is optimized for the image reconstruction task (multi-objective training). 

\subsubsection{On-the-fly adaptation to unseen multimodal MRI contrasts}
To verify that KM-MAML can balance flexibility and robustness, we assess the capabilities of on-the-fly adaptation without fine-tuning to unseen multi-contrast MRI datasets.
In this experiment, we consider 24 tasks combining unseen T1, T2, and PD contrasts and  unseen acceleration factors, 6x, 7x, 8x, and 9x, with Cartesian and Gaussian mask patterns. Table \ref{tab:expr2} shows the performance for twelve of them. Our observations are as follows. (i) Both MMAML and KM-MAML exhibit better generalization than vanilla MAML and joint training. This observation indicates that task-aware modulation is an essential principle in improving the performance of MAML on heterogeneous tasks. (ii) MAML and MMAML show better generalization compared to joint training with respect to SSIM. (iii) KM-MAML can encompass a variety of distribution shifts in contrasts, acceleration factors, and mask patterns with higher improvement margins of over 0.1dB in PSNR and 0.01 SSIM for most tasks. 
These observations indicate that the KM hypernetworks exhibit an improved representational capacity to learn discriminative features effectively. At the same time, the hypernetworks learn task-to-task similarities and re-use the information for related tasks with drifts in degradation levels and contrast types.
\begin{table}[]
\scriptsize
\centering
\caption{Quantitative comparison of SGD, MAML, MMAML, and KM-MAML for on-the-fly adaptation to various unseen tasks combining multiple contrasts, acceleration factors, and mask types with deviated acquisition settings. Green and blue colors indicate the best and the second best metrics respectively. The tasks are denoted in short as contrast type - mask type - acceleration factor}
\label{tab:expr2}
\begin{tabular}{|l|l|l|l|l|}
\hline
                                & \multicolumn{1}{c|}{\textbf{SGD}} & \multicolumn{1}{c|}{\textbf{MAML}} & \multicolumn{1}{c|}{\textbf{MMAML}} & \multicolumn{1}{c|}{\textbf{KM-MAML}}         \\ \cline{2-5} 
\multirow{-2}{*}{\textbf{Task}} & PSNR / SSIM                       & PSNR / SSIM                        & PSNR / SSIM                         & PSNR / SSIM                                   \\ \hline
T1C6                            & \color[HTML]{3531FF}31.33 / 0.837                     & 31.20 / \color[HTML]{3531FF} 0.837                      & 31.23 / 0.832                       & {\color[HTML]{009901} 31.42 / 0.842} \\ \hline
T1C7                            & \color[HTML]{3531FF}32.65 / \color[HTML]{000000}0.846                     & 32.60 / \color[HTML]{3531FF}0.849                      & 32.50 / 0.843                       & {\color[HTML]{009901} 32.79 / 0.851} \\ \hline
T1C8                            & \color[HTML]{3531FF}30.63 / 0.833                     & 30.61 / \color[HTML]{3531FF}0.834                      & 30.60 / 0.830                       & {\color[HTML]{009901} 30.87 / 0.842} \\ \hline
T1C9                            & 30.07 / 0.824                     & \color[HTML]{3531FF}30.16 / 0.826                      & 30.08 / 0.822                       & {\color[HTML]{009901} 30.32 / 0.834} \\ \hline
T2G6                            & \color[HTML]{3531FF}35.20 / \color[HTML]{000000}0.866                     & 34.72 / 0.855                      & 35.10 / \color[HTML]{3531FF}0.869                       & {\color[HTML]{009901} 35.27 / 0.870} \\ \hline
T2G7                            & \color[HTML]{3531FF}33.98 / \color[HTML]{000000}0.851                     & 33.50 / 0.839                      & 33.82 / \color[HTML]{009901}0.855                       & {\color[HTML]{009901} 34.18 / 0.855} \\ \hline
T2G8                            & \color[HTML]{3531FF}33.37 / \color[HTML]{000000}0.837                     & 33.00 / 0.827                      & 33.23 / \color[HTML]{3531FF}0.838                       & {\color[HTML]{009901} 33.57 / 0.845} \\ \hline
T2G9                            & \color[HTML]{3531FF}32.31 / \color[HTML]{000000}0.819                     & 31.78 / 0.804                      & 32.06 / \color[HTML]{3531FF}0.820                       & {\color[HTML]{009901} 32.41 / 0.823} \\ \hline
PDC6                            & 31.70 / 0.862                     & \color[HTML]{3531FF}31.77 / \color[HTML]{000000}0.864                      & 31.70 / \color[HTML]{3531FF}0.867                       & {\color[HTML]{009901} 31.95 / 0.871} \\ \hline
PDC7                            & 30.45 / 0.843                     & 30.83 / 0.852                      & \color[HTML]{3531FF}30.86 / 0.855                       & {\color[HTML]{009901} 31.11 / 0.861} \\ \hline
PDC8                            & 31.30 / 0.853                     & 31.43 / 0.857                      & \color[HTML]{3531FF}31.53 / 0.862                       & {\color[HTML]{009901} 31.85 / 0.868} \\ \hline
PDC9                            & 28.67 / 0.806                     & \color[HTML]{3531FF}28.93 / 0.814                      & 28.81 / 0.812                       & {\color[HTML]{009901} 29.20 / 0.827} \\ \hline
\end{tabular}
\end{table}

\begin{figure}[t!]
    \centering
    \includegraphics[width=\linewidth]{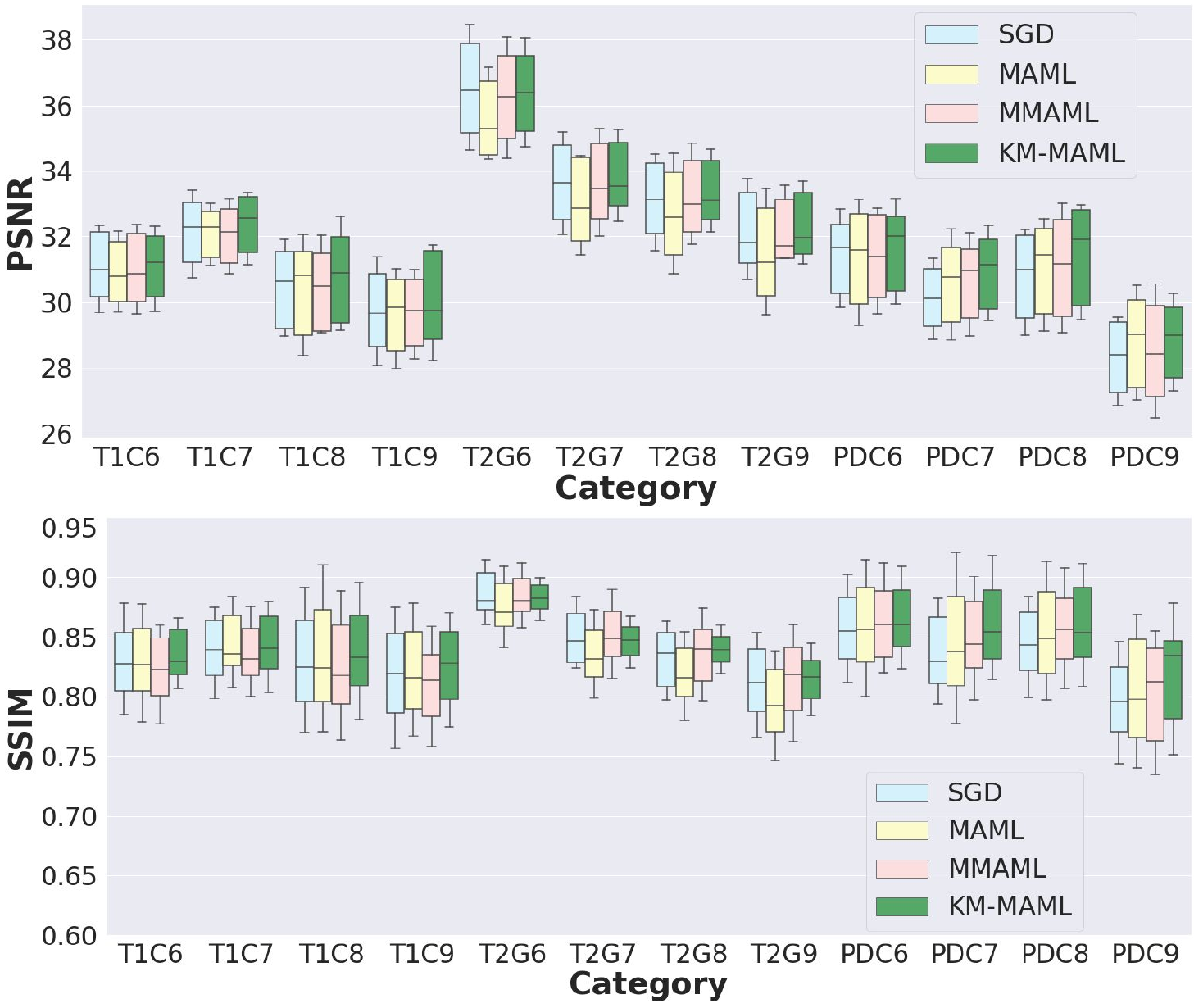}
    \caption{PSNR and SSIM plots comparing on-the-fly adaptation performance of joint training, MAML, MMAML, and KM-MAML for 12 unseen tasks for T1, T2,  and PD MRI data shown in Table \ref{tab:expr2} with deviations from training data. C denotes the Cartesian mask type, and G denotes the Gaussian mask type. 
    }
    \label{fig:expr2_boxplot}
\end{figure}

The box plots in Figure \ref{fig:expr2_boxplot} show the comparative study of PSNR and SSIM metrics for the twelve unseen tasks specified in Table \ref{tab:expr2}. We see that joint training, MAML, and MMAML exhibit a drop in performance for PD, T2, and T1, respectively while KM-MAML gives the highest scores in all unseen contrasts. Out of the 24 unseen tasks overall, our method gives the highest scores for 80\% of the tasks in terms of PSNR and 92\% of the tasks in SSIM.

\begin{figure}[t!]
    \centering
    \includegraphics[width=\linewidth]{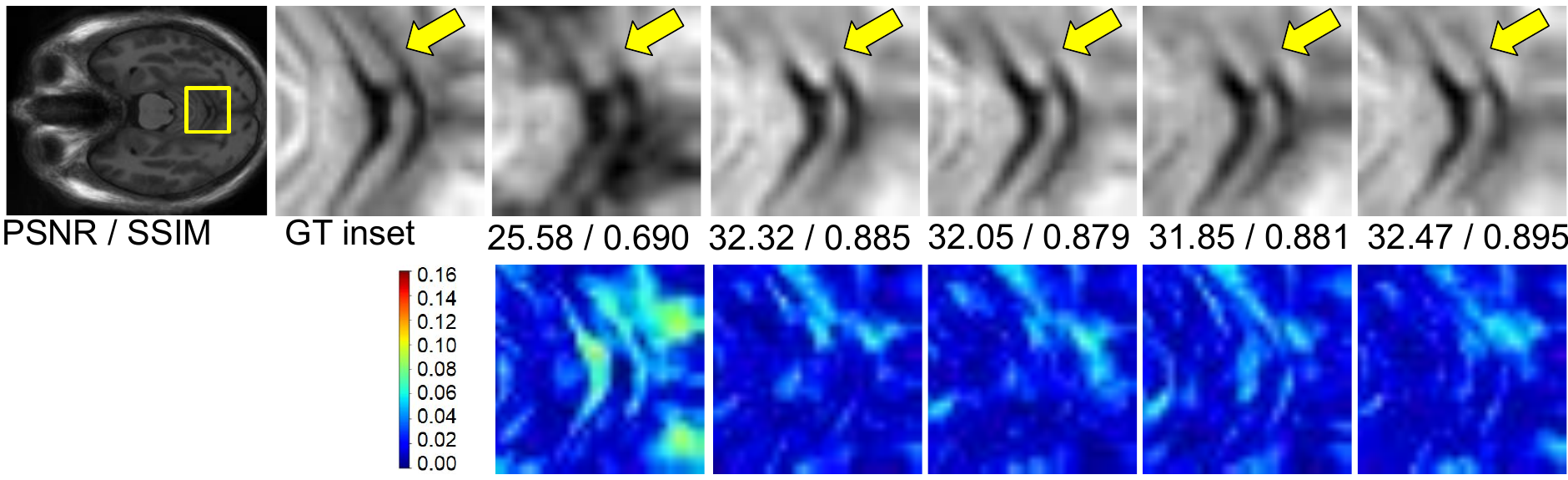}
    \caption{Qualitative comparison of the on-the-fly adaptation capabilities of the methods to unseen contrasts (T1 MRI). From the left, we have GT image, GT ROI, ZF, joint training, MAML, MMAML, and KM-MAML. The figure shows the superior recovery of a pair of structures in the culmen region of the axial T1 MRI for KM-MAML while other methods exhibit relatively more blur in the region.
    }
    \label{fig:expr2_visual_t1}
\end{figure}


\begin{figure}[!]
    \centering
    \includegraphics[width=\linewidth]{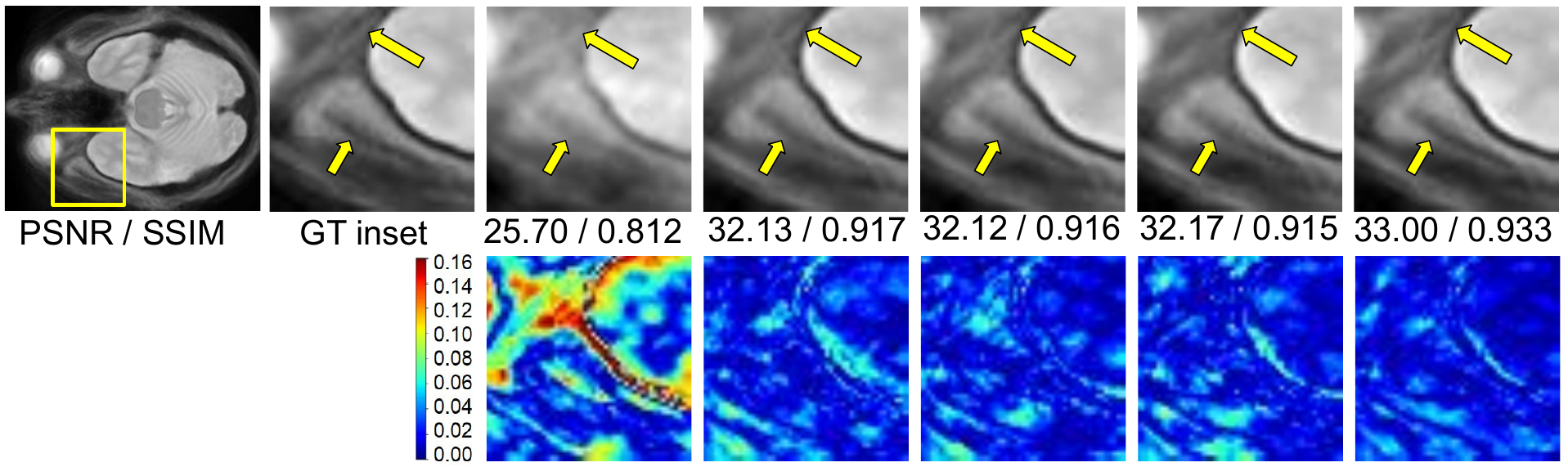}
    \caption{Qualitative Results for PD MRI comparing the target image, zero-filled reconstruction, joint training, MAML, MMAML, and KM-MAML with respect to the on-the-fly adaptation capabilities of the models to multiple contrasts. 
    The two region pointed out by the yellow arrows in the predicted images show that KM-MAML exhibits better recovery of fine details over other methods.
    }
    \label{fig:expr2_visual_pd}
\end{figure}

Figures \ref{fig:expr2_visual_t1}, and \ref{fig:expr2_visual_pd} show the visual results for the unseen contrasts highlighting the regions around the central lobule of the brain image. In both T1 and PD, the recovery of repeated patterns is much closer with respect to the target image than in other methods. Secondly, in the case of PD (Figure \ref{fig:expr2_visual_pd}), the highlighted region indicated by a yellow arrow in the zero-filled input image (third from left) shows a  structure with discontinuity due to aliasing. We notice that other methods still show discontinuity in the image structure while KM-MAML exhibits superior structure recovery. These observations are on par with the least residual error for KM-MAML.

From the quantitative and qualitative results, we see that (i) the meta-initializations of the KM hypernetworks evolve a structure of weights for continuous image generation by extrapolating and interpolating between various contextual settings. (ii) kernel modulation provides a way to re-calibrate the weights by learning global information of multi-modal data and dynamically emphasizing informative features in the base network at each subsampling level. This aspect is very similar to the attention and gating mechanisms \cite{cSE, miccan} applied on CNN features to focus on important regions of the image features. 

\subsubsection{Adaptation via fine-tuning to unseen multimodal MRI contrasts}
\begin{table}[]
\scriptsize
\centering
\caption{Quantitative comparison of SGD, MAML, MMAML, and KM-MAML for adaptation via fine-tuning with 10 gradient steps for various unseen contrasts, acceleration factors, and mask types. Ablative studies for adapting either the base or the modulation network for MMAML and KM-MAML}
\label{tab:expr3}
\begin{tabular}{|ccc|cc|cc|}
\hline
\multicolumn{3}{|c|}{}                                                                                                                                                                       & \multicolumn{2}{c|}{\textbf{Adapt base n/w}}                                                                                         & \multicolumn{2}{c|}{\textbf{Adapt modulation n/w}}                                                                                   \\ \hline
\multicolumn{1}{|c|}{}                                & \multicolumn{1}{c|}{\textbf{SGD}}                                           & \textbf{MAML}                                          & \multicolumn{1}{c|}{\textbf{MMAML}}                                         & \textbf{KM-MAML}                                       & \multicolumn{1}{c|}{\textbf{MMAML}}                                         & \textbf{KM-MAML}                                       \\ \cline{2-7} 
\multicolumn{1}{|c|}{\multirow{-2}{*}{\textbf{Task}}} & \multicolumn{1}{c|}{\begin{tabular}[c]{@{}c@{}}PSNR / \\ SSIM\end{tabular}} & \begin{tabular}[c]{@{}c@{}}PSNR / \\ SSIM\end{tabular} & \multicolumn{1}{c|}{\begin{tabular}[c]{@{}c@{}}PSNR / \\ SSIM\end{tabular}} & \begin{tabular}[c]{@{}c@{}}PSNR / \\ SSIM\end{tabular} & \multicolumn{1}{c|}{\begin{tabular}[c]{@{}c@{}}PSNR / \\ SSIM\end{tabular}} & \begin{tabular}[c]{@{}c@{}}PSNR / \\ SSIM\end{tabular} \\ \hline
\multicolumn{1}{|c|}{T1C5}                             & \multicolumn{1}{c|}{\color[HTML]{3531FF}33.77/ \color[HTML]{000000}.872}                                            & \color[HTML]{009901}33.94/ \color[HTML]{009901}.876                                            & \multicolumn{1}{c|}{33.48/ .867}                                            & 33.67/ \color[HTML]{3531FF}.874                                            & \multicolumn{1}{c|}{33.44/ .866}                                            & 33.65/ \color[HTML]{3531FF}.874                                            \\ \hline
\multicolumn{1}{|c|}{T1C6}                            & \multicolumn{1}{c|}{\color[HTML]{3531FF}31.33/ \color[HTML]{3531FF}.837}                                            & 31.21/ \color[HTML]{3531FF}.837                                            & \multicolumn{1}{c|}{31.30/ \color[HTML]{3531FF}.837}                                            & {\color[HTML]{009901} 31.45/ .844}                     & \multicolumn{1}{c|}{31.25/ .835}                                            & {\color[HTML]{009901} 31.43/ .844}                     \\ \hline
\multicolumn{1}{|c|}{T1C7}                            & \multicolumn{1}{c|}{\color[HTML]{3531FF}32.67/ \color[HTML]{3531FF}.847}                                            & 32.62/ \color[HTML]{009901}.850                                            & \multicolumn{1}{c|}{32.47/ .843}                                            & {\color[HTML]{009901} 32.90/ .854}                     & \multicolumn{1}{c|}{32.39/ .840}                                            & {\color[HTML]{009901} 32.82/ .851}                     \\ \hline
\multicolumn{1}{|c|}{T1C9}                            & \multicolumn{1}{c|}{30.08/ .824}                                            & \color[HTML]{3531FF}30.18/ \color[HTML]{3531FF}.827                                            & \multicolumn{1}{c|}{30.05/ .821}                                            & {\color[HTML]{009901} 30.34/ .834}                     & \multicolumn{1}{c|}{30.05/ .821}                                            & {\color[HTML]{009901} 30.33/ .834}                     \\ \hline
\multicolumn{1}{|c|}{T1G5}                            & \multicolumn{1}{c|}{\color[HTML]{3531FF}35.88/ .852}                                            & 35.86/ \color[HTML]{3531FF}.854                                            & \multicolumn{1}{c|}{35.75/ .852}                                            & {\color[HTML]{009901} 36.01/ .858}                     & \multicolumn{1}{c|}{35.75/ .852}                                            & {\color[HTML]{009901} 35.97/ .857}                     \\ \hline
\multicolumn{1}{|c|}{T1G6}                            & \multicolumn{1}{c|}{\color[HTML]{009901}34.72/ \color[HTML]{009901}.841}                                            & 34.57/ .837                                            & \multicolumn{1}{c|}{34.41/ .834}                                            & {\color[HTML]{3531FF} 34.73/ \color[HTML]{009901}.843}                     & \multicolumn{1}{c|}{34.26/ .831}                                            & {\color[HTML]{3531FF} 34.71/ \color[HTML]{009901}.841}                     \\ \hline
\multicolumn{1}{|c|}{T1G7}                            & \multicolumn{1}{c|}{\color[HTML]{009901}32.45/ \color[HTML]{3531FF}.794}                                            & \color[HTML]{3531FF}32.38/ .792                                            & \multicolumn{1}{c|}{32.15/ .789}                                            & {\color[HTML]{3531FF}32.56/ \color[HTML]{009901}.801}                     & \multicolumn{1}{c|}{32.11/ .788}                                            & {32.44/ \color[HTML]{009901}.798}                     \\ \hline
\multicolumn{1}{|c|}{T1G9}                            & \multicolumn{1}{c|}{\color[HTML]{009901}30.99/ \color[HTML]{000000}.779}                                            & \color[HTML]{009901}30.99/ \color[HTML]{3531FF}.781                                            & \multicolumn{1}{c|}{30.83/ .780}                                            & {\color[HTML]{3531FF} 30.90/ \color[HTML]{009901}.785}                     & \multicolumn{1}{c|}{30.70/ .775}                                            & {\color[HTML]{3531FF} 30.89/ \color[HTML]{009901}.785}                     \\ \hline
\multicolumn{1}{|c|}{PDC5}                            & \multicolumn{1}{c|}{32.22/ .869}                                            & \color[HTML]{3531FF}32.54/ \color[HTML]{3531FF}.876                                            & \multicolumn{1}{c|}{32.39/ .875}                                            & {\color[HTML]{009901} 32.72/ .883}                     & \multicolumn{1}{c|}{32.37/ .875}                                            & {\color[HTML]{009901} 32.70/ .883}                     \\ \hline
\multicolumn{1}{|c|}{PDC6}                            & \multicolumn{1}{c|}{31.71/ .862}                                            & \color[HTML]{3531FF}31.77/ .864                                            & \multicolumn{1}{c|}{31.76/ \color[HTML]{3531FF}.868}                                            & {\color[HTML]{009901} 32.00/ .872}                     & \multicolumn{1}{c|}{31.74/ .867}                                            & {\color[HTML]{009901} 31.99/ .872}                     \\ \hline
\multicolumn{1}{|c|}{PDC7}                            & \multicolumn{1}{c|}{30.49/ .844}                                            & 30.85/ .852                                            & \multicolumn{1}{c|}{\color[HTML]{3531FF}30.90/ \color[HTML]{3531FF}.856}                                            & {\color[HTML]{009901} 31.15/ .861}                     & \multicolumn{1}{c|}{\color[HTML]{3531FF}30.89/ \color[HTML]{3531FF}.856}                                            & {\color[HTML]{009901} 31.15/ .861}                     \\ \hline
\multicolumn{1}{|c|}{PDC9}                            & \multicolumn{1}{c|}{28.72/ .807}                                            & \color[HTML]{3531FF}28.95/ .815                                            & \multicolumn{1}{c|}{28.86/ .813}                                            & {\color[HTML]{009901} 29.20/ .827}                     & \multicolumn{1}{c|}{28.83/ .812}                                            & {\color[HTML]{009901} 29.20/ .827}                     \\ \hline
\multicolumn{1}{|c|}{PDG5}                            & \multicolumn{1}{c|}{36.46/ .894}                                            & 36.35/ .895                                            & \multicolumn{1}{c|}{\color[HTML]{3531FF}36.50/ .900}                                            & {\color[HTML]{009901} 37.10/ .908}                     & \multicolumn{1}{c|}{\color[HTML]{3531FF}36.50/ .900}                                            & {\color[HTML]{009901} 37.03/ .907}                     \\ \hline
\multicolumn{1}{|c|}{PDG6}                            & \multicolumn{1}{c|}{33.84/ .852}                                            & 33.57/ .848                                            & \multicolumn{1}{c|}{\color[HTML]{3531FF}33.87/ .858}                                            & {\color[HTML]{009901} 34.26/ .865}                     & \multicolumn{1}{c|}{\color[HTML]{3531FF}33.81/ .856}                                            & {\color[HTML]{009901} 34.25/ .865}                     \\ \hline
\multicolumn{1}{|c|}{PDG7}                            & \multicolumn{1}{c|}{\color[HTML]{3531FF}32.99/ \color[HTML]{000000}.838}                                            & 32.72/ .832                                            & \multicolumn{1}{c|}{32.85/ \color[HTML]{3531FF}.840}                                            & {\color[HTML]{009901} 33.29/ .852}                     & \multicolumn{1}{c|}{32.72/ .836}                                            & {\color[HTML]{009901} 33.27/ .851}                     \\ \hline
\multicolumn{1}{|c|}{PDG9}                            & \multicolumn{1}{c|}{\color[HTML]{3531FF}31.50/ \color[HTML]{000000}.804}                                            & 31.32/ .804                                            & \multicolumn{1}{c|}{31.38/ \color[HTML]{3531FF}.809}                                            & {\color[HTML]{009901} 31.91/ .824}                     & \multicolumn{1}{c|}{31.30/ \color[HTML]{3531FF}.805}                                            & {\color[HTML]{009901} 31.90/ .824}                     \\ \hline
\end{tabular}
\end{table}
The purpose of this experiment (Table \ref{tab:expr3}) is twofold. (i) To compare the adaptation capabilities of the models via fine-tuning to a few gradient steps (ii) To understand whether the weights of the KM hypernetworks are extensible and favorable for rapid adaptation to unseen tasks. We analyze the second aspect with an ablative study of adapting either the base network (columns 4 and 5 in Table \ref{tab:expr3} and Refer Algorithm \ref{kMAMLalgoadapt} in Appendix) or the KM hypernetworks (columns 6 and 7 in Table \ref{tab:expr3}) in MMAML and KM-MAML. A possible scenario wherein adapting the KM hypernetworks would be faster and more efficient is when the base network is very deep with more parameters while the hypernetworks are lightweight with relatively lesser weights. We evaluate the adaptation performance on 16 unseen tasks with T1 and PD contrasts with unseen acceleration factors 5x, 6x, 7x, and 9x and unseen mask patterns. 

\begin{figure}[t!]
    \centering
    \includegraphics[width=\linewidth]{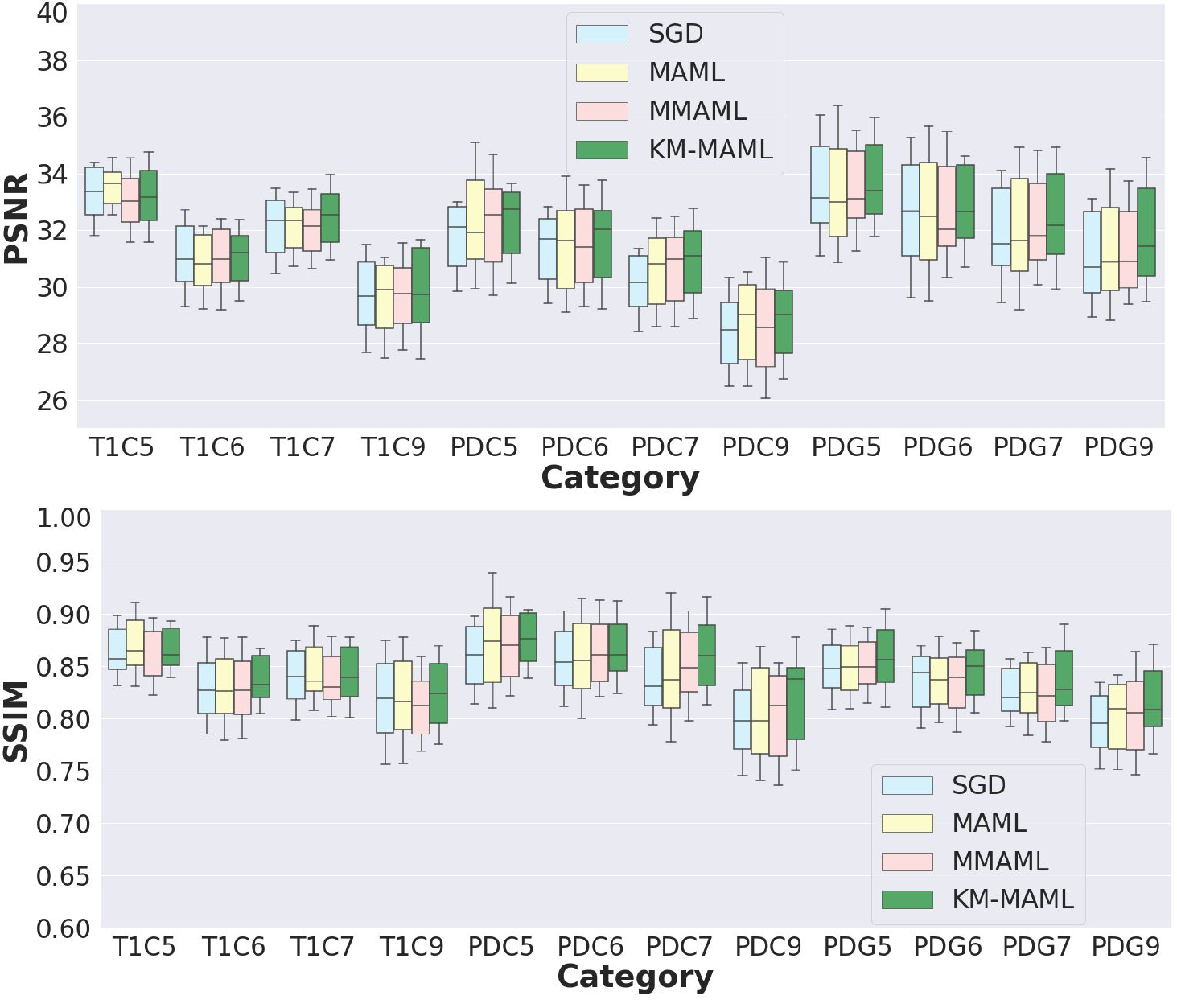}
    \caption{PSNR and SSIM plots comparing generalization through adaptation in few gradient steps for joint training, MAML, MMAML, and KM-MAML for 12 tasks with deviated contrasts, acceleration factors, and mask types for T1, T2, and PD respectively deviated from training data. C denotes the Cartesian mask type, and G denotes the Gaussian mask type. For instance, T2C7 indicates an unseen task with T2 MRI contrast with a Cartesian mask pattern with 7x acceleration. 
    }
    \label{fig:expr3_boxplot}
\end{figure}

\begin{figure}[t!]
    \centering
    \includegraphics[width=\linewidth]{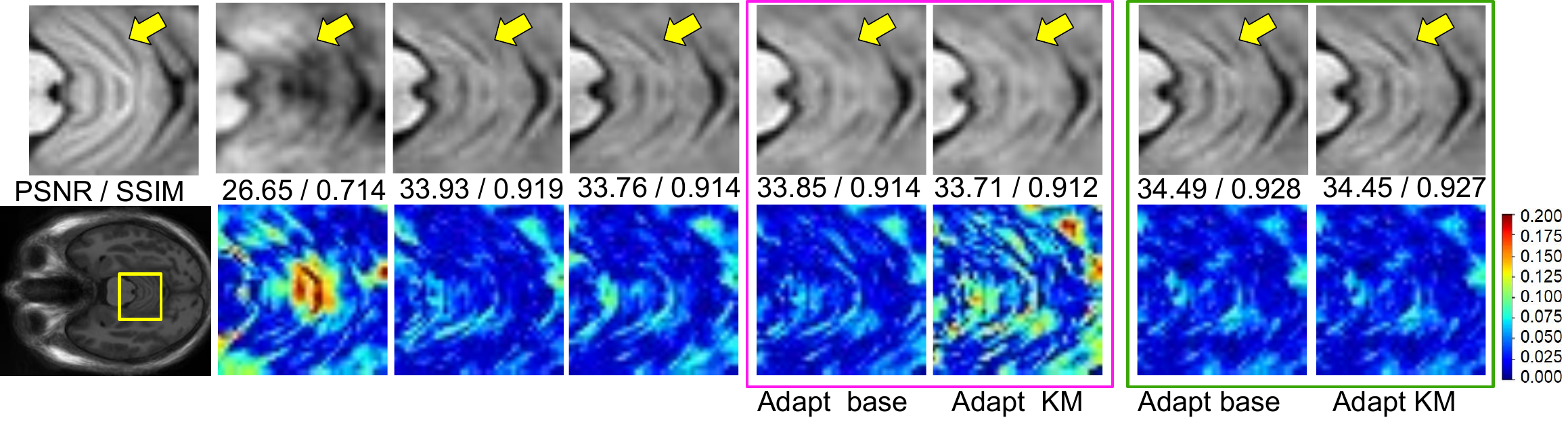}
    \caption{Qualitative Results for T1 MRI comparing (from the top left) the target image, ZF image, joint training, MAML, MMAML (pink box), and KM-MAML (green box) with respect to the adaptation capabilities in a few gradient steps for multiple contrasts. The structure pointed out by the yellow arrow shows that the missing structure in the under-sampled input is inadequately captured by other methods while KM-MAML is able to recover the structure with improved accuracy. In the ablative studies, the pink box shows the visual results of MMAML, and the green box shows the results for KM-MAML.
    }
    \label{fig:expr3_visual_t1}
\end{figure}

\begin{figure}[t!]
    \centering
    \includegraphics[width=\linewidth]{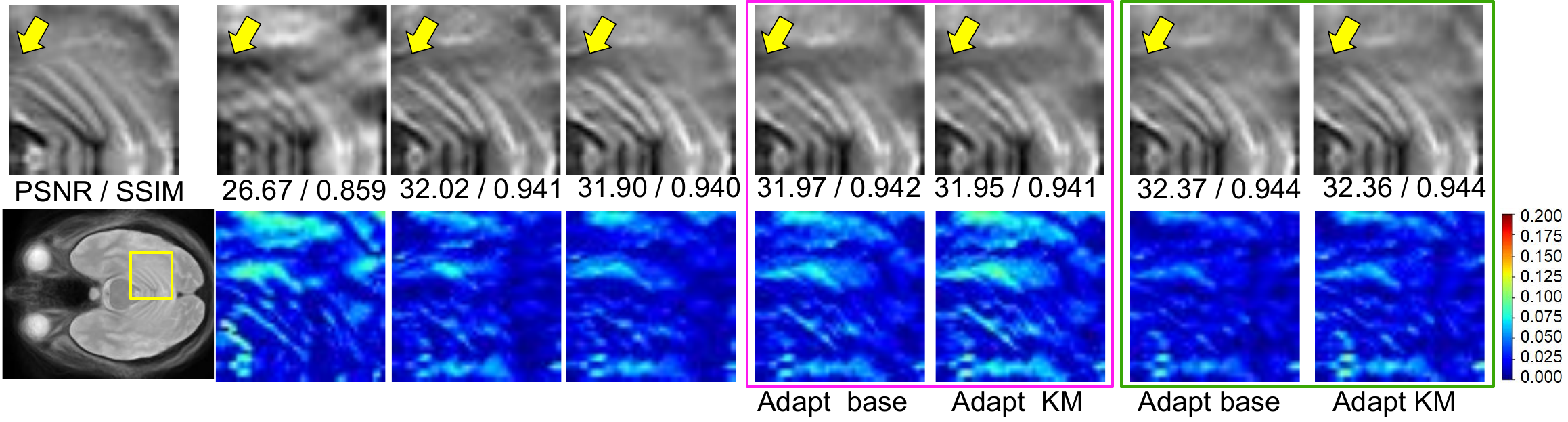}
    \caption{Qualitative Results for PD MRI comparing the target image, ZF image, joint training, MAML, MMAML, and KM-MAML with respect to the adaptation capabilities in a few gradient steps for multiple contrasts. The structure pointed out by the yellow arrow shows that other methods exhibit artifacts present in the ZF image while in KM-MAML the reconstruction is closer to the target. In the ablative studies, the pink box shows the visual results of MMAML, and the green box shows the results for KM-MAML.
    }
    \label{fig:expr3_visual_pd}
\end{figure}

Our observations from the table are as follows:  (i) Our first objective is met, wherein KM-MAML exhibits better adaptation over other methods, improving further upon KM (refer to box plots in Figure \ref{fig:expr3_boxplot}). This observation shows that the proposed model provides stronger mode-specific meta-initializations to adapt in a few gradient steps. (ii) Our ablative study shows that KM-MAML outperforms MMAML in both cases of adapting either the base or the modulation network. Also, comparing columns 4 and 7, KM-MAML is better than MMAML. This observation shows that fine-tuning the KM hypernetworks enables adaptation to unseen related tasks in the contextual space while preserving the image reconstruction features learned by the base network. (iii) Adapting the KM hypernetworks gives a competitive performance with adapting the base network.

The qualitative results for adaptation through fine-tuning are shown in Figures \ref{fig:expr3_visual_t1} and \ref{fig:expr3_visual_pd}. Figures show that KM-MAML provides better reconstructions when compared with other methods. The improvement is consistent as compared to MMAML in both cases of adapting either the base network or the modulation network to unseen tasks. 

\subsubsection{Interpreting the kernel modulation-based meta-learning}
To gain insights into kernel modulation, we analyze KM-MAML by comparing the nearness between the on-the-fly adaptation versus adaptation in few gradient steps. Furthermore, we perform a representational analysis using Centered Kernel Alignment (CKA) \cite{cka, featurereuse} metric to understand the ability of the kernel modulation network in learning mode-specific discriminative features. The CKA gives a number between 0 and 1 to quantify the correlation between representations. For instance, a CKA value of 0.7 between pre and post modulation features of a base network layer means that 30\% of discriminative knowledge associated with a task is induced post-modulation. 

\paragraph{Comparing on-the-fly adaptation with fine-tuning}

We compare the on-the-fly adaptation performance with adaptation via fine-tuning of the KM network to understand the contribution of gradient-based meta-learning for fine-tuning. 
Table \ref{tab:expr4} compares on-the-fly adaptation (or the post modulation step) and adaptation via fine-tuning to 10 and 30 gradient steps further after post-modulation. 
The post-modulation performance improves with adaptation (highlighted in green). 
The results show that with mode-specific initializations provided by the KM, gradient-based meta-learning can further improve the performance on unseen tasks. 
We also note that the SSIM values post-modulation exhibit closeness to post-adaptation (metrics highlighted in blue), implying that context-aware meta-initializations have gained highly reusable features for new MRI contrasts with deviated acquisition settings.

\begin{table}[t!]
\scriptsize
\centering
\caption{Quantitative comparison of on-the-fly adaptation and adaptation through fine-tuning to 10 and 30 gradient steps. The table shows the adaptation performance for various configurations of T2 and PD MRI contrasts. Note that blue indicates the similarity between on-the-fly and fine-tuning performance. Green showcases the benefits of gradient-based meta-learning by improving the accuracy via fine-tuning in few gradient steps.}
\label{tab:expr4}
\begin{tabular}{|c|c|c|c|}
\hline
\multicolumn{1}{|l|}{}                                & \textbf{\begin{tabular}[c]{@{}c@{}}KM-MAML \\ (On-the-fly)\end{tabular}} & \textbf{\begin{tabular}[c]{@{}c@{}}KM-MAML \\ (10 STEPS)\end{tabular}} & \textbf{\begin{tabular}[c]{@{}c@{}}KM-MAML \\ (30 STEPS)\end{tabular}} \\ \cline{2-4} 
\multicolumn{1}{|l|}{\multirow{-2}{*}{\textbf{TASK}}} & \textbf{PSNR / SSIM}                                                     & \textbf{PSNR / SSIM}                                                   & \textbf{PSNR / SSIM}                                                   \\ \hline
T2C4                                                  & 36.20 / \color[HTML]{3531FF} 0.929                                                            & 36.25 / 0.928                                                          & { 36.31 / \color[HTML]{3531FF} 0.929}                                   \\ \hline
T2C5                                                  & 35.58 / \color[HTML]{3531FF} 0.919                                                            & 35.66 / 0.918                                                          & { 35.69 / \color[HTML]{3531FF} 0.919}                                   \\ \hline
T2C7                                                  & 31.73 / \color[HTML]{3531FF} 0.857                                                            & 31.80 / \color[HTML]{3531FF} 0.858                                                          & { 31.84 / \color[HTML]{3531FF} 0.858}                                   \\ \hline
T2C9                                                  & 30.60 / \color[HTML]{3531FF} 0.860                                                             & 30.61 / 0.859                                                          & {\color[HTML]{009901} 30.66 / \color[HTML]{3531FF} 0.861}                                   \\ \hline
T2G4                                                  & 39.07 / 0.921                                                            & 39.15 / 0.922                                                          & {\color[HTML]{009901} 39.19 / 0.923}                                   \\ \hline
T2G5                                                  & 36.66 / 0.889                                                            & 36.79 / 0.891                                                          & {\color[HTML]{009901} 36.84 / 0.892}                                   \\ \hline
T2G7                                                  & 34.18 / 0.855                                                            & 34.27 / 0.857                                                          & {\color[HTML]{009901} 34.30 / 0.857}                                   \\ \hline
T2G8                                                  & 33.57 / 0.845                                                            & 33.65 / 0.847                                                          & {\color[HTML]{009901} 33.70 / 0.848}                                   \\ \hline
PDC4                                                  & 33.95 / \color[HTML]{3531FF} 0.904                                                            & { 33.98 / \color[HTML]{3531FF}0.904}                                   & 33.95 / \color[HTML]{3531FF}0.904                                                          \\ \hline
PDC5                                                  & 32.70 / \color[HTML]{3531FF} 0.883                                                            & 32.71 / \color[HTML]{3531FF} 0.883                                                          & {\color[HTML]{009901} 32.72 / \color[HTML]{3531FF} 0.883}                                   \\ \hline
PDC6                                                  & 31.95 / \color[HTML]{3531FF} 0.871                                                            & 32.00 / \color[HTML]{3531FF} 0.872                                                          & {\color[HTML]{009901} 32.02 / 0.873}                                   \\ \hline
PDC7                                                  & 31.11 / \color[HTML]{3531FF} 0.861                                                            & 31.15 / \color[HTML]{3531FF} 0.861                                                          & {\color[HTML]{009901} 31.17 / 0.862}                                   \\ \hline
PDG4                                                  & 39.31 / \color[HTML]{3531FF} 0.934                                                            & 39.34 / \color[HTML]{3531FF} 0.935                                                          & {\color[HTML]{009901} 39.40 / 0.936}                                   \\ \hline
PDG5                                                  & 37.02 / 0.906                                                            & 36.97 / 0.905                                                          & {\color[HTML]{009901} 37.16 / 0.910}                                   \\ \hline
PDG6                                                  & 34.23 / \color[HTML]{3531FF} 0.864                                                            & {\color[HTML]{009901} 34.26 / \color[HTML]{3531FF} 0.865}                                   & 34.24 / \color[HTML]{3531FF}0.864                                                          \\ \hline
PDG7                                                  & 33.26 / \color[HTML]{3531FF} 0.851                                                            & 33.29 / \color[HTML]{3531FF} 0.852                                                          & {\color[HTML]{009901} 33.32 / 0.853}                                   \\ \hline
\end{tabular}
\end{table}

\paragraph{Representational Similarity Analysis}
\begin{figure}[t!]
    \centering
    \includegraphics[width=0.8\linewidth]{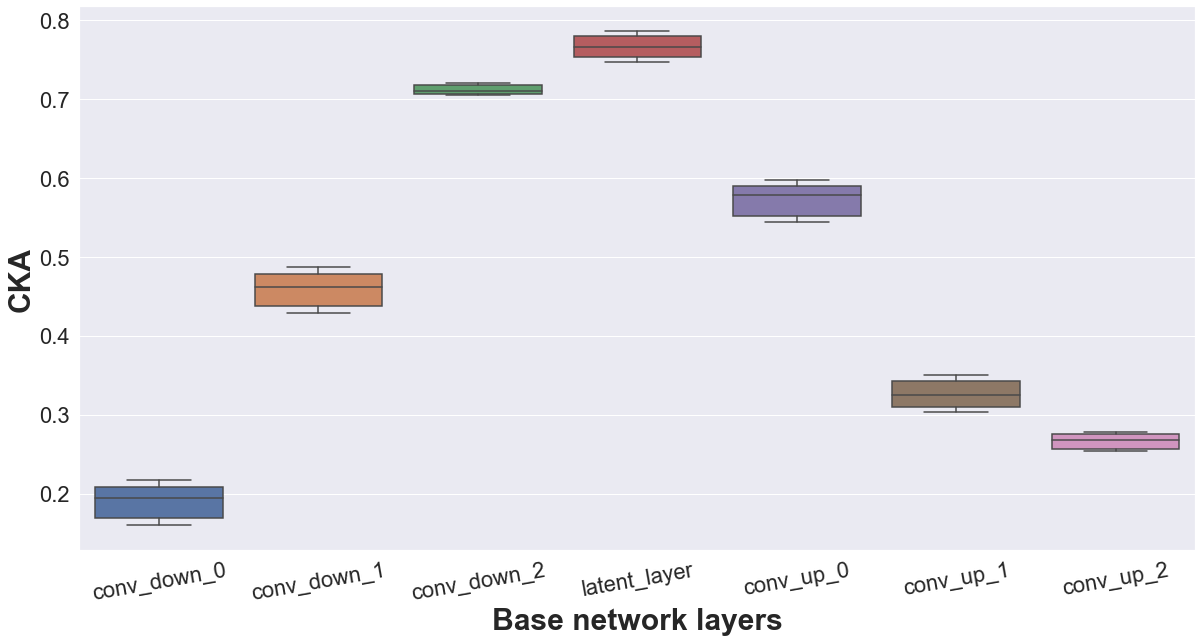}
    \caption{CKA similarity scores between the representations of the seven layers of the encoder-decoder base reconstruction network pre and post kernel modulation. Low CKA similarity at the top layers of the encoder and decoder shows that these layers must change significantly to recover structures according to different modes of the MRI contrasts. The CKA similarity gradually increases with its peak in the latent layer which shows that low frequency details of related multi-contrast images exhibit hight similarity. In the figure $conv\_down\_0$ and $conv\_up\_2$ are the top most levels of the encoder-decoder network. $conv\_down\_1$ and $conv\_up\_1$ form the second sub-sampling level and so on. The $latent\_layer$ is the bottleneck convolution layer. Note that the variations in the box plots are small indicating that hill-like profile is consistent across all the unseen contrasts (T1, T2, and PD).
    }
    \label{fig:cka}
\end{figure}
We study the extent to which the base neural network's latent representations (activations) change based on task-specific kernel modulation. Following the recent works \cite{featurereuse,cka}, we measure the changes in the representations before and after KM based on the CKA metrics to compare similarities in patterns. Figure \ref{fig:cka} shows the CKA plots for the three down-sampling layers, the bottleneck layer, and the three upsampling layers of the encoder-decoder base network. The plots show the mean CKA values taken across the 24 unseen MRI tasks. 
From the plot, our observations are: (i) The CKA metrics are the least, with around 0.2 to 0.3 in the highest levels of the encoder and decoder. This observation shows that these layers, which play a crucial role in recovering fine-grained image details \cite{understandEDCNN} for reconstruction, obtain around 70 to 80\% mode-specific discriminative knowledge post KM. (ii) Representational similarity increases over sub-sampling levels and reaches a maximum of about 0.78 at the bottleneck layer from the encoder. This observation indicates that lower resolution layers capture low-frequency details that are mode-invariant. 
Each KM hypernetwork of the corresponding base layer takes varying roles of learning mode-specific and mode-invariant knowledge wherein the top layers learn more mode-specific features while the bottom layer learns more mode-invariant features. (iii) CKA values gradually decrease in the decoder layers as we move towards the output layer to around 28\%. This observation is very similar to the CKA analysis in Almost no inner loop (ANIL)\cite{featurereuse} for classification tasks wherein the head of the network exhibits the least CKA similarity while the intermediate layers exhibit higher CKA scores. This analysis provides an architecture-level interpretation of kernel modulation of the base network. 



\begin{figure}[t!]
    \centering
    \includegraphics[width=\linewidth]{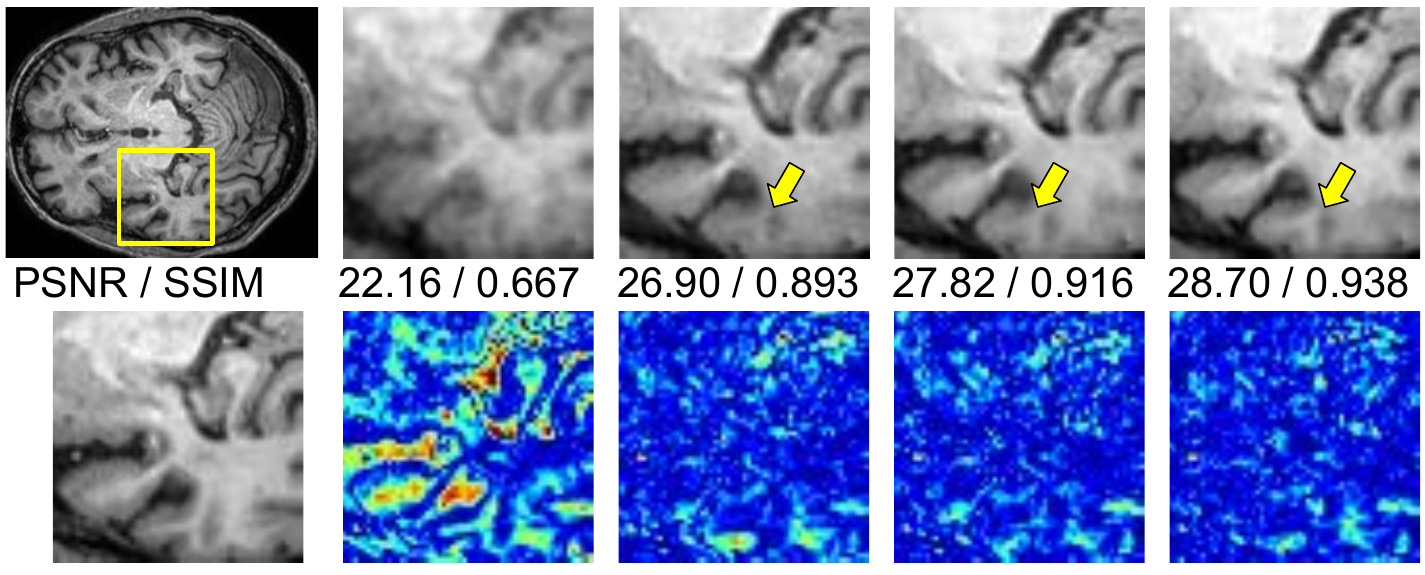}
    \caption{Qualitative results for the ablative study. Top: From the left are ZF image, and the predictions of U-Net, U-Net + GBML, and U-Net + KM + GBML. Bottom: From the left, target inset, residual error images with respect to the target for ZF, U-Net, U-Net + GBML, and U-Net + KM + GBML}
    
    \label{fig:ablative}
\end{figure}

\begin{table}[t!]
\scriptsize
\centering
\caption{Ablative study showing the role of the base network alone with SGD (single level of optimization), base network alone with gradient-based meta-learning which has two levels of optimization, and the proposed model with base network and kernel modulation network as the architecture and gradient-based meta-learning as the learning process. T1 denotes T1 MRI contrast, C - Cartesian mask pattern, and G - Gaussian mask pattern. For example, T1C8 indicates T1 MRI image with Cartesian mask under-sampling at 5x acceleration.}
\label{tab:ablative}
\begin{tabular}{|c|c|c|c|}
\hline
                       & \begin{tabular}[c]{@{}c@{}}U-Net w/o \\ GBML\end{tabular} & \begin{tabular}[c]{@{}c@{}}U-Net + \\ GBML\end{tabular} & \begin{tabular}[c]{@{}c@{}}U-Net + KM \\ + GBML\end{tabular} \\ \cline{2-4} 
\multirow{-2}{*}{Task} & PSNR / SSIM                                               & PSNR / SSIM                                             & PSNR / SSIM                                                  \\ \hline
T1C4                   & 36.09 / 0.889                                             & 36.45 / 0.900                                           & 36.75 / 0.915                        \\ \hline
T1C5                   & 34.78 / 0.861                                             & 35.22 /  0.875                                          & 35.58 / 0.895                        \\ \hline
T1C8                   & 32.52 / 0.817                                             & 32.99 /  0.839                                          & 33.37 / 0.864                        \\ \hline
T1G4                   & 40.20 / 0.913                                             & 40.59 / 0.922                                           & 41.02 / 0.935                        \\ \hline
T1G5                   & 38.20 / 0.874                                             & 38.67 / 0.887                                           & 39.05 / 0.906                        \\ \hline
T1G8                   & 34.85 / 0.802                                             & 35.33 / 0.820                                           & 35.96 / 0.859                        \\ \hline
\end{tabular}
\end{table}

\subsubsection{Ablative study of the model and the learning process}
We perform an ablative study to understand the role of meta-learning and kernel modulation. We consider 6 acquisition contexts with T1 MRI images, Cartesian and Gaussian mask patterns, and acceleration factors 4x, 5x, and 8x.  Table \ref{tab:ablative}  and Figure \ref{fig:ablative} show the quantitative and qualitative results, respectively, for the ablative study considering three cases, 1) Only the base network with conventional joint training involving a single level of optimization, 2) Only the base network trained with gradient-based meta-learning involving two-level optimization at the task level, and 3) With both base and kernel modulation networks within the gradient-based meta-learning process. From the results, it is clear that under variations in the contexts, GBML can provide better meta-initializations as compared to conventional joint training. Furthermore, combining model-based meta-learning via the context-specific kernel modulation using hypernetworks and optimization-based meta-learning enhances the learning with improvement margins of $\sim$1 dB in PSNR and $\sim$0.01 in SSIM.


\section{Model-based Comparative Studies}
\label{sec:modelcompare}

\subsubsection{Comparison against context-specific MRI reconstruction architectures}

\begin{table}[]
\scriptsize
\centering
\caption{Quantitative comparison with context-specific deep cascaded MRI reconstruction architectures for fixed anatomy under study (cardiac), acceleration factors 4x and 5x, and Cartesian mask type. Green indicates that our model can operate on multiple acquisition contexts while matching the performance of context-specific models.}
\label{tab:sotasc}
\begin{tabular}{|c|c|c|}
\hline
                                  & \textbf{4x}          & \textbf{5x}                           \\ \cline{2-3} 
{\textbf{Method}} & \textbf{PSNR / SSIM} & \textbf{PSNR / SSIM}                  \\ \hline
ZF    & 24.27 / 0.699        & 23.82 / 0.674 \\ \hline
DAGAN \cite{dagan}    & 28.52 / 0.841        & 28.02 / 0.825 \\ \hline
DC-CNN \cite{dc_cnn}    & 32.75 / 0.920        & 31.75 / 0.905 \\ \hline
DC-DEN \cite{dc-ensemble}    & 33.22 / 0.925        & 32.30 / 0.913 \\ \hline
DC-RDN \cite{recursive_dilated}    & 32.95 / 0.923        & 32.09 / 0.912 \\ \hline
DC-UNet \cite{dc_unet}   & 33.17 / 0.928        & 32.55 / \color[HTML]{009901}0.919 \\ \hline
MICCAN \cite{miccan}                           & 33.38 / \color[HTML]{009901}0.930        & 32.52 / 0.918
\\ \hline
MAC-ReconNet \cite{mac}                           & 32.98 / 0.923        & 32.07 / 0.910                         \\ \hline
OUCR \cite{oucr}                           & 32.98 / 0.923        & 32.07 / 0.910                         \\ \hline
ViT \cite{vit}                           & 28.55 / 0.828        & 27.86 / 0.806                         \\ \hline
SWIN \cite{swin}                           & 30.15 / 0.869        & 29.24 / 0.848                         \\ \hline
SFT-KD-Recon \cite{mac}                           & 32.03 / 0.907        & 30.93 / 0.888                         \\ \hline
KM-MAML                           & \color[HTML]{009901}33.40 / 0.930        & \color[HTML]{009901}32.64 / 0.919                         \\ \hline
\end{tabular}
\end{table}

\begin{figure}[t!]
    \centering
    \includegraphics[width=0.85\linewidth]{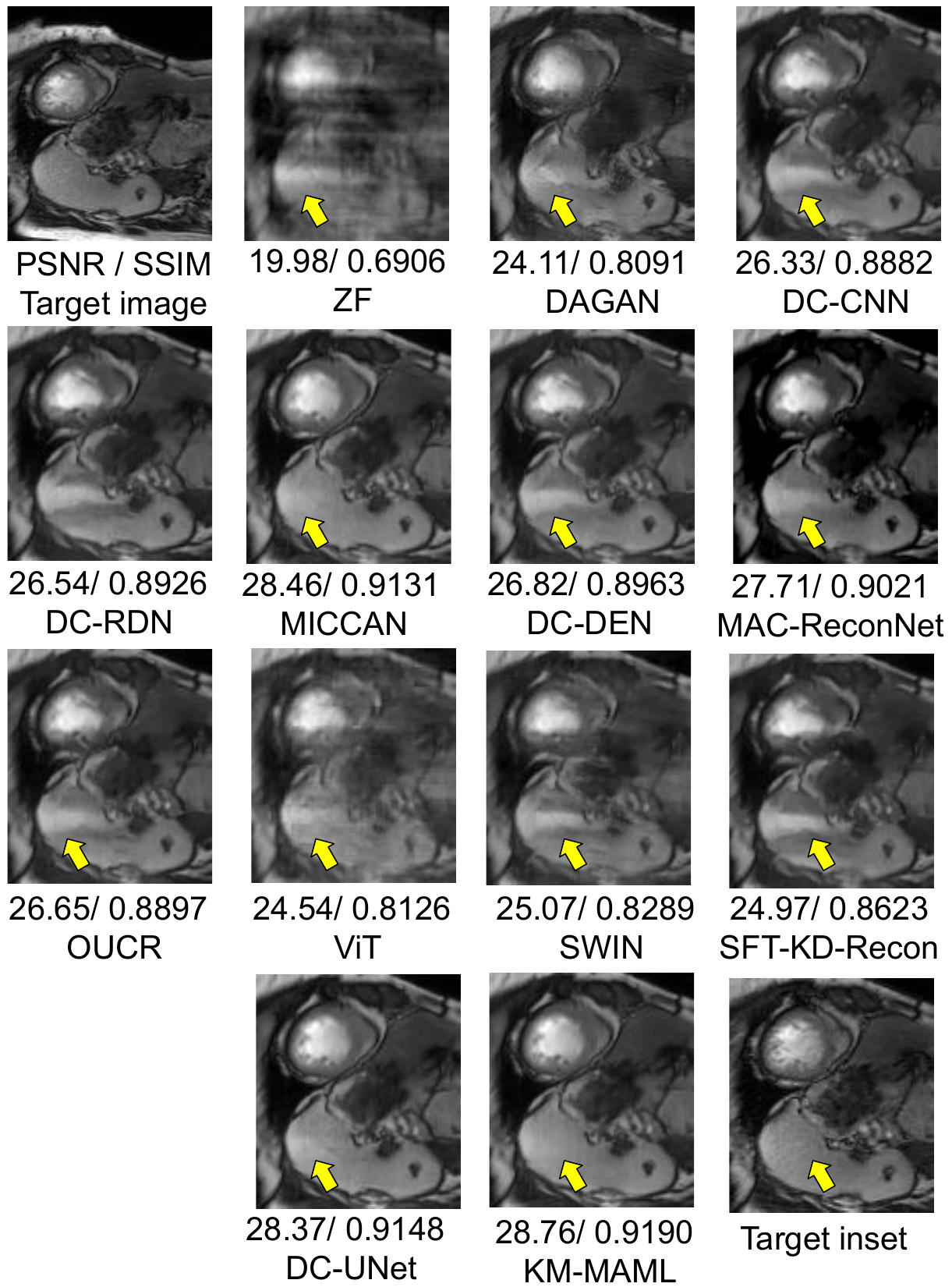}
    \caption{Qualitative comparison of KM-MAML in deep-cascaded mode against ZF, context-specific and adaptive MRI reconstruction architectures. The yellow arrows in the figures point below the ventricle region of the cardiac MRI anatomy with a heavy aliasing on the ZF image. KM-MAML recovers finer details much closer to the target as compared to other methods where the aliasing artifacts are still present.
    }
    \label{fig:sotasc}
\end{figure}

We provide architecture-based performance comparisons to show that our method can operate in multiple acquisition context-based configurations as well as match the reconstruction performance of other context-specific networks. We perform kernel modulation using a single set of KM hypernetworks on the base network (U-Net) in deep cascaded mode \cite{dc_cnn}. We compare our network against DAGAN, and various deep cascaded CNNs that are trained for a specific context - cardiac anatomy, a fixed Cartesian under-sampling pattern, and a specific acceleration factor (4x or 5x).  We train the adaptive MRI reconstruction models, MAC-ReconNet and KM-MAML on ten acquisition configurations with the cardiac anatomy, varying mask types - Gaussian and Cartesian, and varying acceleration factors - 2x, 3.3x, 4x, 5x, and 8x.

Table \ref{tab:sotasc} shows that our model can operate in multiple configurations and outperform the context-specific models in terms of PSNR and SSIM. We also note that as compared to MAC-ReconNet our model performs better when trained in multiple contexts. Unlike our network which has both context-specific weights provided by the KM hypernetworks and context-invariant weights of the base network, MAC-ReconNet lacks context-invariant weights of the base network, as all the weights of the base network are predicted by the hypernetworks. The qualitative results (Figure. \ref{fig:sotasc}) reveal that KM-MAML is able to reconstruct images closer to the target image as compared to other methods.

\subsubsection{Multiple Acquisition Contexts - Multiple Anatomies and Image Resolution levels}
\label{sec:mulanatomy}

As the main goal of this work is to enable more capabilities to approximate diverse contexts, we compare the scalability of our network with other networks when combining multiple anatomies, contrasts, and image resolution levels ( $320 \times 320$ and $240 \times 240$), mask pattern and acceleration factors, making to 12 contexts in a single training. For our comparative study, we take two of the top-performing architectures from Table \ref{tab:sotasc}, the DC-Unet (encoder-decoder CNN like MICCAN) and DC-DEN (DenseNet CNN configuration). Table \ref{tab:mulanatomy} shows the comparative study of the scalability of these models and our method to multiple acquisition contexts. Under diverse data settings, the modulated weights in the proposed method provide context-adaptive reconstruction and exhibit better metrics than DC-UNet and DC-DEN. 
\begin{table}[t!]
\scriptsize
\centering
\caption{Comparative study of scalability of context-specific networks (DC-Unet and DC-DEN) and the proposed network when combining multiple anatomies with different image resolution levels, mask patterns, and acceleration factors. T1 denotes T1 contrast, and PD denotes proton density-weighted contrast. B and K denote brain and knee anatomies. C and G denote Cartesian and Gaussian mask patterns. For example, the task PDKC5 denotes the context with PD knee, Cartesian, 5x acceleration.}
\label{tab:mulanatomy}
\begin{tabular}{|c|c|c|c|}
\hline
\multirow{2}{*}{Task} & DC-Unet        & DC-DEN        & KM-MAML       \\ \cline{2-4} 
                      & PSNR / SSIM    & PSNR / SSIM   & PSNR / SSIM   \\ \hline
T1BC5                 & 42.67 / 0.984  & 42.87 / 0.987 & 43.64 / 0.990 \\ \hline
T1BC8                 & 38.81 /  0.967 & 38.25 / 0.967 & 39.33 / 0.974 \\ \hline
T1BG5                 & 49.10 / 0.990  & 50.48 / 0.996 & 51.50 / 0.998 \\ \hline
T1BG8                 & 46.27 / 0.986  & 46.44 / 0.991 & 47.99 / 0.994 \\ \hline
PDKC5                 & 35.64 / 0.921  & 35.39 / 0.916 & 35.83 / 0.923 \\ \hline
PDKC8                 & 32.81 / 0.877  & 32.18 / 0.865 & 32.96 / 0.880 \\ \hline
PDKG5                 & 37.42 / 0.935  & 37.20 / 0.929 & 37.82 / 0.940 \\ \hline
PDKG8                 & 34.58 / 0.895  & 33.00 / 0.869 & 35.00 / 0.902 \\ \hline
\end{tabular}
\end{table}

\begin{figure}[t!]
    \centering
    \includegraphics[width=\linewidth]{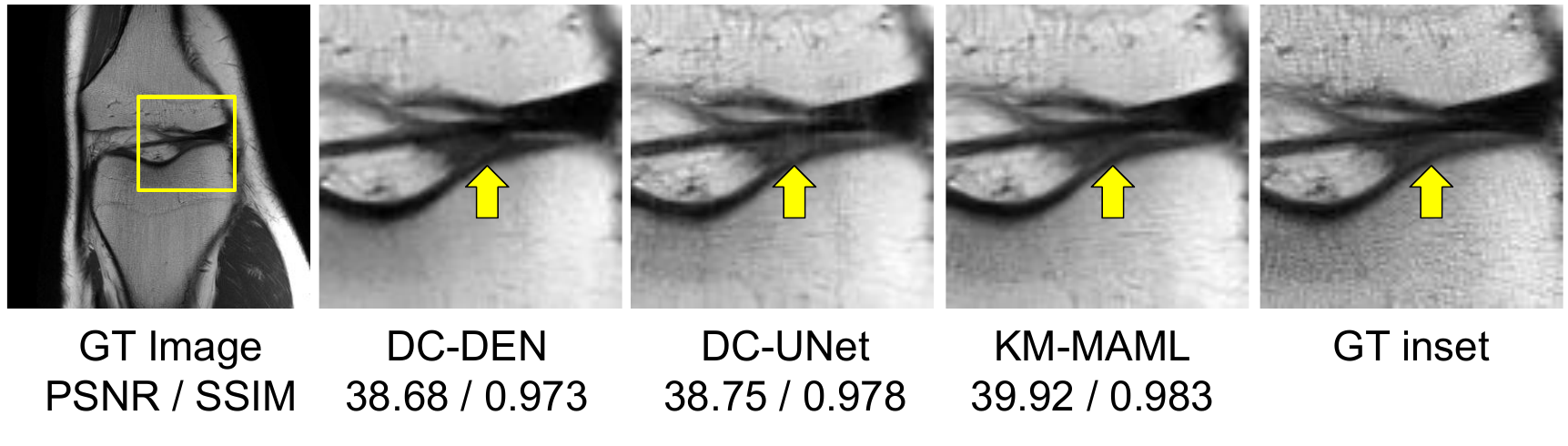}
    \caption{Qualitative comparison of KM-MAML with DC-DEN and DC-UNet when combining multiple anatomies, mask patterns, and acceleration factors. The yellow arrows pointing at the medial meniscus and ligament regions of the knee MRI anatomy show that KM-MAML reconstructs the image much closer to the target as compared to DC-DEN and DC-UNet under heterogeneous data scenario.
    }
    \label{fig:knee}
\end{figure}
The improvement margins over DC-UNet are $\sim$ 0.8 dB in PSNR and $\sim$ 0.006 in SSIM, while the improvement margins over DC-DEN are $\sim$ 1 dB in PSNR and $\sim$ 0.01 in SSIM. We believe the main reason for this observation is that DC-DEN and DC-UNet learn a single set of shared weights that pull the reconstruction towards an average of the possible reconstructions equidistant from all diverse acquisition contexts considered at train time. These weights are inadequate to scale to multiple diverse contexts. On the other hand, KM-MAML exhibits dynamic context-specific weights at inference time. The qualitative results for the knee context in Figure \ref{fig:knee} show that the reconstructed image corresponding to DC-DEN shows missing structures and that corresponding to DC-UNet shows smudged regions, while the predicted image of KM-MAML is much closer to the target image.

\begin{table}[t!]
\scriptsize
\centering
\caption{Comparative study of context-specific networks (DC-Unet and DC-DEN) and the proposed network when transfer learning from knee $320\times320$ and brain $240\times240$ contexts to cardiac $150\times150$ with Gaussian, radial and Cartesian mask patterns and acceleration factors 3.3x, 5x and 10x. CC, CR, and CG indicate the cardiac anatomy with Cartesian, radial, and Gaussian mask under-sampling patterns, respectively.  For instance, task CR5 denotes cardiac anatomy undersampled with radial masks with 5x under-sampling. The numbers highlighted in green show that the fine-tuning performance (Epoch 1 and 2) of KM-MAML matches the transfer learned performance (Epoch 10) of context-specific networks. }
\label{tab:transfer}
\begin{tabular}{|c|cc|cc|ccc|}
\hline
                                & \multicolumn{2}{c|}{\textbf{DC-DEN}}                                                                                                                 & \multicolumn{2}{c|}{\textbf{DC-UNet}}                                                                                                                & \multicolumn{3}{c|}{\textbf{KM-MAML}}                                                                                                                                                                                                                                                   \\ \cline{2-8} 
                                & \multicolumn{1}{c|}{\textbf{\begin{tabular}[c]{@{}c@{}}Epoch \\ 1\end{tabular}}}    & \textbf{\begin{tabular}[c]{@{}c@{}}Epoch \\ 10\end{tabular}}   & \multicolumn{1}{c|}{\textbf{\begin{tabular}[c]{@{}c@{}}Epoch \\ 1\end{tabular}}}    & \textbf{\begin{tabular}[c]{@{}c@{}}Epoch \\ 10\end{tabular}}   & \multicolumn{1}{c|}{\textbf{\begin{tabular}[c]{@{}c@{}}Epoch \\ 1\end{tabular}}}                   & \multicolumn{1}{c|}{\textbf{\begin{tabular}[c]{@{}c@{}}Epoch \\ 2\end{tabular}}}                   & \textbf{\begin{tabular}[c]{@{}c@{}}Epoch\\ 10\end{tabular}}                   \\ \cline{2-8} 
\multirow{-3}{*}{\textbf{Task}} & \multicolumn{1}{c|}{\textbf{\begin{tabular}[c]{@{}c@{}}PSNR/ \\ SSIM\end{tabular}}} & \textbf{\begin{tabular}[c]{@{}c@{}}PSNR/ \\ SSIM\end{tabular}} & \multicolumn{1}{c|}{\textbf{\begin{tabular}[c]{@{}c@{}}PSNR/ \\ SSIM\end{tabular}}} & \textbf{\begin{tabular}[c]{@{}c@{}}PSNR/ \\ SSIM\end{tabular}} & \multicolumn{1}{c|}{\textbf{\begin{tabular}[c]{@{}c@{}}PSNR/ \\ SSIM\end{tabular}}}                & \multicolumn{1}{c|}{\textbf{\begin{tabular}[c]{@{}c@{}}PSNR/ \\ SSIM\end{tabular}}}                & \textbf{\begin{tabular}[c]{@{}c@{}}PSNR/ \\ SSIM\end{tabular}}                \\ \hline
CC3.3                            & \multicolumn{1}{c|}{\begin{tabular}[c]{@{}c@{}}33.58/\\ .934\end{tabular}}          & \begin{tabular}[c]{@{}c@{}}34.30/\\ .941\end{tabular}          & \multicolumn{1}{c|}{\begin{tabular}[c]{@{}c@{}}34.24/\\ .943\end{tabular}}          & \begin{tabular}[c]{@{}c@{}}34.61/\\ .946\end{tabular}          & \multicolumn{1}{c|}{{\color[HTML]{6AA84F} \begin{tabular}[c]{@{}c@{}}34.43/ \\ .945\end{tabular}}} & \multicolumn{1}{c|}{{\color[HTML]{6AA84F} \begin{tabular}[c]{@{}c@{}}34.58/ \\ .946\end{tabular}}} & {\color[HTML]{333333} \begin{tabular}[c]{@{}c@{}}34.82/ \\ .948\end{tabular}} \\ \hline
CC5                             & \multicolumn{1}{c|}{\begin{tabular}[c]{@{}c@{}}30.73/\\ .888\end{tabular}}          & \begin{tabular}[c]{@{}c@{}}31.60/\\ .903\end{tabular}          & \multicolumn{1}{c|}{\begin{tabular}[c]{@{}c@{}}31.83/\\ .909\end{tabular}}          & \begin{tabular}[c]{@{}c@{}}32.23/\\ .915\end{tabular}          & \multicolumn{1}{c|}{\begin{tabular}[c]{@{}c@{}}32.01/\\ .910\end{tabular}}                         & \multicolumn{1}{c|}{{\color[HTML]{6AA84F} \begin{tabular}[c]{@{}c@{}}32.16/\\ .913\end{tabular}}}  & \begin{tabular}[c]{@{}c@{}}32.45/\\ .917\end{tabular}                         \\ \hline
CC10                            & \multicolumn{1}{c|}{\begin{tabular}[c]{@{}c@{}}24.98/\\ .727\end{tabular}}          & \begin{tabular}[c]{@{}c@{}}25.54/\\ .749\end{tabular}          & \multicolumn{1}{c|}{\begin{tabular}[c]{@{}c@{}}25.93/\\ .765\end{tabular}}          & \begin{tabular}[c]{@{}c@{}}26.31/\\ .778\end{tabular}          & \multicolumn{1}{c|}{\begin{tabular}[c]{@{}c@{}}26.12/\\ .770\end{tabular}}                         & \multicolumn{1}{c|}{{\color[HTML]{6AA84F} \begin{tabular}[c]{@{}c@{}}26.26/ \\ .775\end{tabular}}} & \begin{tabular}[c]{@{}c@{}}26.49/ \\ .783\end{tabular}                        \\ \hline
CR3.3                           & \multicolumn{1}{c|}{\begin{tabular}[c]{@{}c@{}}37.47/\\ .961\end{tabular}}          & \begin{tabular}[c]{@{}c@{}}38.31/\\ .966\end{tabular}          & \multicolumn{1}{c|}{\begin{tabular}[c]{@{}c@{}}38.22/\\ .967\end{tabular}}          & \begin{tabular}[c]{@{}c@{}}38.90/\\ .971\end{tabular}          & \multicolumn{1}{c|}{{\color[HTML]{6AA84F} \begin{tabular}[c]{@{}c@{}}38.71/\\ .970\end{tabular}}}  & \multicolumn{1}{c|}{{\color[HTML]{6AA84F} \begin{tabular}[c]{@{}c@{}}38.90/ \\ .971\end{tabular}}} & \begin{tabular}[c]{@{}c@{}}39.18/ \\ .973\end{tabular}                        \\ \hline
CR5                             & \multicolumn{1}{c|}{\begin{tabular}[c]{@{}c@{}}33.86/\\ .923\end{tabular}}          & \begin{tabular}[c]{@{}c@{}}34.66/\\ .933\end{tabular}          & \multicolumn{1}{c|}{\begin{tabular}[c]{@{}c@{}}34.72/\\ .936\end{tabular}}          & \begin{tabular}[c]{@{}c@{}}35.14/\\ .941\end{tabular}          & \multicolumn{1}{c|}{{\color[HTML]{6AA84F} \begin{tabular}[c]{@{}c@{}}34.92/\\ .939\end{tabular}}}  & \multicolumn{1}{c|}{{\color[HTML]{6AA84F} \begin{tabular}[c]{@{}c@{}}35.07/\\ .941\end{tabular}}}  & \begin{tabular}[c]{@{}c@{}}35.34/\\ .944\end{tabular}                         \\ \hline
CR10                            & \multicolumn{1}{c|}{\begin{tabular}[c]{@{}c@{}}27.95/\\ .794\end{tabular}}          & \begin{tabular}[c]{@{}c@{}}28.64/\\ .816\end{tabular}          & \multicolumn{1}{c|}{\begin{tabular}[c]{@{}c@{}}29.02/\\ .832\end{tabular}}          & \begin{tabular}[c]{@{}c@{}}29.43/\\ .842\end{tabular}          & \multicolumn{1}{c|}{\begin{tabular}[c]{@{}c@{}}29.15/\\ .834\end{tabular}}                         & \multicolumn{1}{c|}{{\color[HTML]{6AA84F} \begin{tabular}[c]{@{}c@{}}29.33/\\ .839\end{tabular}}}  & \begin{tabular}[c]{@{}c@{}}29.57/\\ .846\end{tabular}                         \\ \hline
CG3.3                           & \multicolumn{1}{c|}{\begin{tabular}[c]{@{}c@{}}39.16/\\ .975\end{tabular}}          & \begin{tabular}[c]{@{}c@{}}39.94/\\ .978\end{tabular}          & \multicolumn{1}{c|}{\begin{tabular}[c]{@{}c@{}}39.85/\\ .978\end{tabular}}          & \begin{tabular}[c]{@{}c@{}}40.34/\\ .980\end{tabular}          & \multicolumn{1}{c|}{{\color[HTML]{6AA84F} \begin{tabular}[c]{@{}c@{}}40.29/\\ .980\end{tabular}}}  & \multicolumn{1}{c|}{{\color[HTML]{6AA84F} \begin{tabular}[c]{@{}c@{}}40.42/ \\ .981\end{tabular}}} & \begin{tabular}[c]{@{}c@{}}40.62/\\ .982\end{tabular}                         \\ \hline
CG5                             & \multicolumn{1}{c|}{\begin{tabular}[c]{@{}c@{}}35.89/\\ .951\end{tabular}}          & \begin{tabular}[c]{@{}c@{}}36.63/\\ .957\end{tabular}          & \multicolumn{1}{c|}{\begin{tabular}[c]{@{}c@{}}36.72/\\ .958\end{tabular}}          & \begin{tabular}[c]{@{}c@{}}37.27/\\ .962\end{tabular}          & \multicolumn{1}{c|}{{\color[HTML]{6AA84F} \begin{tabular}[c]{@{}c@{}}37.05/\\ .961\end{tabular}}}  & \multicolumn{1}{c|}{{\color[HTML]{6AA84F} \begin{tabular}[c]{@{}c@{}}37.21/\\ .962\end{tabular}}}  & \begin{tabular}[c]{@{}c@{}}37.46/\\ .963\end{tabular}                         \\ \hline
CG10                            & \multicolumn{1}{c|}{\begin{tabular}[c]{@{}c@{}}29.20/\\ .845\end{tabular}}          & \begin{tabular}[c]{@{}c@{}}30.21/\\ .868\end{tabular}          & \multicolumn{1}{c|}{\begin{tabular}[c]{@{}c@{}}30.73/\\ .880\end{tabular}}          & \begin{tabular}[c]{@{}c@{}}31.19/\\ .888\end{tabular}          & \multicolumn{1}{c|}{\begin{tabular}[c]{@{}c@{}}30.89/\\ .880\end{tabular}}                         & \multicolumn{1}{c|}{{\color[HTML]{6AA84F} \begin{tabular}[c]{@{}c@{}}31.03/ \\ .884\end{tabular}}} & \begin{tabular}[c]{@{}c@{}}31.37/\\ .890\end{tabular}                         \\ \hline
\end{tabular}
\end{table}

\subsubsection{Fine-tuning Vs. Transfer learning with Context-specific networks}
\label{sec:transfer}

\begin{figure}[t!]
    \centering
    \includegraphics[width=0.9\linewidth]{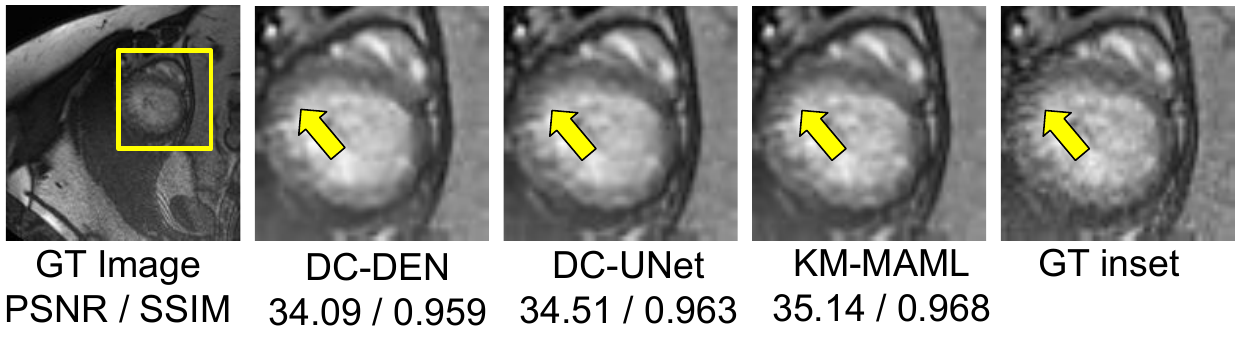}
    \caption{Qualitative comparison of KM-MAML with DC-DEN and DC-UNet when adapting from knee and brain to cardiac contexts with 5x acceleration. The yellow arrows pointing at the ventricle regions show that KM-MAML is able to recover fine structures and textures with less blur than other methods.
    }
    \label{fig:transfer}
\end{figure}

In this experiment, we have taken the models trained on multiple anatomies and contrasts (proton density knee $320\times320$ and T1 brain $240\times240$), mask patterns, and acceleration factors (as shown in Section \ref{sec:mulanatomy}) as the pre-trained models. The models are fine-tuned (one or two epochs i.e. visiting the samples once or a maximum of two times) and transfer learned (ten epochs) to unseen contexts covering cardiac anatomy with Cartesian, radial (unseen mask pattern) and Gaussian mask patterns with acceleration factors 3.3x, 5x, and 10x and image resolution level of $150\times150$. 
From the quantitative results shown in Table \ref{tab:transfer}, our observations are as follows. 1) The performance of KM-MAML after adaptation is, in general, better than the other two methods in both fine-tuning (Epoch 1 and 2) and transfer learning cases (Epoch 10) for all mask patterns and acceleration factors. 2) When the fine-tuning (Epoch 1 and Epoch 2) performance of KM-MAML is compared with DC-UNet, and DC-DEN after transfer learning (Epoch 10), we note that KM-MAML is competitive with DC-UNet models for 5 out of 8 cases at epoch 1 and almost all the cases at epoch 2. These cases are highlighted in green in Table \ref{tab:transfer}. 3) KM-MAML fine-tuning shows significant improvement over DC-DEN after transfer learning. These observations reveal superior meta-learning capabilities of KM-MAML with context-specific learning using kernel modulation as compared to other models. The qualitative results in Figure \ref{fig:transfer} comparing the transfer learning performance of DC-DEN and DC-UNet with the Epoch 2 fine-tuning performance of KM-MAML show that our model has the capabilities to generalize via faster adaptation to unseen data domains. The visual quality of KM-MAML prediction is much closer to the target with respect to the recovered textures and patterns as compared to DC-DEN and DC-UNet.

\subsubsection{Comparative Study using large scale clinical data}
\begin{figure}[t!]
    \centering
    \includegraphics[width=0.9\linewidth]{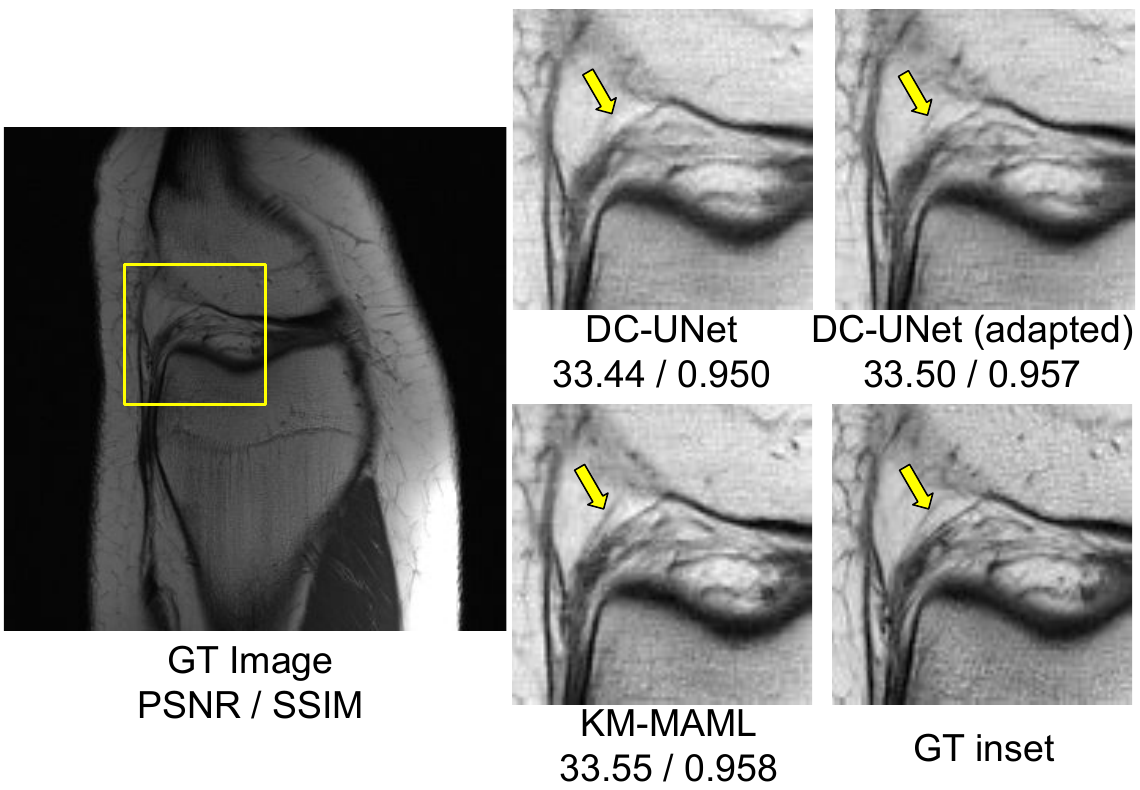}
    \caption{Qualitative comparison of KM-MAML with DC-UNet when trained on large scale data (fastMRI) and adapting to few samples of images under-sampled on unseen mask patterns and image resolution levels of knee MRI.
    }
    \label{fig:transferfastmri}
\end{figure}

We demonstrate our comparative study of the transfer learning performance of DC-UNet and the on-the-fly adapted performance of KM-MAML on the fastMRI dataset \cite{fastmri}, a large-scale collection of clinical MR images. The dataset consists of coronal PD and PDFS images with varying Cartesian mask patterns for 4x and 8x accelerations. The validation set consists of mask patterns and varying image resolution levels different from the training dataset. The quantitative results for DC-UNet are PSNR / SSIM: 31.11 / 0.852, and DC-UNet + transfer learned to 10 epochs are, PSNR / SSIM: 31.14 / 0.854. The quantitative results for KM-MAML in the one-the-fly adaptation setting (i.e. without fine-tuning) are, PSNR / SSIM: 31.16 / 0.855.
The qualitative results are shown in Figure \ref{fig:transferfastmri}. 
The results demonstrate the consistency in generalization capabilities of dynamic weight prediction and kernel modulation of the base network across varying mask patterns and resolution levels when trained on large-scale training data.

\section{Summary and Conclusion}
\label{sec:concl}
In this work, we adaptively learn the structure of multimodal data distributions using KM-MAML, a learning model that combines the strengths of model-based evolutionary deep learning and optimization-based meta-learning. The hypernetworks output low-rank approximation weights to modulate the base network based on different modalities of the acquired image data. The training objectives include learning the inductive bias of multiple modes of heterogeneous data and low-level image features. To achieve the multi-objective training, our model optimally utilizes the hypernetworks via kernel modulation and gradient-based meta-learning (Figure \ref{fig:graphabsmaml} and \ref{fig:arch}). 

We have demonstrated the efficacy of our method for MC-MRI reconstruction considering the two essential fastMRI research directions: multimodal and transfer learning across MRI scanners. We considered benchmark MC-MRI datasets, such as MRBrainS, SRI24 Atlas, and IXI datasets, to compare the scalability of the learning methods and the cardiac ACDC dataset, and fastMRI knee datasets to compare the reconstruction performance against various acquisition context-specific MRI reconstruction networks.
Scalability to numerous multimodal acquisition settings for clinical use is facilitated through hypernetworks (Figure \ref{fig:expr1_boxplot}) that impart the necessary context-aware bias into base CNN and is further enhanced by discriminatively fine-tuning to the shifted target distribution. Representation similarity analysis provides insights into mode-specific knowledge in the high-resolution layers of the base network (Figure \ref{fig:cka}). 
Our proposed KM-based architecture outperforms various context-specific MRI reconstruction architectures (Tables \ref{tab:sotasc} and \ref{tab:mulanatomy}) and the adaptive MRI reconstruction architecture, MAC-ReconNet in the multiple context-based setting quantitatively and qualitatively.

Our future research directions include analyzing the model in terms of better context embedding, extending the work for multi-coil reconstruction with self-supervision, other regression tasks like image imputation in MRI, and contrast augmentation techniques to improve the generalization capabilities of the model.

\section{Acknowledgement}

This work is supported by IITM Pravartak Technologies Foundation and R Jhunjhunwala Foundation. 

\appendix


 \bibliographystyle{elsarticle-num} 
 \bibliography{cas-refs}





\section{Appendix}
\label{sec:appendix}

\subsection{Data Fidelity Operation in MRI reconstruction}
\label{appendix:datafidelity}
\begin{figure}[t!]
    \centering
    \includegraphics[width=0.5\linewidth]{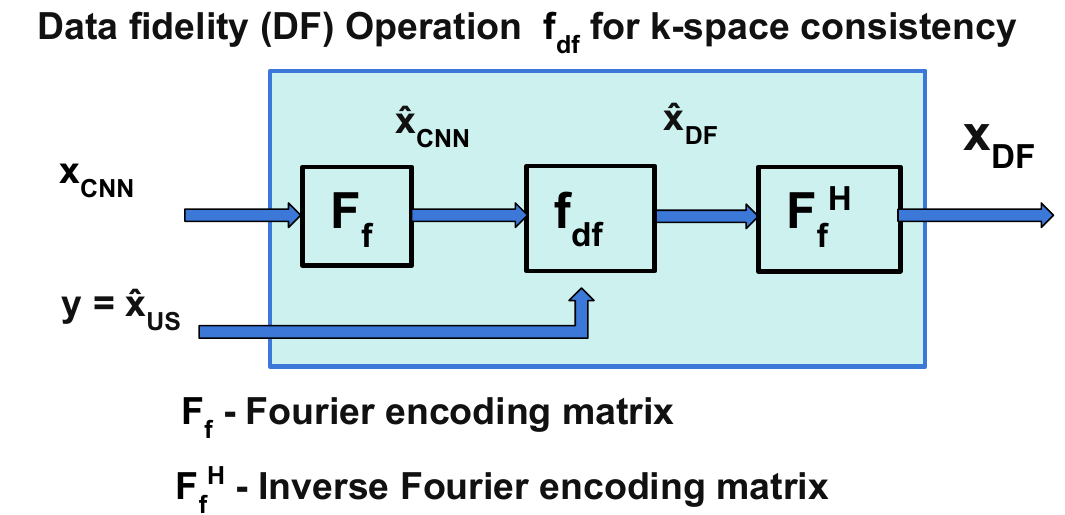}
    \caption{K-space Data fidelity for MRI reconstruction}
    \label{fig:df}
\end{figure}

The proposed architecture of KM-MAML has a data fidelity (DF) block \cite{dc_cnn} in k-space domain after the CNN base network to ensure that the CNN reconstruction is consistent with the acquired k-space measurements. The data fidelity operation $f_{df}$ can be expressed as,
  \begin{equation}
    \hat{x}_{df}=
    \begin{cases}
      \hat{x}_{CNN}(k)  & \ k\notin\Omega \\
      \frac{\hat{x}_{CNN}(k) + \lambda \hat{x}_{u}(k)}{1+\lambda} & k\in\Omega \\
    \end{cases}
  \end{equation}
Here, $\hat{x}_{CNN} = F_{f} x_{CNN}$, $\hat{x}_{u}= F_{f} x_{u}$, $\Omega$ is the index set of sampled k-space data, $F_{f}$ is the Fourier encoding matrix,  and $\hat{x}_{df}$ is the corrected k-space and the data fidelity weight $\lambda\to\infty$. The reconstructed image is obtained by inverse Fourier encoding of $\hat{x}_{df}$, i.e. ${x}_{df} = F_{f}^{H}\hat{x}_{df}$ (Figure \ref{fig:df}).

\subsection{Glossary of Technical Terms}

\textbf{Meta-initialization or meta-parameters $\theta$} - globally shared initialization point of parameters (weights) of a task-oriented base neural network trained on various tasks (for example, different classes in classification tasks) using gradient-based meta-learning, such that a few gradient steps from the initialization parameters can generalize to new related tasks (new classes in classification tasks).

\textbf{Task-specific parameters $\phi_{M}$} - In meta-learning, task-specific parameters refer to the parameters of a neural network that are obtained after rapidly adapting the network to a specific task M encountered during the inner loop of the meta-testing phase. These parameters are different from the meta-initialization parameters that are optimized in the outer loop across tasks \cite{maml}.

\textbf{Multimodal MRI data} - Multimodal data refers to the data collected through different acquisition technologies. The output of each acquisition technology is represented as a mode in the form of a dataset associated with a medium of expression, such as vision, audio or text. Multimodal MRI data refers to the use of multiple imaging techniques within a single MRI examination to gather complementary information about the structure and function of interest.

\textbf{Hypernetworks} - Hypernetworks are used to adaptively generate weights to initialize or update the parameters of another network, called a base or backbone network \cite{hypernetworks}. Hypernetworks form the basis for model-based meta-learning \cite{metal_survey}.

\textbf{Inductive bias} - Inductive bias refers to the prior knowledge or assumptions that a learning algorithm uses to make predictions or generalize from training data to unseen data. The meta-learning process learns the inductive bias in the form of the meta-initializations, which allow the model to generalize by adapting quickly to new tasks with limited data \cite{cavia}.

\textbf{Mode-specific inductive bias} -  Meta-initializations closer to the given target data mode in multimodal data.

\textbf{Rank-1 kernel modulation} - Element-wise multiplication of the base network kernels with the predicted weights of the hypernetwork. The predicted weights are two vectors corresponding to the number of input channels $C_{in}$ and output channels $C_{out}$ of the corresponding base network layer. For example, if the base network layer has $C_{out} \times C_{in}$ kernels each of size $k \times k$, then the hypernetwork predicts two vectors of sizes $C_{out} \times 1$ and $1 \times C_{in}$, such that the outer product of these two vectors gives a matrix of size $C_{out} \times C_{in}$. The hypernetwork predicts $C_{out} + C_{in}$ weights rather than $C_{out} \times C_{in}$ weights to modulate each kernel of the convolution layer differently.

\begin{table}[t]
\centering
\scriptsize
\caption{Quantitative comparison of SGD, MAML, MMAML and KM-MAML on 24 training tasks combining four MRI contrasts - T1, FLAIR, T2 and PD, two under-sampling mask types - Cartesian and Gaussian and three acceleration factors 4x, 5x and 8x. Tasks are encoded in contrast type - mask type - acceleration factor value. FL denotes FLAIR MRI contrast}
\label{tab:suppl_table1}
\begin{tabular}{|c|c|c|c|c|}
\hline
                                & \textbf{SGD}  & \textbf{MAML}  & \textbf{MMAML} & \textbf{KM-MAML}                              \\ \cline{2-5} 
\multirow{-2}{*}{\textbf{Task}} & PSNR / SSIM   & PSNR / SSIM    & PSNR / SSIM    & PSNR / SSIM                                   \\ \hline
T1C4                            & 36.14 / 0.891 & 36.45 / 0.900  & 36.45  / 0.903 & \textbf{36.75 / 0.915}                        \\ \hline
T1C5                            & 35.03 / 0.866 & 35.22 /  0.875 & 35.14 / 0.876  & \textbf{35.58 / 0.895}                        \\ \hline
T1C8                            & 32.69 / 0.822 & 32.99 /  0.839 & 32.96 / 0.842  & \textbf{33.37 / 0.864}                        \\ \hline
T1G4                            & 40.07 / 0.913 & 40.59 / 0.922  & 40.71 / 0.927  & {\color[HTML]{000000} \textbf{41.02 / 0.935}} \\ \hline
T1G5                            & 38.18 / 0.877 & 38.67 / 0.887  & 38.70 / 0.894  & {\color[HTML]{000000} \textbf{39.05 / 0.906}} \\ \hline
T1G8                            & 35.03 / 0.805 & 35.33 / 0.820  & 35.51 / 0.841  & {\color[HTML]{000000} \textbf{35.96 / 0.859}} \\ \hline
FLC4                            & 34.12 / 0.879 & \textbf{34.40} / 0.885  & 34.23 / 0.882  & {\color[HTML]{000000} \textbf{34.40 / 0.889}} \\ \hline
FLC5                            & 32.87 / 0.849 & \textbf{33.18} / 0.860  & 32.96 / 0.854  & {\color[HTML]{000000} 33.16 / \textbf{0.864}} \\ \hline
FLC8                            & 30.83 / 0.813 & 31.02 / 0.828  & 30.89 / 0.824  & {\color[HTML]{000000} \textbf{31.20 / 0.841}} \\ \hline
FLG4                            & 39.25 / 0.911 & \textbf{39.42} / 0.907  & 39.23 / 0.908  & {\color[HTML]{000000} 39.30 / \textbf{0.911}} \\ \hline
FLG5                            & 36.95 / 0.875 & \textbf{37.19} / 0.873  & 36.97 / 0.876  & {\color[HTML]{000000} \textbf{37.15 / 0.881}} \\ \hline
FLG8                            & 33.40 / 0.802 & 33.73 / 0.811  & 33.56 / 0.820  & {\color[HTML]{000000} \textbf{33.93 / 0.832}} \\ \hline
PDC4                            & 32.76 / 0.877 & 33.06 / 0.882  & 32.96 / 0.879  & {\color[HTML]{000000} \textbf{33.24 / 0.888}} \\ \hline
PDC5                            & 31.17 / 0.835 & 31.60 / \textbf{0.845}  & 31.39 / 0.840  & \textbf{31.66 / 0.845}                                 \\ \hline
PDC8                            & 29.07 / 0.803 & 29.48 / 0.814  & 29.32 / 0.804  & {\color[HTML]{000000} \textbf{29.64 / 0.819}} \\ \hline
PDG4                            & 37.78 / 0.908 & 37.98 / 0.911  & 37.81 / 0.908  & {\color[HTML]{000000} \textbf{38.03 / 0.914}} \\ \hline
PDG5                            & 35.14 / 0.857 & \textbf{35.39 / 0.863}  & 35.21 / 0.857  & 35.37 / \textbf{0.863}                                 \\ \hline
PDG8                            & 31.27 / 0.783 & \textbf{31.72 / 0.791}  & 31.51 / 0.780  & 31.69 / 0.789                                 \\ \hline
T2C4                            & 32.01 / 0.873 & 32.43 / 0.880  & 32.26 / 0.875  & {\color[HTML]{000000} \textbf{32.55 / 0.884}} \\ \hline
T2C5                            & 30.58 / 0.825 & 30.96 / 0.834  & 30.74 / 0.825  & 31.07 / \textbf{0.839}                                 \\ \hline
T2C8                            & 28.24 / 0.787 & 28.56 / 0.797  & 28.45 /  0.791 & {\color[HTML]{000000} \textbf{28.71 / 0.803}} \\ \hline
T2G4                            & 37.27 / 0.910 & 37.48 / 0.912  & 37.21 / 0.906  & {\color[HTML]{000000} \textbf{37.59 / 0.916}} \\ \hline
T2G5                            & 34.83 / 0.860 & 35.06 / 0.863  & 34.82 / 0.855  & {\color[HTML]{000000} \textbf{35.24 / 0.870}} \\ \hline
T2G8                            & 31.06 / 0.779 & 31.50 / 0.788  & 31.22 / 0.774  & {\color[HTML]{000000} \textbf{31.61 / 0.795}} \\ \hline
\end{tabular}
\end{table}

\begin{table}[t!]
\centering
\scriptsize
\caption{Quantitative comparison of on-the-fly adaptation performance of SGD, MAML, MMAML, and KM-MAML to 24 unseen tasks with deviated acquisition settings}
\label{tab:suppl_table2}
\begin{tabular}{|c|c|c|c|c|}
\hline
                                & \textbf{SGD}  & \textbf{MAML}                         & \textbf{MMAML}                        & \textbf{KM-MAML}                                                      \\ \cline{2-5} 
\multirow{-2}{*}{\textbf{Task}} & PSNR / SSIM   & PSNR / SSIM                           & PSNR / SSIM                           & PSNR / SSIM                                                           \\ \hline
T1C6                            & 31.33 / 0.837 & 31.20 / 0.837                         & 31.23 / 0.832                         & {\color[HTML]{333333} \textbf{31.42 / 0.842}} \\ \hline
T1C7                            & 32.65 / 0.846 & 32.60 / 0.849                         & 32.50 / 0.843                         & {\color[HTML]{333333} \textbf{32.79 / 0.851}} \\ \hline
T1C8                            & 30.63 / 0.833 & 30.61 / 0.834                         & 30.60 / 0.830                         & {\color[HTML]{333333} \textbf{30.87 / 0.842}} \\ \hline
T1C9                            & 30.07 / 0.824 & 30.16 / 0.826                         & 30.08 / 0.822                         & {\color[HTML]{333333} \textbf{30.32 / 0.834}} \\ \hline
T1G6                            & \textbf{34.72} / 0.841 & 34.57 / 0.837                         & 34.34 / 0.832                         & {\color[HTML]{000000} 34.70 / \textbf{0.842}}         \\ \hline
T1G7                            & \textbf{32.44} / 0.794 & 32.37 / 0.792                         & 32.12 / 0.788                         & {\color[HTML]{000000} 32.41 / \textbf{0.797}}          \\ \hline
T1G8                            & 31.69 / 0.785 & 31.68 / 0.786                         & 31.28 / 0.774                         & {\color[HTML]{333333} \textbf{31.69 / 0.791}} \\ \hline
T1C9                            & 30.98 / 0.779 & 30.96 / 0.780 & 30.68 / 0.773                         & {\color[HTML]{333333} \textbf{30.89 / 0.785}}                         \\ \hline
T2C6                            & 32.00 / 0.872 & 31.94 / 0.871                         & 31.97 / 0.875 & {\color[HTML]{333333} \textbf{32.16 / 0.877}} \\ \hline
T2C7                            & 31.68 / 0.848 & 31.68 / 0.850                         & \textbf{31.83 / 0.857} & {\color[HTML]{000000} 31.73 / \textbf{0.857}}          \\ \hline
T2C8                            & 30.63 / 0.833 & 30.67 / 0.833                         & \textbf{30.88 / 0.844} & {\color[HTML]{000000} 30.76 / 0.842}          \\ \hline
T2C9                            & 30.47 / 0.854 & 30.46 / 0.855                         & \textbf{30.66 / 0.862}                         & {\color[HTML]{000000} 30.6 /  0.860}                                  \\ \hline
T2G6                            & 35.20 / 0.866 & 34.72 / 0.855                         & 35.10 / 0.869 & {\color[HTML]{333333} \textbf{35.27 / 0.870}} \\ \hline
T2G7                            & 33.98 / 0.851 & 33.50 / 0.839                         & 33.82 / 0.855 & {\color[HTML]{333333} \textbf{34.18 / 0.855}} \\ \hline
T2G8                            & 33.37 / 0.837 & 33.00 / 0.827                         & 33.23 / 0.838 & {\color[HTML]{333333} \textbf{33.57 / 0.845}} \\ \hline
T2G9                            & 32.31 / 0.819 & 31.78 / 0.804                         & 32.06 / 0.820 & {\color[HTML]{333333} \textbf{32.41 / 0.823}} \\ \hline
PDC6                            & 31.70 / 0.862 & 31.77 / 0.864                         & 31.70 / 0.867 & {\color[HTML]{333333} \textbf{31.95 / 0.871}} \\ \hline
PDC7                            & 30.45 / 0.843 & 30.83 / 0.852                         & 30.86 / 0.855                         & {\color[HTML]{333333} \textbf{31.11 / 0.861}} \\ \hline
PDC8                            & 31.30 / 0.853 & 31.43 / 0.857                         & 31.53 / 0.862                         & {\color[HTML]{333333} \textbf{31.85 / 0.868}} \\ \hline
PDC9                            & 28.67 / 0.806 & 28.93 / 0.814                         & 28.81 / 0.812                         & {\color[HTML]{333333} \textbf{29.20 / 0.827}} \\ \hline
PDG6                            & 33.80 / 0.851 & 33.51 / 0.846                         & 33.77 / 0.854                         & {\color[HTML]{333333} \textbf{34.23 / 0.864}} \\ \hline
PDG7                            & 32.98 / 0.838 & 32.71 / 0.831                         & 32.72 / 0.836                         & {\color[HTML]{333333} \textbf{33.26 / 0.851}} \\ \hline
PDG8                            & 32.51 / 0.831 & 32.27 / 0.828                         & 32.52 / 0.833                         & {\color[HTML]{333333} \textbf{32.97 / 0.845}} \\ \hline
PDG9                            & 31.46 / 0.803 & 31.26 / 0.802                         & 31.28 / 0.805                         & {\color[HTML]{333333} \textbf{31.88 / 0.824}} \\ \hline
\end{tabular}
\end{table}

\begin{table}[]
\scriptsize
\centering
\caption{Quantitative comparison of SGD, MAML, MMAML, and KM-MAML in fine-tuning to 10 gradient steps from the meta-initializations to unseen contrasts.}
\label{tab:suppl_adapt_finetune}
\begin{tabular}{|c|cc|cc|cc|}
\hline
                       & \multicolumn{2}{c|}{}                                       & \multicolumn{2}{c|}{\textbf{adapt base n/w}}                                                                                       & \multicolumn{2}{c|}{\textbf{adapt modulation n/w}}           \\ \cline{2-7} 
                       & \multicolumn{1}{c|}{\textbf{SGD}}  & \textbf{MAML}          & \multicolumn{1}{c|}{\textbf{MMAML}}                        & \textbf{KM-MAML}                                                      & \multicolumn{1}{c|}{\textbf{MMAML}} & \textbf{KM-MAML}       \\ \cline{2-7} 
\multirow{-3}{*}{Task} & \multicolumn{1}{c|}{PSNR / SSIM}   & PSNR / SSIM            & \multicolumn{1}{c|}{PSNR / SSIM}                           & PSNR / SSIM                                                           & \multicolumn{1}{c|}{PSNR / SSIM}    & PSNR / SSIM            \\ \hline
T1C4                   & \multicolumn{1}{c|}{34.91 / 0.872} & \textbf{34.94 / 0.875} & \multicolumn{1}{c|}{34.73 / 0.869}                         & 34.90 / 0.874                                                         & \multicolumn{1}{c|}{34.71 / 0.869}  & 34.90 / 0.874          \\ \hline
T1C5                   & \multicolumn{1}{c|}{33.77 / 0.872} & \textbf{33.94 / 0.876} & \multicolumn{1}{c|}{33.48 / 0.867}                         & 33.62 / 0.874                                                         & \multicolumn{1}{c|}{33.44 / 0.866}  & 33.61 / 0.873          \\ \hline
T1C6                   & \multicolumn{1}{c|}{31.33 / 0.837} & 31.21 / 0.837          & \multicolumn{1}{c|}{31.30 / 0.837} & {\color[HTML]{333333} \textbf{31.42 / 0.844}} & \multicolumn{1}{c|}{31.25 / 0.835}  & \textbf{31.41 / 0.844} \\ \hline
T1C7                   & \multicolumn{1}{c|}{32.67 / 0.847} & 32.62 / \textbf{0.850}          & \multicolumn{1}{c|}{32.47 / 0.843}                         & {\color[HTML]{000000} \textbf{32.75 / 0.850}}                         & \multicolumn{1}{c|}{32.39 / 0.840}  & \textbf{32.73} / 0.849 \\ \hline
T1C8                   & \multicolumn{1}{c|}{30.64 / 0.834} & 30.62 / 0.835          & \multicolumn{1}{c|}{30.62 / 0.832}                         & {\color[HTML]{000000} \textbf{30.82 / 0.842}}                         & \multicolumn{1}{c|}{30.62 / 0.832}  & \textbf{30.82 / 0.842} \\ \hline
T1C9                   & \multicolumn{1}{c|}{30.08 / 0.824} & 30.18 / 0.827          & \multicolumn{1}{c|}{30.05 / 0.821}                         & {\color[HTML]{000000} \textbf{30.24 / 0.832}}                         & \multicolumn{1}{c|}{30.05 / 0.821}  & \textbf{30.24 / 0.832} \\ \hline
T1G4                   & \multicolumn{1}{c|}{37.87 / 0.892} & 37.82 / 0.893          & \multicolumn{1}{c|}{37.90 / \textbf{0.897}}                         & {\color[HTML]{000000} \textbf{37.91} / 0.895}                         & \multicolumn{1}{c|}{37.84 / 0.895}  & 37.88 / 0.895 \\ \hline
T1G5                   & \multicolumn{1}{c|}{35.88 / 0.852} & 35.86 / 0.854          & \multicolumn{1}{c|}{35.75 / 0.850}                         & {\color[HTML]{000000} \textbf{35.92 / 0.856}}                         & \multicolumn{1}{c|}{35.75 / 0.852}  & \textbf{35.92 / 0.856} \\ \hline
T1G6                   & \multicolumn{1}{c|}{34.72 / 0.841} & 34.57 / 0.837          & \multicolumn{1}{c|}{34.41 / 0.834}                         & {\color[HTML]{000000} \textbf{34.69 / 0.842}}                         & \multicolumn{1}{c|}{34.26 / 0.831}  & \textbf{34.65 / 0.841} \\ \hline
T1G7                   & \multicolumn{1}{c|}{32.45 / 0.794} & 32.38 / 0.792          & \multicolumn{1}{c|}{32.15 / 0.789}                         & {\color[HTML]{000000} \textbf{32.38 / 0.796}}                         & \multicolumn{1}{c|}{32.11 / 0.788}  & \textbf{32.37 / 0.796} \\ \hline
T1G8                   & \multicolumn{1}{c|}{31.72 / 0.786} & \textbf{31.72 / 0.788} & \multicolumn{1}{c|}{31.24 / 0.772}                         & 31.56 / 0.786                                                         & \multicolumn{1}{c|}{31.18 / 0.770}  & 31.56 / 0.785          \\ \hline
T1G9                   & \multicolumn{1}{c|}{\textbf{30.99} / 0.779} & \textbf{30.99} / 0.781          & \multicolumn{1}{c|}{30.83 / 0.780}                         & {\color[HTML]{000000} 30.87 / \textbf{0.785}}                         & \multicolumn{1}{c|}{30.70 / 0.775}  & 30.84 / \textbf{0.784} \\ \hline
PDC4                   & \multicolumn{1}{c|}{33.64 / 0.897} & 33.71 / 0.899          & \multicolumn{1}{c|}{33.68 / 0.900}                         & {\color[HTML]{000000} \textbf{33.98 / 0.904}}                         & \multicolumn{1}{c|}{33.66 / 0.899}  & \textbf{33.98 / 0.904} \\ \hline
PDC5                   & \multicolumn{1}{c|}{32.22 / 0.869} & 32.54 / 0.876          & \multicolumn{1}{c|}{32.39 / 0.875}                         & {\color[HTML]{000000} \textbf{32.71 / 0.882}}                         & \multicolumn{1}{c|}{32.37 / 0.875}  & \textbf{32.70 / 0.882} \\ \hline
PDC6                   & \multicolumn{1}{c|}{31.71 / 0.862} & 31.77 / 0.864          & \multicolumn{1}{c|}{31.76 / 0.868}                         & {\color[HTML]{000000} \textbf{32.00 / 0.872}}                         & \multicolumn{1}{c|}{31.74 / 0.867}  & \textbf{31.99 / 0.872} \\ \hline
PDC7                   & \multicolumn{1}{c|}{30.49 / 0.844} & 30.85 / 0.852          & \multicolumn{1}{c|}{30.90 / 0.856}                         & {\color[HTML]{000000} \textbf{31.15 / 0.861}}                         & \multicolumn{1}{c|}{30.89 / 0.856}  & \textbf{31.15 / 0.861} \\ \hline
PDC8                   & \multicolumn{1}{c|}{31.30 / 0.853} & 31.43 / 0.857          & \multicolumn{1}{c|}{31.57 / 0.863}                         & {\color[HTML]{000000} \textbf{31.89 / 0.868}}                         & \multicolumn{1}{c|}{31.54 / 0.863}  & \textbf{31.88 / 0.868} \\ \hline
PDC9                   & \multicolumn{1}{c|}{28.72 / 0.807} & 28.95 / 0.815          & \multicolumn{1}{c|}{28.86 / 0.813}                         & {\color[HTML]{000000} \textbf{29.20 / 0.827}}                         & \multicolumn{1}{c|}{28.83 / 0.812}  & \textbf{29.20 / 0.826} \\ \hline
PDG4                   & \multicolumn{1}{c|}{38.74 / 0.925} & 38.91 / 0.928          & \multicolumn{1}{c|}{39.06 / 0.931}                         & {\color[HTML]{000000} \textbf{39.34 / 0.935}}                         & \multicolumn{1}{c|}{38.93 / 0.929}  & \textbf{39.30 / 0.934} \\ \hline
PDG5                   & \multicolumn{1}{c|}{36.46 / 0.894} & 36.35 / 0.895          & \multicolumn{1}{c|}{36.50 / 0.900}                         & {\color[HTML]{000000} \textbf{36.97 / 0.905}}                         & \multicolumn{1}{c|}{36.50 / 0.900}  & \textbf{36.97 / 0.905} \\ \hline
PDG6                   & \multicolumn{1}{c|}{33.84 / 0.852} & 33.57 / 0.848          & \multicolumn{1}{c|}{33.87 / 0.858}                         & {\color[HTML]{000000} \textbf{34.26 / 0.865}}                         & \multicolumn{1}{c|}{33.81 / 0.856}  & \textbf{34.25 / 0.864} \\ \hline
PDG7                   & \multicolumn{1}{c|}{32.99 / 0.838} & 32.72 / 0.832          & \multicolumn{1}{c|}{32.85 / 0.840}                         & {\color[HTML]{000000} \textbf{33.29 / 0.852}}                         & \multicolumn{1}{c|}{32.72 / 0.836}  & \textbf{33.27 / 0.851} \\ \hline
PDG8                   & \multicolumn{1}{c|}{32.53 / 0.832} & 32.30 / 0.829          & \multicolumn{1}{c|}{32.48 / 0.832}                         & {\color[HTML]{000000} \textbf{32.98 / 0.846}}                         & \multicolumn{1}{c|}{32.48 / 0.832}  & \textbf{32.96 / 0.845} \\ \hline
PDG9                   & \multicolumn{1}{c|}{31.50 / 0.804} & 31.32 / 0.804          & \multicolumn{1}{c|}{31.38 / 0.809}                         & {\color[HTML]{000000} \textbf{31.91 / 0.824}}                         & \multicolumn{1}{c|}{31.30 / 0.805}  & \textbf{31.90 / 0.824} \\ \hline
\end{tabular}
\end{table}

\begin{table}[t!]
\scriptsize
\centering
\caption{Quantitative comparison of MAML, MMAML and KM-MAML for five-layer CNN. Meta-training is done on two contrasts (two modes) consisting of T1 and FLAIR contrasts.}
\label{tab:suppl_5layercnn}
\begin{tabular}{|c|c|c|c|}
\hline
                                 & \textbf{MAML}        & \textbf{MMAML}       & \textbf{KM-MAML}                      \\ \cline{2-4} 
\multirow{-2}{*}{\textbf{Tasks}} & \textbf{PSNR / SSIM} & \textbf{PSNR / SSIM} & \textbf{PSNR / SSIM}                  \\ \hline
T1C4                             & 35.97 / 0.9132       & 35.98 / 0.9115       & {\color[HTML]{000000} \textbf{36.38 / 0.9189}} \\ \hline
T1C5                             & 33.81 / 0.8569       & 33.91 / 0.8568       & {\color[HTML]{000000} \textbf{34.16 / 0.8660}} \\ \hline
T1C8                             & 31.86 / 0.8248       & 32.33 / 0.8402       & {\color[HTML]{000000} \textbf{32.38 / 0.8417}} \\ \hline
T1G4                             & 40.16 / 0.9357       & 40.80 / 0.9441       & {\color[HTML]{000000} \textbf{40.93 / 0.9473}} \\ \hline
T1G5                             & 37.96 / 0.8967       & 38.21 / 0.8964       & {\color[HTML]{000000} \textbf{38.30 / 0.9024}} \\ \hline
T1G8                             & 33.92 / 0.8232       & 34.24 / 0.8330       & {\color[HTML]{000000} \textbf{34.56 / 0.8380}} \\ \hline
FLC4                             & 33.32 / 0.8827       & 33.35 / 0.8829       & {\color[HTML]{000000} \textbf{33.46 / 0.8832}} \\ \hline
FLC5                             & 31.15 / 0.8213       & 31.23 / 0.8230       & {\color[HTML]{000000} \textbf{31.30 / 0.8253}} \\ \hline
FLC8                             & 29.21 / 0.7947       & 29.69 / \textbf{0.8051}       & {\color[HTML]{000000} \textbf{29.74} / 0.8004} \\ \hline
FLG4                             & 37.97 / 0.8972       & 38.50 / 0.9129       & {\color[HTML]{000000} \textbf{38.79 / 0.9158}} \\ \hline
FLG5                             & 35.77 / 0.8611       & 35.62 / 0.8607       & {\color[HTML]{000000} \textbf{35.80 / 0.8643}} \\ \hline
FLG8                             & 31.85 / 0.7875       & 31.82 / 0.7869       & \textbf{32.20 / 0.7914}                        \\ \hline
\end{tabular}
\end{table}

\begin{table}[]
\scriptsize
\centering
\caption{Quantitative comparison of SGD, MAML, MMAML, and KM-MAML for DenseNet. Meta-training is done on four contrasts  consisting of T1, FLAIR, T2, and PD contrasts.}
\begin{tabular}{|c|c|c|c|c|}
\hline
\multirow{2}{*}{Task} & SGD            & MAML           & MMAML          & KM-MAML        \\ \cline{2-5} 
                      & PSNR/SSIM      & PSNR/SSIM      & PSNR/SSIM      & PSNR/SSIM      \\ \hline
T1C5                  & 32.55 / 0.8014 & 32.15 / 0.7868 & 33.15 / 0.8271 & \textbf{33.66 / 0.8384} \\ \hline
T1C8                  & 31.14 / 0.7809 & 30.80 / 0.7594 & 31.25 / 0.7880 & \textbf{31.51 / 0.7975} \\ \hline
T1G5                  & 36.61 / 0.8467 & 35.98 / 0.8279 & 37.51 / \textbf{0.8771} & \textbf{37.80} / 0.8763 \\ \hline
T1G8                  & 32.59 / 0.7560 & 32.09 / 0.7316 & 33.35 / 0.7899 & \textbf{33.93 / 0.8091} \\ \hline
FLC5                  & 30.80 / 0.7947 & 31.12 / 0.8073 & 31.25 / \textbf{0.8199} & \textbf{31.33} / 0.8149 \\ \hline
FLC8                  & 28.99 / 0.7767 & 29.09 / 0.7734 & \textbf{29.26 / 0.7881} & 29.17 / 0.7762 \\ \hline
FLG5                  & 35.64 / 0.8385 & 35.95 / 0.8488 & 36.01 / \textbf{0.8513} & \textbf{36.06} / 0.8483 \\ \hline
FLG8                  & 31.49 / 0.7672 & 31.82 / 0.7757 & 32.14 / 0.7927 & \textbf{32.46 / 0.7949} \\ \hline
PDC5                  & 28.90 / 0.7565 & 29.47 / 0.7865 & 29.34 / 0.7720 & \textbf{30.01 / 0.7898} \\ \hline
PDC8                  & 26.08 / 0.7106 & 26.48 / 0.7276 & 26.47 / 0.7214 & \textbf{27.15 / 0.7371} \\ \hline
PDG5                  & 33.16 / 0.7729 & 33.90 / \textbf{0.8105} & 33.69 / 0.7958 & \textbf{34.00} / 0.7998 \\ \hline
PDG8                  & 28.40 / 0.6787 & 29.08 / 0.7154 & 28.98 / 0.6939 & \textbf{30.05 / 0.7222} \\ \hline
T2C5                  & 28.68 / 0.7595 & 29.08 / 0.7780 & 28.85 / 0.7669 & \textbf{29.30 / 0.7801} \\ \hline
T2C8                  & 26.23 / 0.7124 & 26.47 / 0.7248 & 26.43 / 0.7225 & \textbf{26.73 / 0.7306} \\ \hline
T2G5                  & 33.26 / 0.7883 & 33.80 / 0.8113 & 33.51 / 0.7992 & \textbf{33.86 / 0.8038} \\ \hline
T2G8                  & 28.83 / 0.6983 & 29.38 / \textbf{0.7239} & 29.14 / 0.6986 & \textbf{29.87} / 0.7203 \\ \hline
\end{tabular}
\end{table}

\begin{figure}[b!]
    \centering
    \includegraphics[width=\linewidth]{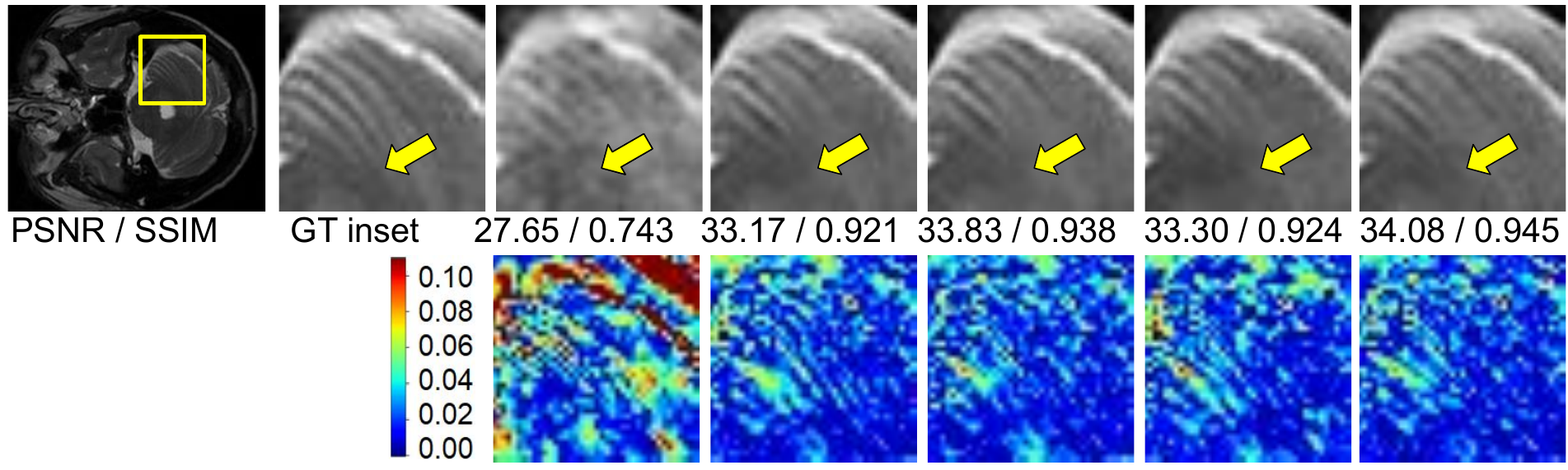}
    \caption{Qualitative Comparison of the reconstruction performance for  T2 MRI. From the left: the ground truth (GT) image with the region of interest (ROI), GT inset, ZF image, joint training, MAML, MMAML, and KM-MAML. The yellow arrows and the residual images indicate that KM-MAML can recover repeating patterns much better than other methods.  
    }
    \label{fig:expr1_visual_t2}
\end{figure}

\begin{figure}[t!]
    \centering
    \includegraphics[width=\linewidth]{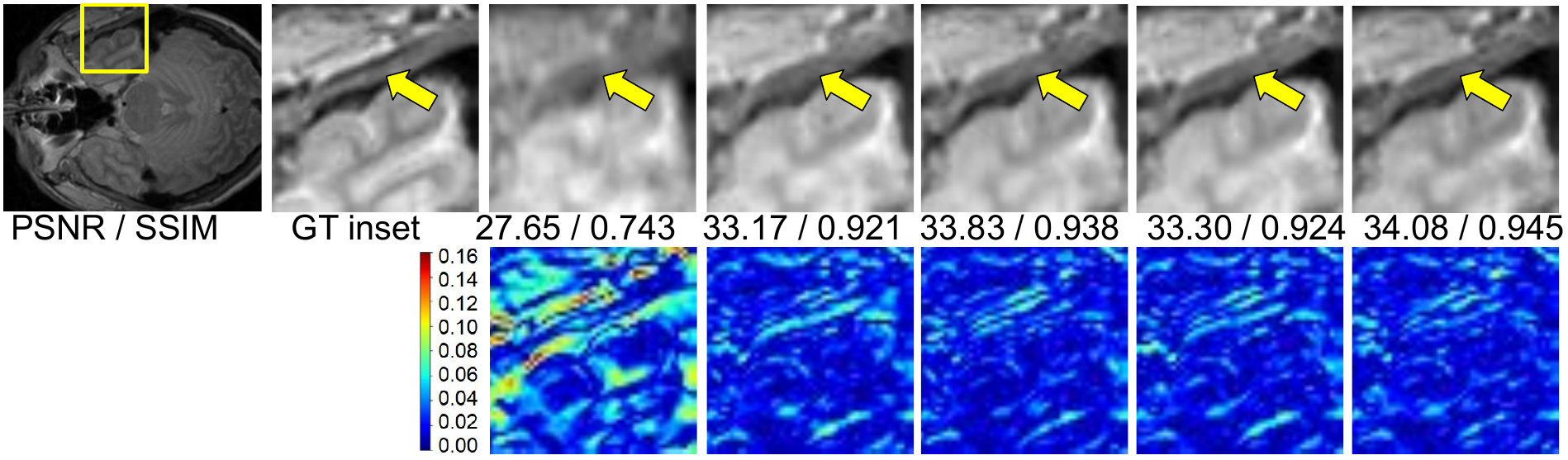}
    \caption{Qualitative Comparison of the reconstruction performance for PD MRI. From the left: the ground truth (GT) image with the region of interest (ROI), GT inset, ZF image, joint training, MAML, MMAML, and KM-MAML. The yellow arrows and the residual images indicate that KM-MAML can recover details better than other  methods.  
    }
    \label{fig:expr1_visual_t2}
\end{figure}

\begin{figure}[t!]
    \centering
    \includegraphics[width=\linewidth]{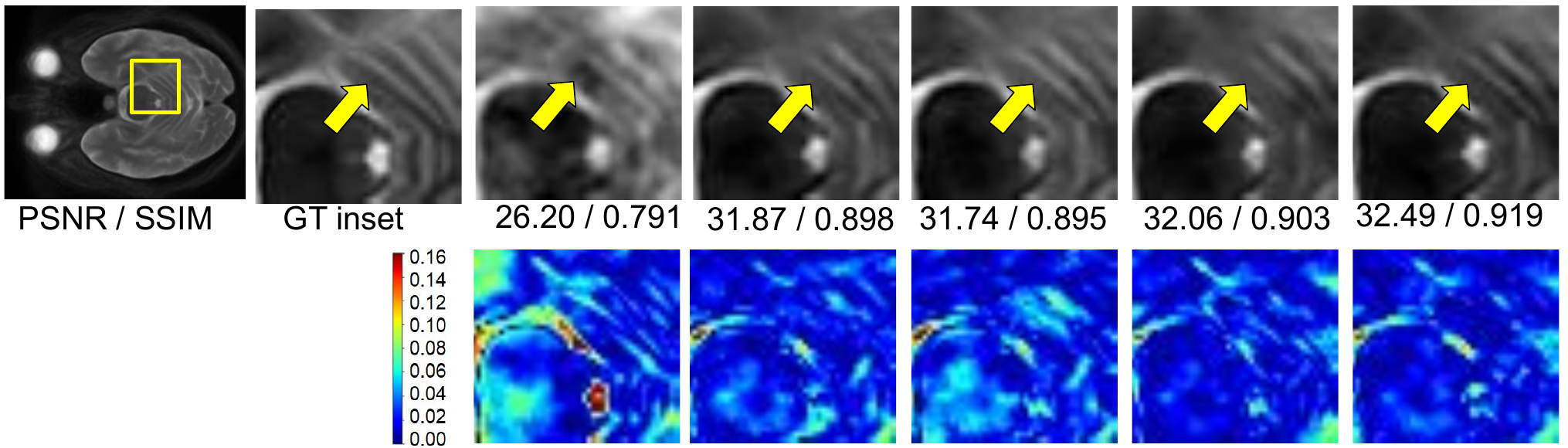}
    \caption{Qualitative Results for T2 MRI comparing the target image, ZF image, joint training, MAML, MMAML, and KM-MAML with respect to the on-the-fly adaptation capabilities of the models to multiple contrasts. The images show that the recovery of repeated image patterns in the predictions of KM-MAML is better than other methods.
    }
    \label{fig:expr2_visual_t2}
\end{figure}

\begin{figure}[!]
    \centering
    \includegraphics[width=\linewidth]{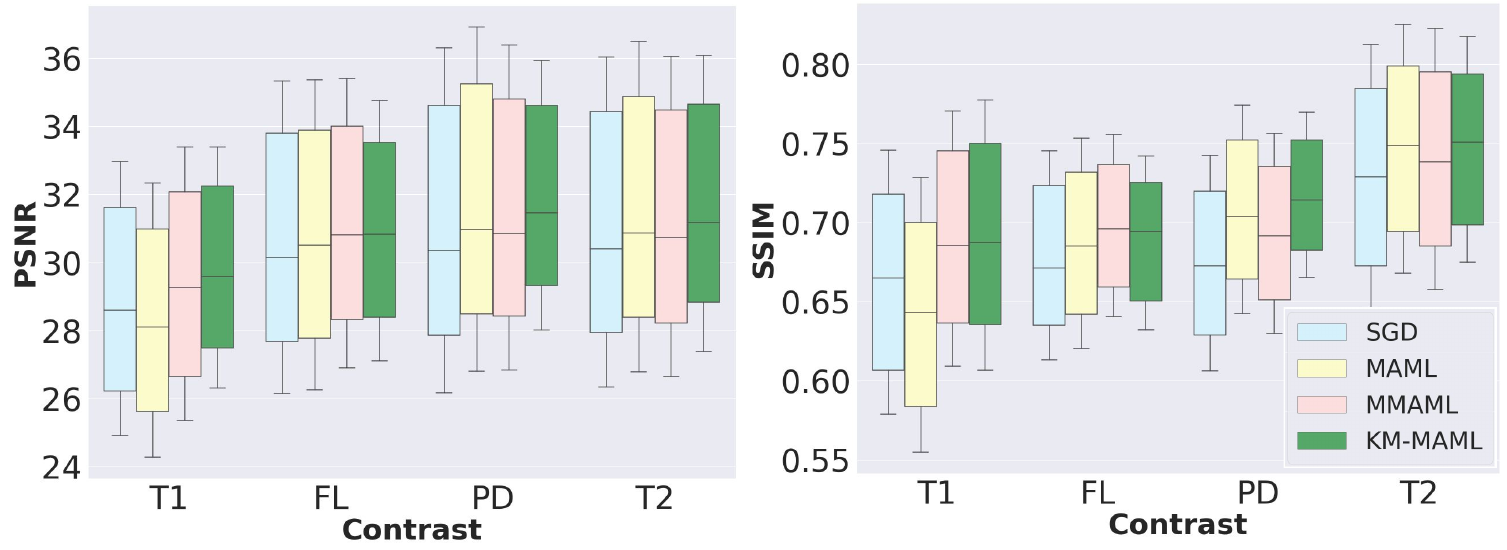}
    \caption{PSNR and SSIM plots comparing joint training, MAML, MMAML, and KM-MAML with DenseNet as the base network for four MRI contrasts - T1, FLAIR, PD, and T2. FL denotes FLAIR
    }
    \label{fig:den}
\end{figure}


\begin{algorithm}[t!]
\caption{KM-MAML Fine-tuning the base network for test-time tasks}
\label{kMAMLalgoadapt}
\begin{algorithmic}[1]
\Require Learning rate $\eta_{2}$ and \( p(\mathcal{T}) \): Multimodal test task distribution
\Require \( CE(.) \): Context encoder
\Require Meta-trained model,    $\omega$: modulation network weights and $\theta$: base network weights

\State Sample a test task mini-batch: $\mathcal{T}^{test}_{batch}$ $\sim$ \( p(\mathcal{T}) \)
\For{each test task $M$ in \( \mathcal{T}^{test}_{batch} \)}
\State Sample a mini-batch of test support data: $D^{M, test}_{spt}=\{ x_{US,n}^{M, test},x_{FS,n} \}_{n=1}^{N^{test}_{spt}}$
\For{$u = 0 \ to \ U-1$}
\State Context embedding: $\gamma$ = $CE(x_{US}^{M, test})$
\Comment{Support data mini-batch 

\hskip18.7eminput
}
\State $\alpha$, $\beta$ = KM Hypernetwork($\gamma; \omega$)
\State $W_{M, test}$ = $\beta \otimes_{outer} \alpha$
\Comment{Modulation weights of task M
}
\State $\theta_{mod}$ = $\theta \odot W_{M, test}$
\Comment{Mode-specific initialization (KM)
}
\State Initialize $\phi_{0}^{M, test} \leftarrow \theta_{mod}$

\State $\mathcal{L}_{1} = L([\phi_{u}^{M, test}, \omega], D^{M, test}_{spt})$
\State $\phi_{u+1}^{M, test} \leftarrow \phi_{u}^{M, test}-\eta_{2} \nabla_{\phi_{u}^{M, test}}\mathcal{L}_{1}$ \Comment{mode-specific weight updates 
}

\EndFor
\EndFor
\end{algorithmic}
\end{algorithm}

\begin{figure}[]
    \centering
    \includegraphics[width=0.7\linewidth]{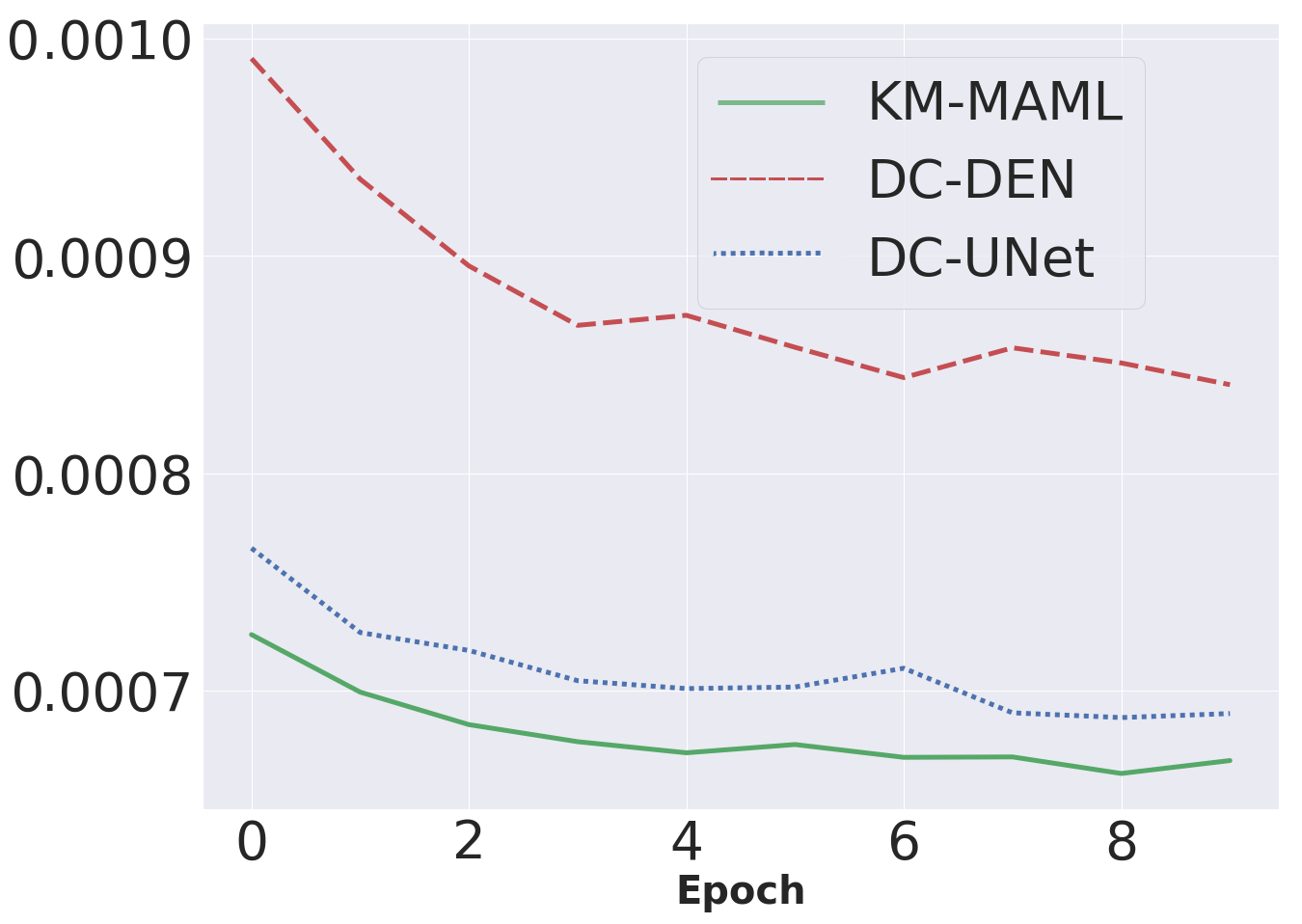}
    \caption{Validation Mean Square Error (MSE) plots comparing the transfer learning performance of DC-DEN, DC-UNet, with the fine-tuning performance of KM-MAML. Plots show that the validation error values at Epochs 1 and 2 of KM-MAML are comparable with Epochs 8 to 10 of DC-UNet.
    }
    \label{fig:valid}
\end{figure}

\begin{table}[t!]
\scriptsize
\centering
\caption{List of factors influencing the computation overhead in the training process. For meta-learning methods, the task mini-batch size is set to 3 out of a total of 24 tasks.
NA stands for 'Not Applicable' in SGD due to single-loop optimization. 'ms' and 'sec' denote milliseconds and seconds, respectively. For SGD, there is only a single loop for processing the mini-batch of samples. For the meta-learning methods, MAML, MMAML, and KM-MAML, there is a task mini-batch and for each task, there are inner and outer loop iterations. Also, we see that KM-MAML is slightly faster than MMAML in terms of epoch time, as the hypernetworks are meta-learning instead of the base network}
\label{tab:computationoverheadfactors}
\begin{tabular}{|c|c|c|c|c|}
\hline
\textbf{Overhead type}                                                                                                     & \textbf{SGD}                                                 & \textbf{MAML}                                                                                                             & \textbf{MMAML}                                                                                                           & \textbf{KM-MAML}                                                                                                          \\ \hline
\begin{tabular}[c]{@{}c@{}}Inner loop \\ optimization\end{tabular}                                                         & NA                                                           & Yes                                                                                                                       & Yes                                                                                                                      & Yes                                                                                                                       \\ \hline
\begin{tabular}[c]{@{}c@{}}Outer loop \\ optimization\end{tabular}                                                         & NA                                                           & Yes                                                                                                                       & Yes                                                                                                                      & Yes                                                                                                                       \\ \hline
\begin{tabular}[c]{@{}c@{}}Outer product\\ computation\\ (Rank 1 \\ approximation) \\ of predicted \\ weights\end{tabular} & No                                                           & No                                                                                                                        & No                                                                                                                       & Yes                                                                                                                       \\ \hline
\begin{tabular}[c]{@{}c@{}}Kernel \\ modulation \\ (model-based \\ meta-learning)?\end{tabular}                            & No                                                           & No                                                                                                                        & Yes                                                                                                                      & Yes                                                                                                                       \\ \hline
\begin{tabular}[c]{@{}c@{}}Task mini \\ batch?\end{tabular}                                                                & NA                                                           & Yes                                                                                                                       & Yes                                                                                                                      & Yes                                                                                                                       \\ \hline
\begin{tabular}[c]{@{}c@{}}No. of \\ optimization loops\end{tabular}                                                       & 1                                                            & 2                                                                                                                         & 2                                                                                                                        & 2                                                                                                                         \\ \hline
\begin{tabular}[c]{@{}c@{}}Which network \\ weights are \\ optimized in \\ two loops?\end{tabular}                         & \begin{tabular}[c]{@{}c@{}}Base \\ network\end{tabular}      & \begin{tabular}[c]{@{}c@{}}Base\\ network\end{tabular}                                                                    & \begin{tabular}[c]{@{}c@{}}Base \\ network\end{tabular}                                                                  & Hypernetworks                                                                                                             \\ \hline
\begin{tabular}[c]{@{}c@{}}Time taken \\ per iteration\\ (approximately)\end{tabular}                                      & \begin{tabular}[c]{@{}c@{}}5ms per \\ iteration\end{tabular} & \begin{tabular}[c]{@{}c@{}}Inner loop:\\ 4ms per \\ task\\ \\ Outer loop:\\ 1.8 sec\\ per task \\ mini-batch\end{tabular} & \begin{tabular}[c]{@{}c@{}}Inner loop:\\ 7 ms per\\ task\\ \\ Outer loop:\\ 3 sec \\ per task \\ mini-batch\end{tabular} & \begin{tabular}[c]{@{}c@{}}Inner loop:\\ 8 ms per \\ task\\ \\ Outer loop:\\ 2.2 sec\\ per task\\ mini-batch\end{tabular} \\ \hline
\begin{tabular}[c]{@{}c@{}}Time taken \\ per epoch in sec\\ (approaximately)\end{tabular}                                  & 32                                                           & 39                                                                                                                        & 59                                                                                                                       & 47                                                                                                                        \\ \hline
\end{tabular}
\end{table}



\end{document}